\documentclass[11pt]{article}
\usepackage{mathrsfs}
\usepackage{amsfonts}
\usepackage{amsmath,amsthm,hyperref}
\usepackage{amssymb}
\usepackage{hyperref,color}
\usepackage{exscale,relsize}
\def\cur{\mathcal{K}_n}

\numberwithin{equation}{section}
\begin{document}

% define the title
\author{ Peng Zhao   \and  Engui Fan\footnote{Corresponding
author and  e-mail address:
      faneg@fudan.edu.cn} \and  Yu Hou}
\date{   \small{ School of Mathematical Sciences, Institute of Mathematics \\
 and Key Laboratory of Mathematics for Nonlinear Science, \\ Fudan
University, Shanghai 200433, P.R. China}}
\title{\bf \Large{Algebro-Geometric Solutions and Their Reductions
for the Fokas-Lenells Hierarchy} }
\maketitle
\begin{abstract}
This paper is dedicated to provide theta function representations
of algebro-geometric solutions for the Fokas-Lenells (FL) hierarchy
through studying an algebro-geometric initial value problem. Further,
we reduce these solutions into $n$-dark solutions through the degeneration
of associated Riemann surfaces.

\end{abstract}
\section{Introduction}
In the past few decades, the celebrated nonlinear Schr$\ddot{o}$dinger (NLS)
equation has been widely studied in various of aspects.
In ref. \cite{1}, Fokas proposed
an integrable generalization of the NLS equation,
\begin{equation}\label{1.2}
\begin{split}
  &iu_t-\nu u_{tx}+\gamma u_{xx}+\sigma|u|^2(u+i\nu u_x)=0,\,\, \sigma=\pm 1,\,\, x,\in\mathbb{R},t>0\\ &~~~~~~~~~~~~~~~~~~~~~~~~~~~~~~~~~~~~~~~~~~~~~~~~~~~~~~~\nu,\gamma,\rho\equiv \textrm{constant}\in\mathbb{R},
\end{split}
\end{equation}
which is known as Fokas equation later,
%where $\nu,\gamma,\rho$ are real parameters and $u=u(x,t)$
%is a complex-valued function,
 %(or more strictly speaking, Forkas equation,
%(see (\ref{1.2}))
using bi-Hamiltionian methods.
Just like the bi-Hamiltionian structure of the well-known Korteweg-de Vries equation can
be perturbed to yield the integrable Camassa-Holm equation, the same mathematical trick
applied to the two Hamiltionian operators associated with the NLS equation yields the Forkas equation.
Under the simple transformation
$$u\rightarrow \beta\sqrt{\alpha}e^{i\beta x}u,\,\alpha=\gamma/\nu, \,\beta= 1/\nu,\,\sigma=-\sigma,$$
the Fokas equation (\ref{1.2})
changes to
the Forkas-Lenells (FL) equation
\begin{equation}\label{1.3}
  u_{tx}+\alpha\beta^2u-2i\alpha\beta u_x-\alpha u_{xx}+\sigma i\alpha\beta^2|u|^2u_x=0,\,\,\sigma=\pm 1.
\end{equation}
In the context of nonlinear optics, the FL equation models the propagation of nonlinear
light pulses in monomode optical fibers when certain higher-oder nonlinear effects are taken into
account \cite{2}. In contrast to the case of the NLS equation which comes in a focusing as well as in
a defocusing version depending on the values of the parameters, and solitons only
exist in the focusing regime, all versions of equation (\ref{1.3}) are mathematically
equivalent up to a change of variables.
The transformation
\begin{align*}
&
u\rightarrow \sqrt{a}be^{i(bx+2abt)}u,\,\, a=\gamma/\nu>0,\\
&
\xi=x+at,\,\,\tau=-ab^2t.
\end{align*}
transforms the Forkas equation (\ref{1.2}) into
\begin{equation}\label{1.4}
  u_{\tau \xi}=u-i\sigma|u|^2u_{\xi},\,\,\sigma=\pm 1,
\end{equation}
the so-called Forkas-Lenells derivative
nonlinear Schr$\ddot{o}$dinger equation (FDNS) in some references \cite{9,8}.
An important
feature of equation (\ref{1.4}) is that
it describes the first negative flow of the integrable hierarchy associated with the
derivative nonlinear Schr$\ddot{o}$dinger (DNLS) equation \cite{1,5,6,2}.

In this paper,
we start from the following coupled
form
\begin{equation}\label{1.1}\begin{cases}
\begin{split}
q_{xt}-q_{xx}+iqq_xr-2i q_x+q=0,\\
r_{xt}-r_{xx}-iqrr_x+2i r_x+r=0,
\end{split}
\end{cases}
\end{equation}
which are exactly reduced to the FL equation ($\alpha=\beta=1$ in (\ref{1.3}))
\begin{equation}\label{1.1a}
  q_{xt}-q_{xx}\mp i|q|^2q_x-2i q_x+q=0,
\end{equation}
for $r=\pm \overline{q}.$
Related results can also be directly applied to (\ref{1.2}) and (\ref{1.4}) since
the existence of these simple transformations among them.
It is shown that (\ref{1.1})/(\ref{1.1a}) is a completely integrable nonlinear partial differential equation possessing
Lax pair, bi-Hamiltonian structure, and soliton solutions \cite{1,2,3,4}. One of the most remarkable feature of the FL equation is that it possesses various kinds
of exact solutions such as solitons, breathers, etc.. The bright solitons
under vanishing boundary condition
have been constructed
by inverse scattering transform (IST) method \cite{3}, dressing method \cite{4}
and Hirota method \cite{7}.
The lattice representation and the $n$-dark solitons of the FL equation have
been presented in \cite{8}, where a relationship is also established between the FL equation and other
integrable models including the NLS equation, the Merola-Ragnisco-Tu equations and the Ablowitz-Ladik equation. In \cite{9}, the author has dealt with a sophisticated problem on the dark
soliton solutions with a plane wave boundary condition using Hirota method.
The breather
solutions of the FL equation have also been constructed via a dressing-B$\ddot{a}$cklund transformation related
to the Riemann-Hilbert problem formulation of the inverse scattering theory \cite{10}.
Recently, the authors of \cite{11a} has investigated $n$-order rogue waves solutions of FL equation using
Darboux transformation method.

The algebro-geometric solution, parameterized
by compact Riemann surface of finite genus, is a kind of important solutions
in soliton theory.
This kind of solutions was originally studied on the KdV
equation based on the inverse spectral theory, developed by pioneers such as the authors in \cite{01,16,19,16a,18,17,11}
and further developed
by the authors in \cite{20,12,21,22}, etc.
In a degenerated case of the
algebro-geometric solution, the multi-soliton solution and periodic solution
in elliptic function type may be obtained \cite{16}.

The purpose of this paper is to analyze the quasi-periodic solutions and
dark soliton solutions of the FL hierarchy using the algebro-geometric method \cite{15}.
This systematic approach, proposed by Gesztesy and Holden to construct
algebro-geometric solutions for integrable equations, has been extended to
the whole (1+1) dimensional integrable hierarchy, such as the AKNS hierarchy,
the CH hierarchy etc. Recently, we investigated algebro-geometric solutions for the
the Degasperis-Procesi hierarchy and
Hunter-Saxton hierarchy \cite{ii,ij} using this method.

%We investigate the
%Choosing quasi-periodic function (cf. (\ref{solutionq}), (\ref{solutionr}))
 %\begin{align*}
 %       &q(x)=q(x_0)\frac{\theta(\underline{\xi}(P_{0,-},\hat{\underline{\mu}}(x_0)))}
 %      {\theta(\underline{\xi}(P_{0,-},\hat{\underline{\mu}}(x)))}
  %     \frac{\theta(\underline{\xi}(P_{0,+},\hat{\underline{\mu}}(x)))}
  %     {\theta(\underline{\xi}(P_{0,+},\hat{\underline{\mu}}(x_0)))}e^{i(x-x_0)(e_{0,-}-e_{0,+})},
  %    \\
  %      &r(x)=r(x_0)\frac{\theta(\underline{\xi}(P_{0,-},\hat{\underline{\nu}}(x)))}
  %     {\theta(\underline{\xi}(P_{0,-},\hat{\underline{\nu}}(x_0)))}
  %     \frac{\theta(\underline{\xi}(P_{0,+},\hat{\underline{\nu}}(x_0)))}
  %     {\theta(\underline{\xi}(P_{0,+},\hat{\underline{\nu}}(x)))}e^{-i(x-x_0)(e_{0,-}-e_{0,+})},
 %\end{align*}
%as the initial problem
%by has been extended to the whole hierarchies of nonlinear
%integrable equations by Gesztesy and Holden.
 In the present paper, we consider a Cauchy problem (\ref{4.1}), (\ref{4.2}) of FL hierarchy
with a quasi-periodic initial condition $q,r$ (cf. (\ref{solutionq}), (\ref{solutionr}))
and search for its exact solutions. We will prove the solution of
this cauchy problem is unique (cf. Lemma 4.3) and give the explicit form of $q,r$ (cf. Theorem 4.6).
We also find that the quasi-periodic solutions obtained in Theorem 4.6
can be linked with the dark solitons
of FL hierarchy. Especially,
for the FL equation (\ref{1.1})/(\ref{1.1a}),
the results of \cite{8} about the $n$-dark solitons can be obtained from a different standpoint.
As shown in \cite{7,9}, the bright solitons and dark solitons correspond to
the vanishing boundary condition and non-vanishing boundary condition, respectively.
Hence the authors are confident that there exists another kind of quasi-periodic solutions
which may degenerate to the bright solitons. Obviously, this depends on
what kinds of Cauchy problem we will investigate.

%It is worth notice that our results in the present paper can not applied to analyze the
%$n$-bright soliton of the FL equation since these two kinds of soliton
%belong to different boundary conditions and its may
%be discussed in detail elsewhere
%if necessary.

%(i)

This paper is organized as follows. In section 2, we construct the FL hierarchy using
a zero-curvature approach and a polynomial recursion formalism.
 %As a byproduct, we gives the conservation law of this newly established hierarchy.
 Moreover, the
hyperelliptic curve $\cur$ of genus $n$ associated with the FL zero-curvature pairs is introduced
with
the help of the characteristic polynomial of Lax matrix $V_{\underline{n}}$ for the
stationary FL hierarchy.
In section 3, we treat the stationary FL hierarchy and its quasi-periodic solutions.
Using these stationary quasi-periodic solutions as initial values,
we solve the Cauchy problem and obtain
the quasi-periodic solutions of FL hierarchy In section 4. In section 5,
we consider the soliton limit of these quasi-periodic solutions given in section 4
and finally derive the $n$-dark solitons of the FL hierarchy.

\section{The Fokas-Lenells Hierarchy, Recursion Relations, and Hyperelliptic Curves}
In this section, we provide the construction of FL hierarchy and derive the
corresponding
sequence of zero-curvature pairs using a polynomial recursion formalism.
 %As a byproduct, we obtain the first conservation law of the FL hierarchy.
Moreover, we introduce the underlying hyperelliptic curve in connection with
the stationary FL hierarchy.

Throughout this section, we make the following hypothesis.
\newtheorem{hyp1}{Hypothesis}[section]
 \begin{hyp1}
      In the stationary case we assume that
  \begin{equation}\label{2.1}
    \begin{split}
    & q, r\in C^\infty(\mathbb{R}),\ \
     q(x),q_x(x),r_x(x)\neq 0,\,x\in\mathbb{R}.\\
    \end{split}
  \end{equation}
    In the time-dependent case we suppose
  \begin{equation}
    \begin{split}\label{2.2}
    & q(\cdot,t), r(\cdot,t)\in C^\infty (\mathbb{R}), t\in\mathbb{R}, \ \
    q(x,\cdot), r(x,\cdot)\in C^1(\mathbb{R}), x\in\mathbb{R},
      \\
    & q(x,t),q_x(x,t),r_x(x,t)\neq 0, (x,t)\in\mathbb{R}^2.\\
  \end{split}
  \end{equation}
\end{hyp1}
We first introduce the basic polynomial recursion formalism. Define $\{f_{\ell,\pm}\}_{\ell\in\mathbb{N}_0},$ $\{g_{\ell,\pm}\}_{\ell\in\mathbb{N}_0}$ and
$\{h_{\ell,\pm}\}_{\ell\in\mathbb{N}_0}$ recursively by
\begin{align}
g_{0,+}&=-1,\,h_{0,+}=r_x,\, f_{0,+}=-q_x,\label{2.3}\\
   ig_{\ell,+,x}&=r_{x}f_{\ell,+}+q_{x}h_{\ell,+},\,\ell\in\mathbb{N}_0,\label{2.4}\\
  f_{\ell,+,x}&=2iq_xg_{\ell+1,+}-2if_{\ell+1,+},\,\ell\in\mathbb{N}_0,\label{2.5}\\
  h_{\ell,+,x}&=2ih_{\ell+1,+}+2ir_xg_{\ell+1,+},\,\ell\in\mathbb{N}_0,\label{2.6}
\end{align}
and
\begin{align}
g_{0,-}&=-1/4,\, h_{0,-}=-ir/2,\,f_{0,-}=-iq/2,\label{2.7}\\
ig_{\ell+1,-,x}&=q_xh_{\ell,-}+r_xf_{\ell,-},\,\ell\in\mathbb{N}_0,\label{2.8}\\
f_{\ell+1,-,x}&=2iq_{x}g_{\ell+1,-}-2if_{\ell,-},\,\ell\in\mathbb{N}_0,\label{2.9}\\
h_{\ell+1,-,x}&=2ih_{\ell,-}+2ir_xg_{\ell+1,-},\,\ell\in\mathbb{N}_0, \,i=\sqrt{-1},\label{2.10}
\end{align}
where $f_{\ell,\pm,x}$, $g_{\ell,\pm,x}$ and $h_{\ell,\pm,x},\ell\in\mathbb{N}_0,$ denote
the derivative of $f_{\ell,\pm},$ $g_{\ell,\pm},$ $h_{\ell,\pm}$ with respect to the space variable $x,$ respectively.
Explicitly, one obtains
\begin{equation}\label{2.11}
  \begin{split}
  g_{0,+}&=-1,\\
  g_{1,+}&=-\frac{1}{2}q_xr_x-c_{1,+},\\
  f_{0,+}&=-q_x,\\
  f_{1,+}&=\frac{1}{2i}q_{xx}-\frac{1}{2}q_x^2r_x-c_{1,+}q_x,\\
  h_{0,+}&=r_x,\\
  h_{1,+}&=\frac{1}{2i}r_{xx}+\frac{1}{2}q_xr_{x}^2+c_{1,+}r_x,\\
  g_{0,-}&=-\frac{1}{4},\\
  g_{1,-}&=-\frac{1}{2}qr-\frac{1}{4}c_{1,-},\\
  f_{0,-}&=-\frac{1}{2}iq,\\
  f_{1,-}&=-\int_{x_0}^x\left(q+iqq_xr\right)dx-\frac{1}{2}ic_{1,-}q,\\
  h_{0,-}&=-\frac{1}{2}ir,\\
  h_{1,-}&=\int_{x_0}^{x}\left(r-iqrr_x\right)dx-\frac{1}{2}ic_{1,-}r,\ \ \textrm{etc.}
  \end{split}
\end{equation}
Here $\{c_{\ell,\pm}\}_{\ell\in\mathbb{N}}$ denote summation constants which naturally arise
when solving the differential equations for $g_{\ell,+}, f_{\ell,-}, h_{\ell,-}$
in
(\ref{2.3})-(\ref{2.10}).

We first consider the stationary case.
To construct the stationary Fokas-Lenells hierarchy we introduce the following $2\times 2$ matrix
\begin{equation}\label{2.14}
U(\xi,x)=\left(
  \begin{array}{cc}
    -iz & q_x\xi \\
    r_x\xi & iz \\
  \end{array}
\right),\, z=\xi^2
\end{equation}
and make the ansatz
\begin{equation}\label{2.15}
V_{\underline{n}}(\xi,x)=
  \left(
    \begin{array}{cc}
      iG_{\underline{n}}(\xi,x) & -F_{\underline{n}}(\xi,x) \\
      H_{\underline{n}}(\xi,x) & -iG_{\underline{n}}(\xi,x) \\
    \end{array}
  \right),
  \quad   \underline{n}=(n_-,n_+)\in\mathbb{N}_0^2,
\end{equation}
where $G_{\underline{n}}, F_{\underline{n}}$ and $H_{\underline{n}}$ are chosen as Laurent
polynomials, namely
\begin{equation}\label{2.16}
\begin{split}
 &G_{\underline{n}}(\xi,x)=\sum_{\ell=0}^{n_-}\xi^{-2\ell}g_{n_--\ell,-}(x)+\sum_{\ell=1}^{n_+}
  \xi^{2\ell}g_{n_+-\ell,+}(x),\\
 & F_{\underline{n}}(\xi,x)=\sum_{\ell=0}^{n_-}\xi^{-2\ell+1}f_{n_--\ell,-}(x)+\sum_{\ell=0}^{n_+}
  \xi^{2\ell}f_{n_+-\ell,+}(x),\\
 & H_{\underline{n}}(\xi,x)=\sum_{\ell=0}^{n_-}\xi^{-2\ell+1}h_{n_--\ell,-}(x)+\sum_{\ell=0}^{n_+}
  \xi^{2\ell}h_{n_+-\ell,+}(x),
  \end{split}
\end{equation}
and $g_{\ell,\pm}, f_{\ell,\pm},h_{\ell,\pm},$ are defined by (\ref{2.3})-(\ref{2.10}).
The linear system
\begin{equation}\label{2.17}
  \psi_x=U(\xi,x)\psi,\quad \psi_t=V_{\underline{n}}(\xi,x)\psi,\quad \psi=(\psi_1,\psi_2)^{\textrm{T}}
\end{equation}
yields
the stationary zero-curvature equation
\begin{equation}\label{2.18}
   -V_{\underline{n},x}(\xi,x)+[U(\xi,x),V_{\underline{n}}(\xi,x)]=0.
\end{equation}
Inserting (\ref{2.14}) and (\ref{2.15}) into (\ref{2.18}),
one easily finds
 \begin{align}
   \xi q_x(x) H_{\underline{n}}(\xi,x)+\xi r_x(x) F_{\underline{n}}(\xi,x) &= iG_{\underline{n},x}(\xi,x),\label{2.19}\\
   2izF_{\underline{n}}(\xi,x)-2i\xi q_x(x)G_{\underline{n}}(\xi,x)&= -F_{\underline{n},x}(\xi,x),\label{2.20} \\
   2izH_{\underline{n}}(\xi,x)+2ir_x(x)\xi G_{\underline{n}}(\xi,x) &= H_{\underline{n},x}(\xi,x).\label{2.21}
 \end{align}
 Insertion of (\ref{2.16}) into (\ref{2.19})-(\ref{2.21}) then yields
 %the recursion
 %relation (\ref{2.3})-(\ref{2.10}) for $f_{\ell,+}$ and $h_{\ell,+}$ for $\ell=0,\ldots,n_+$,
 %$f_{\ell,-}$ and $h_{\ell,-}$ for $\ell=0,\ldots,n_-$,
 %For fixed $\underline{n}=(n_+,n_-)\in\mathbb{N}_0^2$ we obtain
 %the recursion (\ref{2.3})-(\ref{2.10})for
 %$g_{\ell,\pm}$ for $\ell=0,\ldots,n_{\pm}-1$ and
 \begin{eqnarray}
   f_{n_+-1,+,x}-2iq_xg_{n_-,-}+2if_{n_--1,-} &=& 0 ,\label{2.22}\\
   -h_{n_+-1,+,x}+2ih_{n_--1,-}+2ir_xg_{n_-,-}&=& 0 .\label{2.23}
 \end{eqnarray}
 Thus, varying $n_{\pm}\in\mathbb{N}_0,$ equations (\ref{2.22}) and (\ref{2.23}) give rise
 to the stationary Fokas-Lenells (FL) hierarchy which we introduce as follows
 \begin{equation}\label{2.24}
 \begin{split}
  \textrm{s-FL}_{\underline{n}}(q,r)=
   \left(
     \begin{array}{c}
       f_{n_+-1,+,x}-2iq_xg_{n_-,-}+2if_{n_--1,-}  \\
       -h_{n_+-1,+,x}+2ih_{n_--1,-}+2ir_xg_{n_-,-}  \\
     \end{array}
   \right)=0,\\
   \underline{n}=(n_-,n_+)\in\mathbb{N}_0^2.
   \end{split}
 \end{equation}
 We record the first few equations in FL hierarchy (\ref{2.24}) explicitly,
 \begin{align}
   \textrm{s-FL}_{(0,0)}(q,r)&=\left(
                                \begin{array}{c}
                                  \frac{1}{2}iq_x \\
                                  -\frac{1}{2}ir_x \\
                                \end{array}
                              \right)=0,\label{2.25}
  \\
    \textrm{s-FL}_{(1,1)}(q,r)&=\left(
                                \begin{array}{c}
                                  -q_{xx}+iqq_xr-2ic_{1,-}q_x+q\\
                                   -r_{xx}-iqrr_x+2ic_{1,-}r_x+r\\
                                \end{array}
                              \right)=0.\ \,\label{2.26}
 \end{align}
 In the special case $c_{1,-}=1$ in (\ref{2.26}), one obtains
 the stationary version of the Fokas-Lenells system
 (\ref{1.1}).

 From (\ref{2.19})-(\ref{2.21}) one infers that
 \begin{equation}\label{2.28}
   \frac{d}{dx}\textrm{det}(V_{\underline{n}}(\xi,x))=\frac{d}{dx}\left(G_{\underline{n}}^2(\xi,x)
   +F_{\underline{n}}(\xi,x)H_{\underline{n}}(\xi,x)\right)=0,
 \end{equation}
 and hence
 \begin{equation}\label{2.29}
  G_{\underline{n}}^2(\xi,x)
   +F_{\underline{n}}(\xi,x)H_{\underline{n}}(\xi,x)=R_{\underline{n}}(\xi),
 \end{equation}
where the Laurent polynomial $R_{\underline{n}}$ is $x$-independent. One may write
$R_{\underline{n}}$ as
\begin{equation}\label{2.30}
  \begin{split}
  R_{\underline{n}}(\xi)=z^{-2n_-}\prod_{m=0}^{2n+1}(\xi-E_m),\ \{E_m\}_{m=0}^{2 n+1}
  \subset\mathbb{C},  \\
  \ n=2n_++2n_--1\in\mathbb{N}_0.
  \end{split}
\end{equation}
Moreover, (\ref{2.29}) also implies
\begin{equation*}
  \lim_{\xi\rightarrow 0}z^{n_-}R_{\underline{n}}(\xi)=\prod_{m=0}^{2n+1}E_m
\end{equation*}
and hence
\begin{equation}\label{2.31}
  \prod_{m=0}^{2n+1}E_m=\frac{1}{16}.
\end{equation}
Relation (\ref{2.29}) allows one to introduce a hyperelliptic curve $\mathcal{K}_n$
of arithmetic genus $n=2n_++2n_--1$ (possibly with a singular affine part), where
\begin{eqnarray}\label{2.32}
\mathcal{K}_n: \mathcal{F}(\xi,y)=y^2-z^{2n_-}R_{\underline{n}}(\xi)=y^2-\prod_{m=0}^{2n+1}(\xi-E_m)=0,\nonumber\\
n=2n_++2n_--1\in\mathbb{N}_0.
\end{eqnarray}

Next we turn to the time-dependent Fokas-Lenells hierarchy. For that purpose
the coefficients $q$ and $r$ are now considered as functions of both the space
and time. For each system in this hierarchy, that is, for each $\underline{n},$
we introduce a deformation (time) parameter $t_{\underline{n}}\in\mathbb{R}$
in $q,r,$ replacing $q(x),r(x)$ by $q(x,t_{\underline{n}}),r(x,t_{\underline{n}}).$
Moreover, the definitions (\ref{2.14}), (\ref{2.15}) and (\ref{2.16}) of $U,V$
and $F_{\underline{n}}, G_{\underline{n}}, H_{\underline{n}},$ respectively, still apply
by adding a parameter $t_{\underline{n}}\in\mathbb{R}$, that is,
\begin{flalign}
&U(\xi,x,t_{\underline{n}})=\left(
  \begin{array}{cc}
    -iz & q_x(x,t_{\underline{n}})\xi \\
    r_x(x,t_{\underline{n}})\xi & iz \\
  \end{array}
\right),\,\, z=\xi^2, \,\,\label{2.14aa}\\
&V_{\underline{n}}(\xi,x,t_{\underline{n}})=
  \left(
    \begin{array}{cc}
      iG_{\underline{n}}(\xi,x,t_{\underline{n}}) & -F_{\underline{n}}(\xi,x,t_{\underline{n}}) \\
      H_{\underline{n}}(\xi,x,t_{\underline{n}}) & -iG_{\underline{n}}(\xi,x,t_{\underline{n}}) \\
    \end{array}
  \right),
  \quad   \underline{n}=(n_-,n_+)\in\mathbb{N}_0^2,\label{2.15aa}\\
 &G_{\underline{n}}(\xi,x,t_{\underline{n}})=\sum_{\ell=0}^{n_-}\xi^{-2\ell}g_{n_--\ell,-}
 (x,t_{\underline{n}})+\sum_{\ell=1}^{n_+}
  \xi^{2\ell}g_{n_+-\ell,+}(x,t_{\underline{n}}),\label{2.15aaa}\\
 & F_{\underline{n}}(\xi,x,t_{\underline{n}})=\sum_{\ell=0}^{n_-}\xi^{-2\ell+1}f_{n_--\ell,-}
 (x,t_{\underline{n}})+\sum_{\ell=0}^{n_+}
  \xi^{2\ell}f_{n_+-\ell,+}(x,t_{\underline{n}}),\\
 & H_{\underline{n}}(\xi,x,t_{\underline{n}})=\sum_{\ell=0}^{n_-}\xi^{-2\ell+1}
 h_{n_--\ell,-}(x,t_{\underline{n}})+\sum_{\ell=0}^{n_+}
  \xi^{2\ell}h_{n_+-\ell,+}(x,t_{\underline{n}})\label{2.16aa}
 \end{flalign}
 with $g_{\ell,\pm}, f_{\ell,\pm},h_{\ell,\pm}$ defined by (\ref{2.3})-(\ref{2.10}).
 Equation (\ref{2.18}) now needs to be changed to
\begin{equation}\label{2.33}
U_{t_{\underline{n}}}(\xi,x,t_{\underline{n}})-V_{\underline{n},x}(\xi,x,t_{\underline{n}})+
[U(\xi,x,t_{\underline{n}}),V_{\underline{n}}(\xi,x,t_{\underline{n}})]=0.
\quad\underline{n}\in\mathbb{N}_0^2.
\end{equation}
Insertion of (\ref{2.3})-(\ref{2.10}), (\ref{2.14aa})-(\ref{2.16aa}) into (\ref{2.33}) then yields
 \begin{align}\label{2.34}
  0 &= U_{t_{\underline{n}}}(\xi,x,t_{\underline{n}})-V_{\underline{n},x}(\xi,x,t_{\underline{n}})+
  [U(\xi,x,t_{\underline{n}}),V_{\underline{n}}(\xi,x,t_{\underline{n}})] \nonumber\\
    &= \left(
         \begin{array}{cc}
         \begin{smallmatrix}
           -iG_{\underline{n},x}+q_x\xi H_{\underline{n}}
           +r_x\xi F_{\underline{n}}
          \end{smallmatrix}
            &
           \begin{smallmatrix}
            q_{xt}\xi+F_{\underline{n},x}+2izF_{\underline{n}}-2iq_x\xi G_{\underline{n}}
           \end{smallmatrix}
            \nonumber\\[0.1cm]
           \begin{smallmatrix}
            r_{xt}\xi-H_{\underline{n},x}+2ir_x\xi G_{\underline{n}}+2izH_{\underline{n}}
            \end{smallmatrix}
            &
            \begin{smallmatrix}
            iG_{\underline{n},x}-q_x\xi H_{\underline{n}}
           -r_x\xi F_{\underline{n}}
           \end{smallmatrix} \\
         \end{array}
       \right)\nonumber\\
       &= \left(
           \begin{array}{cc}
           %  \begin{smallmatrix}
             0
            % \end{smallmatrix}
             &
             \begin{smallmatrix}
            \xi ( q_{xt_{\underline{n}}}+ f_{n_{+}-1,+,x}-2iq_xg_{n_-,-}\\
             +2if_{n_--1,-} )
             \end{smallmatrix}\\[0.1cm]
             \begin{smallmatrix}
            \xi ( r_{xt_{\underline{n}}}-h_{n_+-1,+,x}
             +2ih_{n_--1,-}
             \\
             +2ir_xg_{n_-,-}
             )
             \end{smallmatrix}
              &
             %\begin{smallmatrix}
             0
             %\end{smallmatrix} \\
           \end{array}
         \right).
       \end{align}
  Equation (\ref{2.34}) gives rise to two equivalent forms of (\ref{2.33}),
   \begin{align}
     0 & =-iG_{\underline{n},x}(\xi,x,t_{\underline{n}})+q_x(x,t_{\underline{n}})\xi H_{\underline{n}}(\xi,x,t_{\underline{n}})
           +r_x(x,t_{\underline{n}})\xi F_{\underline{n}}(\xi,x,t_{\underline{n}}), \label{2.35a} \\
     q_{xt_{\underline{n}}}(x,t_{\underline{n}})\xi &= -F_{\underline{n},x}(\xi,x,t_{\underline{n}})-2izF_{\underline{n}}
     (\xi,x,t_{\underline{n}})+2iq_x(x,t_{\underline{n}})\xi G_{\underline{n}}(\xi,x,t_{\underline{n}}), \label{2.35b}\\
      r_{xt_{\underline{n}}}(x,t_{\underline{n}})\xi &=H_{\underline{n},x}(\xi,x,t_{\underline{n}})-2ir_x(x,t_{\underline{n}})\xi G_{\underline{n}}(\xi,x,t_{\underline{n}})-2izH_{\underline{n}}(\xi,x,t_{\underline{n}}), \label{2.35c}
   \end{align}
   and
   \begin{align*}
    &  q_{xt_{\underline{n}}}+ f_{n_{+}-1,+,x}-2iq_xg_{n_-,-}
             +2if_{n_--1,-}=0, \\
       &  r_{xt_{\underline{n}}}-h_{n_+-1,+,x}
             +2ih_{n_--1,-}
            +2ir_xg_{n_-,-}=0.
            \end{align*}
   Varying $\underline{n}\in\mathbb{N}_0^2,$ the collection of evolution equations
   \begin{align}
     \textrm{FL}_{\underline{n}}(q,r)&=
     \left(
       \begin{array}{c}
         q_{xt_{\underline{n}}}+ f_{n_{+}-1,+,x}-2iq_xg_{n_-,-}
             +2if_{n_--1,-} \\
         r_{xt_{\underline{n}}}-h_{n_+-1,+,x}
             +2ih_{n_--1,-}
            +2ir_xg_{n_-,-}
       \end{array}
     \right)=0,\nonumber\\
     &~~~~~~~~~~~~~~~~~~~~~~~~~~~~~~
     ~~\quad t_{\underline{n}}\in\mathbb{R},~\underline{n}=(n_-,n_+)\in\mathbb{N}_0^2,\label{2.36}
   \end{align}
   then defines the time-dependent Fokas-Lenells hierarchy.
   Explicitly,
   \begin{align}
   \textrm{FL}_{(0,0)}(q,r)&=\left(
                                \begin{array}{c}
                                  q_{xt_{(0,0)}}+\frac{1}{2}iq_x \\
                                  r_{xt_{(0,0)}}-\frac{1}{2}ir_x \\
                                \end{array}
                              \right)=0,\label{2.37}
  \\
    \textrm{FL}_{(1,1)}(q,r)&=\left(
                                \begin{array}{c}
                                  q_{xt_{(1,1)}}-q_{xx}+iqq_xr-2ic_{1,-}q_x+q\\
                                   r_{xt_{(1,1)}}-r_{xx}-iqrr_x+2ic_{1,-}r_x+r\\
                                \end{array}
                              \right)=0,\ \,\textrm{etc.,}\label{2.38}
 \end{align}
 represent the first few equations of the time-dependent Fokas-Lenells hierarchy.
 The special case $\underline{n}=(1,1),$ and $c_{1,-}=1,$
 that is,
 \begin{equation*}
   \left(
                                \begin{array}{c}
                                  q_{xt_{(1,1)}}-q_{xx}+iqq_xr-2i q_x+q\\
                                   r_{xt_{(1,1)}}-r_{xx}-iqrr_x+2i r_x+r\\
                                \end{array}
                              \right)=0
 \end{equation*}
 represents the Fokas-Lenells system (\ref{1.1}).

 %In addition, combing the recursion relation (\ref{2.9}), (\ref{2.10})
 %with (\ref{2.36}) then yields the first conservation law
 %of the FL hierarchy, that is,
 %\begin{align}
 %& (q_x)_t+(f_{n_+-1,+}-f_{n_-,-})_x=0,\\
% & (r_x)_t+(h_{n_-,-,x}-h_{n_+-1,+})_x=0.
 %\end{align}

Finally, it will also be useful to work with the corresponding homogeneous coefficients
$\hat{f}_{\ell,\pm}, \hat{g}_{\ell,\pm},$ and  $\hat{h}_{\ell,\pm},$ defined by the vanishing of the
integration constants $c_k$ for $k=1,\ldots,\ell,$ and choosing $c_{0,\pm}=1,$
 \begin{equation}\label{2.12}
       \begin{split}
        & \hat{f}_{0,+}=f_{0,+}=-q_x, \quad \hat{f}_{0,-}=f_{0,-}=-\frac{1}{2}iq,
        \quad \hat{f}_{\ell}=f_{\ell}|_{c_k=0,~k=1,\ldots,\ell},
         \quad \ell \in \mathbb{N},\\
        & \hat{g}_{0,+}=g_{0,+}=-1, \quad \hat{g}_{0,-}=g_{0,-}=-\frac{1}{4},\quad
        \hat{g}_{\ell,+}=g_{\ell,+}|_{c_k=0,~k=1,\ldots,\ell},
        \quad  \ell \in \mathbb{N},\\
        &   \hat{h}_{0,+}=h_{0,+}=r_x, \quad \hat{h}_{0,-}=h_{0,-}=-\frac{1}{2}ir,
           \quad \hat{h}_{\ell,+}=h_{\ell,+}|_{c_k=0,~k=1,\ldots,\ell},
           \quad \ell \in \mathbb{N}.
       \end{split}
    \end{equation}
By induction one infers that
    \begin{equation}\label{2.13}
      f_{\ell,\pm}=\sum_{k=0}^{\ell}c_{\ell-k}\hat{f}_{k,\pm}, \quad
      g_{\ell,\pm}=\sum_{k=0}^{\ell}c_{\ell-k}\hat{g}_{k,\pm}, \quad
      h_{\ell,\pm}=\sum_{k=0}^{\ell}c_{\ell-k}\hat{h}_{k,\pm}, \quad
       \ell\in \mathbb{N}_0.
    \end{equation}
In a slight abuse of notation we will occasionally stress the dependence of
$f_{\ell,\pm},g_{\ell,\pm},h_{\ell,\pm}$ on $q,r$ (or $x,t$) by writing $f_{\ell,\pm}(q,r), g_{\ell,\pm}(q,r), h_{\ell,\pm}(q,r)$ (or $f_{\ell,\pm}(x,t), g_{\ell,\pm}(x,t), h_{\ell,\pm}(x,t)$).
Similarly, with $F_{\ell,+}, G_{\ell,+}, H_{\ell,+}$
denoting the polynomial parts of $F_{\underline{\ell}},G_{\underline{\ell}},H_{\underline{\ell}},$
respectively, and $F_{\ell,-}, G_{\ell,-}, H_{\ell,-}$ denoting the Laurant parts of $F_{\underline{\ell}},G_{\underline{\ell}},H_{\underline{\ell}}, \underline{\ell}=(\ell_-,\ell_+)\in\mathbb{N}_0^2,$ such that
\begin{equation*}
\begin{split}
  &F_{\underline{\ell}}(\xi)= F_{\ell,-}(\xi)+F_{\ell,+}(\xi),\quad G_{\underline{\ell}}(\xi)= G_{\ell,-}(\xi)+G_{\ell,+}(\xi),\\
  &H_{\underline{\ell}}(\xi)= H_{\ell,-}(\xi)+H_{\ell,+}(\xi),
  \end{split}
\end{equation*}
one finds that
\begin{equation*}
\begin{split}
&F_{\ell,\pm}=\sum_{k=1}^{\ell_{\pm}}c_{\ell_{\pm}-k,\pm}\widehat{F}_{k,\pm},\quad
G_{\ell,-}=\sum_{k=0}^{\ell_{-}}c_{\ell_{-}-k,\pm}\widehat{G}_{k,+},\\
&G_{\ell,+}=\sum_{k=0}^{\ell_{+}}c_{\ell_{+}-k,\pm}\widehat{G}_{k,-},\quad
H_{\ell,\pm}=\sum_{k=1}^{\ell_{\pm}}c_{\ell_{\pm}-k,\pm}\widehat{H}_{k,\pm},
\end{split}
\end{equation*}
where $\widehat{F}_{k,\pm},\widehat{G}_{k,\pm},\widehat{H}_{k,\pm}$
are corresponding homogeneous quantities of
$F_{k,\pm},G_{k,\pm},H_{k,\pm}.$

\section{Stationary Fokas-Lenells formalism}
This section is devoted to a detailed study of the stationary Fokas-Lenells
hierarchy.
 We first
 define a fundamental meromorphic function $\phi(P,x)$ on the hyperelliptic
 curve $\mathcal{K}_n$, using the polynomial recursion formalism described in section 2,
 and then study the properties of the
 Baker-Akhiezer function $\psi(P,x,x_0)$, Dubrovin-type equations,
trace formulas and theta function representations of $\phi,\psi_1,\psi_2,q,r.$

 For major parts of this section we suppose (\ref{2.1}), (\ref{2.2}), (\ref{2.3})-(\ref{2.10}),
 (\ref{2.14})-(\ref{2.24}), keeping $n\in\mathbb{N}_0$ fixed.

 We recall the hyperelliptic curve
 \begin{align}\label{3.1}
   \mathcal{K}_n: &\,
    \mathcal{F}(\xi,y)=y^2-z^{2n_-}R_{\underline{n}}(\xi)=y^2-\prod_{m=0}^{2n+1}(\xi-E_m)=0, \\
     R_{\underline{n}}(\xi)&=z^{-2n_-}\prod_{m=0}^{2n+1}(\xi-E_m),\, \{E_m\}_{m=0}^{2 n+1}
    \subset\mathbb{C},\, n=2n_++2n_--1\in\mathbb{N}_0,  \nonumber
 \end{align}
as introduced in (\ref{2.32}). Throughout this section we assume $\mathcal{K}_n$
to be nonsingular, that is, we suppose that
\begin{equation}\label{3.2}
  E_m\neq E_{m^\prime}\,
  \, \textrm{for}\,\,m\neq m^\prime,\,\,m,m^\prime=0,1,\cdots,2n+1.
\end{equation}
$\mathcal{K}_n$ is compactified by joining two points at infinity
$P_{\infty_\pm}$, $P_{\infty_+} \neq P_{\infty_-}$, but for
notational simplicity the compactification is also denoted by
$\mathcal{K}_n$. Points $P$ on
      $$\mathcal{K}_{n} \setminus \{P_{\infty_+},P_{\infty_-}\}$$
are represented as pairs $P=(\xi,y(P))$, where $y(\cdot)$ is the
meromorphic function on $\mathcal{K}_{n}$ satisfying
       $$\mathcal{F}_n(\xi,y(P))=0.$$
The complex structure on $\mathcal{K}_{n}$ is defined in the usual
way by introducing local coordinates
$$\zeta_{Q_0}:P\rightarrow(\xi-\xi_0)$$
near points $Q_0=(\xi_0,y(Q_0))\in \mathcal{K}_{n},$ which are neither branch nor singular points of
$\mathcal{K}_{n}$;
near the points $P_{\infty_\pm} \in \mathcal{K}_{n}$, the local
coordinates are
   $$\zeta_{P_{\infty_\pm}}:P \rightarrow \xi^{-1},$$
and similarly at branch and singular points of $\mathcal{K}_{n}.$
Hence $\mathcal{K}_n$ becomes a two-sheeted Riemann surface of
 topological genus $n$ in a standard manner.

The holomorphic map
   $\ast,$ changing sheets, is defined by
       \begin{eqnarray}\label{3.3}
       && \ast: \begin{cases}
                        \mathcal{K}_{n}\rightarrow\mathcal{K}_{n},
                       \\
                       P=(\xi,y_j(\xi))\rightarrow
                       P^\ast=(z,y_{j+1(\mathrm{mod}~
                       2)}(\xi)), \quad j=0,1,
                      \end {cases}
                     \nonumber \\
      && P^{\ast \ast}:=(P^\ast)^\ast, \quad \mathrm{etc}.,
       \end{eqnarray}
       where $y_j(\xi),\, j=0,1,$ denote the two branches of $y(P)$ satisfying
      $\mathcal{F}_{n}(\xi,y)=0$,
  namely
        \begin{equation}\label{3.4}
          (y-y_0(\xi))(y-y_1(\xi))
          =y^2-z^{-2n_-}R_{2n+2}(\xi)=0.
        \end{equation}
        Taking into account (\ref{3.4}), one easily derives
        \begin{equation}\label{3.5}
           \begin{split}
             & y_0(\xi)+y_1(\xi)=0,\\
             & y_0(\xi)y_1(\xi)=-z^{-2n_-}R_{2n+2}(\xi),\\
             & y_0^2(\xi)+y_1^2(\xi)=2z^{-2n_-}R_{2n+2}(\xi).\\
           \end{split}
        \end{equation}
        Positive divisors on
         $\mathcal{K}_{n}$ of degree $n$ are denoted by
        \begin{equation}\label{3.6}
          \mathcal{D}_{P_1,\ldots,P_{n}}:
             \begin{cases}
              \mathcal{K}_{n}\rightarrow \mathbb{N}_0,\\
              P\rightarrow \mathcal{D}_{P_1,\ldots,P_{n}}=
                \begin{cases}
                  \textrm{ $k$ if $P$ occurs $k$
                      times in $\{P_1,\ldots,P_{n}\},$}\\
                   \textrm{ $0$ if $P \notin
                     $$ \{P_1,\ldots,P_{n}\}.$}
                \end{cases}
             \end{cases}
        \end{equation}
        Moreover, for a nonzero, meromorphic function $f$ on
        $\mathcal{K}_n,$ the divisor of $f$ is denoted by $(f).$

        For notational simplicity we will usually
        assume that $n\in\mathbb{N}$ and hence
        $\underline{n}\in\mathbb{N}_0^2\backslash\{(0,0)\}.$
        (The trivial case $\underline{n}=(0,0)$ is excluded in our discussion
        since the "genus" of corresponding curve is $-1<0.$)

        We denote by $\{\mu_j(x)\}_{j=1,\cdots,n}$
        and $\{\nu_j(x)\}_{j=1,\cdots,n}$ the zeros of $(\cdot)^{2n_--1}F_{\underline{n}}(\cdot,x)$
        and $(\cdot)^{2n_--1}H_{\underline{n}}(\cdot,x),$ respectively. Thus we may write
        \begin{align}
          F_{\underline{n}}(\xi,x) & = -q_x(x) \xi^{-2n_-+1}\prod_{j=1}^n\left(\xi-\mu_j(x)\right),\label{3.7a}\\
          H_{\underline{n}}(\xi,x) & =r_x(x) \xi^{-2n_-+1}\prod_{j=1}^n\left(\xi-\nu_j(x)\right).\label{3.7b}
        \end{align}
        We now introduce $\{\hat{\mu}_j\}_{j=1,\ldots,n}\subset\mathcal{K}_n$ and $\{\hat{\nu}_j\}_{j=1,\ldots,n}\subset\mathcal{K}_n$
        by
        \begin{equation}\label{3.8}
          \hat{\mu}_j(x)=(\mu_j(x),-\mu_j(x)^{2n_-}G_{\underline{n}}(\mu_j(x),x)),\quad j=1,\ldots,n,
        \end{equation}
        and
        \begin{equation}\label{3.9}
          \hat{\nu}_j(x)=(\nu_j(x),\nu_j(x)^{2n_-}G_{\underline{n}}(\nu_j(x),x)),\quad j=1,\ldots,n.
        \end{equation}
        We also introduce the points $P_{0,\pm}$ by
        $P_{0,\pm}=\left(0,\pm\frac{1}{4}\right)\in\mathcal{K}_n$
        (cf. (\ref{2.31})).

        Next we define the fundamental meromorphic function on $\mathcal{K}_n$ by
        \begin{align}\label{3.10}
          \phi(P,x)&=-\frac{i\xi^{-2n_-}y-iG_{\underline{n}}(\xi,x)}{F_{\underline{n}}(\xi,x)}\nonumber\\
                   &=\frac{H_{\underline{n}}(\xi,x)}{i\xi^{-2n_-}y+iG_{\underline{n}}(\xi,x)},
        \end{align}
        with divisor of $\phi(\cdot,x)$ given by
        \begin{equation}\label{3.11}
          (\phi(\cdot,x))=\mathcal{D}_{P_{0,-}\underline{\hat{\nu}}(x)}
          -\mathcal{D}_{P_{\infty+}\underline{\hat{\mu}}(x)},
        \end{equation}
        using (\ref{3.7a}) and (\ref{3.7b}).
        Here we abbreviated
        \begin{equation}\label{3.12}
        \begin{split}
         & \underline{\hat{\mu}}(x)=\{\hat{\mu}_1(x)=P_{0,+},\hat{\mu}_2(x),\ldots,\hat{\mu}_n(x)\},\\
         &\underline{\hat{\nu}}(x)=\{\hat{\nu}_1(x)=P_{\infty-},\hat{\nu}_2(x),\ldots,\hat{\nu}_n(x)\}.
         \end{split}
        \end{equation}
        Given $\phi(\cdot,x)$, the stationary Baker-Akhiezer function $\psi$
        is then defined by
        \begin{align}
          \psi(P,x,x_0) &= \left(
                    \begin{array}{c}
                      \psi_1(P,x,x_0) \\
                      \psi_2(P,x,x_0) \\
                    \end{array}
                  \right),
          \nonumber \\
          \psi_1(P,x,x_0)&=\exp\left(\int_{x_0}^xdx^\prime\left(-iz+q_x(x^{\prime})
          \xi\phi(P,x^\prime)\right)\right),\label{3.13a}\\
          \psi_2(P,x,x_0)&=\phi(P,x)\exp\left(\int_{x_0}^xdx^\prime\left(r_x(x^\prime)\xi\phi^{-1}(P,x^\prime)+iz\right)\right).\label{3.13b}
        \end{align}
        Basic properties of $\phi$
        and $\psi$ are summarized in the following result.

        \newtheorem{lem3.1}{Lemma}[section]
        \begin{lem3.1}\label{lemma3.1}
        Assume $(\ref{3.10}), (\ref{3.13a}),(\ref{3.13b})$,
        $P=(z,y)\in\mathcal{K}_n\backslash\{P_{\infty\pm},P_{0,\pm}\},$
        and let $(\xi,x,x_0)\in\mathbb{C}\times\mathbb{R}^2.$ Then

          $(i)$ $\phi(P,x)$ satisfies the Riccati-type equation
        \begin{equation}\label{3.14}
           \phi_x(P,x)=r_x\xi+2iz\phi(P,x)-q_x\xi\phi^2(P,x)
        \end{equation}
        and
        \begin{align}
          \phi(P,x)\phi(P^*,x) & =\frac{H_{\underline{n}}(\xi,x)}{F_{\underline{n}}(\xi,x)}, \label{3.15a}\\
          \phi(P,x)+\phi(P^*,x) &=\frac{2iG_{\underline{n}}(\xi,x)}{F_{\underline{n}}(\xi,x)},\label{3.15b}\\
           \phi(P,x)-\phi(P^*,x) &=-\frac{2i\xi^{-2n_-}y(P)}{F_{\underline{n}}(\xi,x)}.\label{3.15c}
        \end{align}

        $(ii)$ $\psi(P,x,x_0)$ satisfies the first-order system %(cf. (\ref{2.17}))
        \begin{align}
           \psi_x(P,x,x_0)&=U(\xi,x)\psi, \label{3.16a} \\
           V_{\underline{n}}(\xi,x)\psi(P,x,x_0)&=i\xi^{-2n_-}y(P)\psi(P,x,x_0).\label{3.16b}
          \end{align}
        Moreover,
        \begin{equation}\label{3.17}
          \psi_1(P,x,x_0) =\sqrt{\frac{F_{\underline{n}}(\xi,x)}{F_{\underline{n}}(\xi,x_0)}}
          \exp\left(-\int_{x_0}^{x}dx^\prime\left(\frac{q_x(x^\prime)\xi^{-2n_-+1}y(P)}
          {F_{\underline{n}}(\xi,x^\prime)}\right)\right),
        \end{equation}
        and
        \begin{align}
          \psi_1(P,x,x_0)\psi_1(P^*,x,x_0) & =\frac{F_{\underline{n}}(\xi,x)}{F_{\underline{n}}(\xi,x_0)}, \label{3.18a}\\
          \psi_2(P,x,x_0)\psi_2(P^*,x,x_0) & =\frac{H_{\underline{n}}(\xi,x)}{F_{\underline{n}}(\xi,x_0)},
          \label{3.18b}\\
          \psi_1(P,x,x_0)\psi_2(P^*,x,x_0)&+\psi_1(P^*,x,x_0)\psi_2(P,x,x_0)\nonumber\\
          &=\frac{2iG_{\underline{n}}(\xi,x)}{F_{\underline{n}}(\xi,x_0)}.\label{3.18c}
        \end{align}

        $(iii)$ $\phi, \psi$ satisfy
        \begin{equation}\label{3.19}
          \psi_2(P,x,x_0)=\phi(P,x)\psi_1(P,x,x_0).
        \end{equation}
        \end{lem3.1}
        \proof To prove (\ref{3.14}) one uses the definition
        (\ref{3.10}) of $\phi$ and equations (\ref{2.19})-(\ref{2.21})
        to obtain
        \begin{align*}
          \phi_x(P,x)=&-\left(
          \frac{i\xi^{-2n_-}y(P)-iG_{\underline{n}}(\xi,x)}{F_{\underline{n}}(\xi,x)}\right)_x\\
           =&\frac{iG_{\underline{n},x}(\xi,x)}{F_{\underline{n}}(\xi,x)}+\phi(P,x)
          \frac{F_{\underline{n},x}(\xi,x)}{F_{\underline{n}}(\xi,x)}\\
           =&\frac{\xi q_xH_{\underline{n}}(\xi,x)+\xi r_xF_{\underline{n}}(\xi,x)}{F_{\underline{n}}(\xi,x)}+\phi(P,x)
           \frac{2izF_{\underline{n}}(\xi,x)-2i\xi
          q_x G_{\underline{n}}(\xi,x)}{F_{\underline{n}}(\xi,x)}\\
          =&\frac{\xi q_xH_{\underline{n}}(\xi,x)}{i\xi^{-2n_-}y(P)+iG_{\underline{n}}(\xi,x)}
          \frac{i\xi^{-2n_-}y(P)-iG_{\underline{n}}(\xi,x)+2iG_{\underline{n}}(\xi,x)}
          {F_{\underline{n}}(\xi,x)}\\
          &+
          \frac{ -2i\xi
          q_x G_{\underline{n}}(\xi,x)\phi(P,x)}{F_{\underline{n}}(\xi,x)}+2iz\phi(P,x)+ r_x\xi \\
          =&r_x\xi+2iz\phi(P,x)-q_x\xi\phi^2(P,x).
        \end{align*}
        Equations (\ref{3.15a})-(\ref{3.15c}) are clear from the definitions of $\phi$
        and $y$. By definitions of $\psi,$
        \begin{align}
          \psi_{1,x}(P,x,x_0) & = (-iz+q_x\xi\phi(P,x))\psi_{1,x}(P,x,x_0),\label{3.20a}\\
          \psi_{2,x}(P,x,x_0)  & =(iz+r_x\xi\phi(P,x)^{-1})\psi_2(P,x,x_0),\label{3.20b}
        \end{align}
        the function $\psi_2(P,x,x_0)/\psi_1(P,x,x_0)$ satisfies
        the first-order linear equation
        \begin{equation*}
          \frac{dL(P,x,x_0)}{dx}=\left(r_x\xi\phi^{-1}(P,x)+2iz-q_x\xi\phi(P,x)\right)L(P,x,x_0).
        \end{equation*}
        Since $\psi_2(P,x,x_0)/\psi_1(P,x,x_0)$ and $\phi(P,x)$
        take the same value at $x=x_0,$
        that is,
        $\psi_2(P,x_0,x_0)/\psi_1(P,x_0,x_0)=\phi(P,x_0)$, one
        derives (\ref{3.19}).
        (\ref{3.16a}), (\ref{3.16b}) are clear from
        (\ref{3.20a}), (\ref{3.20b}) and (\ref{3.19}).
        (\ref{3.18a})-(\ref{3.18c})
        follow from (\ref{3.13a}), (\ref{3.13b}), (\ref{3.15a})-(\ref{3.15c}).
        Finally, by (\ref{2.20}),(\ref{3.10}), (\ref{3.13a}),
        \begin{align}\label{3.21}
          \psi_1(P,x,x_0) & = \exp\left(\int_{x_0}^xdx^\prime\left(-iz+q_x(x^\prime)\xi\phi(P,x^\prime)\right)\right)\nonumber\\
           & =\exp\left(\int_{x_0}^{x}dx^{\prime}\left(-iz+q_x(x^\prime)\xi\frac{i\xi^{-2n_-}y(P)-
           iG_{\underline{n}}(\xi,x^\prime)}{-F_{\underline{n}}(\xi,x^\prime)}\right)\right)\nonumber\\
           &=\exp\left(\int_{x_0}^xdx^\prime\left(\frac{F_{\underline{n},x}(\xi,x^\prime)}{2
           F_{\underline{n}}(\xi,x^\prime)}-\frac{i\xi^{-2n_-+1}q_x(x^\prime)y(P)}
           {F_{\underline{n}}(\xi,x^\prime)}\right)\right),
        \end{align}
        which proves (\ref{3.17}). \qed\vspace{0.15cm}

        Concerning the dynamics of the zeros $\mu_j(x)$
        and $\nu_j(x)$ of $F_{\underline{n}}(\xi,x)$
        and $H_{\underline{n}}(\xi,x)$ one obtains the following
        Dubrovin-type equations.
        \newtheorem{lem3.3}[lem3.1]{Lemma}
        \begin{lem3.3}\label{lemma3.2}
        Suppose $(\ref{2.1})$ and the $\underline{n}$th stationary Fokas-Lenells
        equation
        $(\ref{2.24})$ holds subject to the constraint $(\ref{3.2})$
        on an open interval $\widetilde{\Omega}_\mu\subseteq\mathbb{R}$.
        Suppose that the zeros $\{\mu_j(x)\}_{j=0,\ldots,n}$
        of $\xi^{2n_--1}F_{\underline{n}}(\xi,x)$ remain distinct and nonzero for $ x \in
        \widetilde{\Omega}_\mu.$ Then
        $\{\hat{\mu}_j(x)\}_{j=0,\ldots,n}$ defined by $(\ref{3.8})$, satisfies the
         following first-order system of differential equations
          \begin{equation}\label{3.23a}
          \mu_{j,x}(x)  =\frac{-2iy(\hat{\mu}_j(x))}{\prod_{k=1,k\neq j}^{n}(\mu_j(x)-\mu_k(x))},\ \ j=1,\ldots,n,\,\, x\in\widetilde{\Omega}_\mu.
          \end{equation}
        %\begin{equation}\label{3.38}
    %       \mu_{j,x}= 2 \frac{ y(\hat{\mu}_j)}{\mu_j}
     %       \prod_{\scriptstyle k=0 \atop \scriptstyle k \neq j }^{n}
      %      (\mu_j(x)-\mu_k(x))^{-1}, \ \ j=0,\ldots,n, \ \ x\in\Omega_\mu.
       % \end{equation}
   Next, assume $\mathcal{K}_n$ to be nonsingular and introduce initial
   condition
   \begin{equation}\label{3.24}
   \{\hat{\mu}_j(x_0)\}_{j=1,\ldots,n}\subset\mathcal{K}_n
   \end{equation}
   for some $x_0\in\mathbb{R},$ where $\mu_j(x_0)\neq 0,j=1,\ldots,n,$
   are assumed to be distinct. Then there exists an open interval
   $ \Omega_{\mu}\subseteq\mathbb{R},$ with $x_0\in \Omega_\mu,$
   such that the initial value problem $(\ref{3.23a}), (\ref{3.24})$
   has a unique solution $\{\hat{\mu}_j\}_{j=1,\ldots,n}\subset\mathcal{K}_n$ satisfying
   \begin{equation}\label{3.25}
      \hat{\mu}_j \in C^\infty(\Omega_\mu,\mathcal{K}_{n}),
      \quad j=0,\ldots,n,
       \end{equation}
     and $\mu_j, j=1,\ldots,n,$ remain distinct and nonzero on $\Omega_{\mu}.$

     \noindent For the zeros $\{\nu_j(x)\}_{j=1,\ldots,n}$
     of $\xi^{2n_--1}H_{\underline{n}}(\xi,x)$
     similar statements hold with $\mu_j$ and $\Omega_\mu$
     replaced by $\nu_j$ and $\Omega_\nu,$ etc. In particular, $\{\hat{\nu}_j\}_{j=1,\ldots,n},$
     defined by $(\ref{3.9})$, satisfies the system
     %\begin{equation}\label{3.41}
      %     \nu_{l,x}=-2\frac{u_{xx}~y(\hat{\nu}_l)}{h_0 ~ \nu_l}
       %     \prod_{\scriptstyle k=1 \atop \scriptstyle k \neq l }^{n}
        %    (\nu_l(x)-\nu_k(x))^{-1}, \quad l=1,\ldots,n,\quad x\in\widetilde{\Omega}_\nu
        %\end{equation}
       \begin{equation} \label{3.23b}
         \nu_{j,x}(x)  =\frac{-2iy(\hat{\mu}_j(x))}{\prod_{k=1,k\neq j}^{n}(\nu_j(x)-\nu_k(x))},\ \ j=1,\ldots,n,\,\, x\in\widetilde{\Omega}_\mu.
        \end{equation}
        % Assume $(\ref{3.7a})$, $(\ref{3.7b})$
        %and let $x\in\mathbb{R}.$ Then

        \end{lem3.3}
        \proof
        We only prove equation (\ref{3.23a})
        since the proof of (\ref{3.23b}) follows in an identical manner.
        Inserting $\xi=\mu_j$ into equation (\ref{2.20}),
        one concludes from (\ref{3.8}),
        \begin{align}
        F_{\underline{n},x}(\mu_j)&=q_x\mu_j^{-2n_-+1}\prod_{\begin{smallmatrix}k=1\\k\neq j\end{smallmatrix}}^n\left(\mu_j-\mu_k\right)=2iq_x\mu_jG_{\underline{n}}(\mu_j)\nonumber\\
         &=-2iq_x\mu_j^{-2n_-+1}y(\hat{\mu}_j),\label{3.26}
        \end{align}
        proving (\ref{3.23a}). The smooth assertion (\ref{3.25})
        is clear as long as $\hat{\mu}_j$
        stays away from the branch points $(E_m,0).$
        In case $\hat{\mu}_j$ hits such a branch point, one can use
        the local chart around $(E_m,0)$ (with the local chart $\zeta=\sigma(\xi-E_m)^{1/2},
        \sigma\in\{1,-1\}$) to verify (\ref{3.25}).

        \qed

 Next, we turn to the trace formulas of the FL invariants, that
is, expressions of $f_{\ell,\pm}$ and $h_{\ell,\pm}$ in terms of symmetric
functions of the zeros $\mu_j$ and $\nu_\ell$ of $(\cdot)^{2n_--1}F_{\underline{n}}(\cdot)$ and $(\cdot)^{2n_--1}H_{\underline{n}}(\cdot)$,
respectively. For simplicity we just record the simplest case.

        \newtheorem{lem3.2}[lem3.1]{Lemma}
        \begin{lem3.2}
        Suppose $(\ref{2.1})$
        and the $\underline{n}$th stationary
        Fokas-Lenells system $(\ref{2.24})$ holds
        and let $x\in\mathbb{R}.$
        Then
        \begin{align}
          &\frac{q_{xx}}{2iq_x} -\frac{1}{2}q_xr_x-c_{1,+} =-\sum_{j=1}^{n}\mu_j,\label{3.22a} \\
          &\frac{r_{xx}}{2ir_x} +\frac{1}{2}q_xr_x+c_{1,+} =-\sum_{j=1}^{n}\nu_j,\label{3.22b}\\
          &\frac{iq}{2q_x}=(-1)^n\prod_{j=1}^n\mu_j,\label{3.22c}\\
          &\frac{ir}{2r_x}=(-1)^{n-1}\prod_{j=1}^n\nu_j.\label{3.22d}
        \end{align}

        \end{lem3.2}
        \proof (\ref{3.22a})-(\ref{3.22d}) follow by comparison powers of $\xi$
         substituting (\ref{3.7a}) and (\ref{3.7b})
         into (\ref{2.16})
         taking into account (\ref{2.11}).\qed\vspace{0.2cm}

         Next we turn to the asymptotic properties of $\phi$
         and $\phi_j, j=1,2.$

         \newtheorem{lem3.4}[lem3.1]{Lemma}
         \begin{lem3.4}
         Suppose $(\ref{2.1})$
         and the $\underline{n}$th stationary
         Fokas-Lenells system $(\ref{2.24})$ holds
         and let $P\in\mathcal{K}_n\backslash\{P_{\infty\pm},P_{0,\pm}\},x\in\mathbb{R}.$
         Then
         \nopagebreak[4]
         \begin{align}
         \phi(P,x) &\underset{\zeta \rightarrow 0}{=}
         \begin{cases} 2i[q_x(x)]^{-1}\zeta^{-1}+O(\zeta),
         \qquad &P \rightarrow P_{\infty+},\cr
         [ir_x(x)/2]\,\zeta+O(\zeta^3),&P \rightarrow P_{\infty-},
         \end{cases}
         \quad\zeta=\xi^{-1},\label{3.27}\\[0.15cm]
         \phi(P,x) &\underset{\zeta \rightarrow 0}{=}
         \begin{cases} q(x)^{-1}\zeta^{-1}+O(\zeta),
         \qquad &P \rightarrow P_{0,+},\cr
         r(x)\,\zeta+O(\zeta^3),&P \rightarrow P_{0,-},
         \end{cases}
         \quad\zeta=\xi,\label{3.28}\\
         \psi_1(P,x) &\underset{\zeta \rightarrow 0}{=}
          \begin{cases} e^{i(x-x_0)\zeta^{-2}+O(1)},
          \qquad &P \rightarrow P_{\infty+},\cr
            e^{-i(x-x_0)\zeta^{-2}+O(1)},&P \rightarrow P_{\infty-},
          \end{cases}
          \quad\zeta=\xi^{-1},\label{3.29}\\[0.15cm]
          \psi_1(P,x) &\underset{\zeta \rightarrow 0}{=}
          \begin{cases} \frac{q(x)}{q(x_0)}(1+O(\zeta)),
          \qquad &P \rightarrow P_{0,+},\cr
          1+O(\zeta),&P \rightarrow P_{0,-},
          \end{cases}
         \quad\zeta=\xi,\label{3.30}\end{align}
         \begin{align}
          \psi_2(P,x) &\underset{\zeta \rightarrow 0}{=}
          \begin{cases}\left(\frac{2i}{q_x(x)}\zeta^{-1}+O(\zeta)\right) e^{i(x-x_0)\zeta^{-2}+O(1)},
          \qquad &P \rightarrow P_{\infty+},\cr
            \left(\frac{ir_x(x)}{2}\,\zeta+O(\zeta^3)\right)e^{-i(x-x_0)\zeta^{-2}+O(1)},&P \rightarrow P_{\infty-},
          \end{cases}
          \quad\zeta=\xi^{-1},\label{3.31}\\[0.15cm]
          \psi_2(P,x) &\underset{\zeta \rightarrow 0}{=}
          \begin{cases} \frac{1}{q(x_0)}\zeta^{-1}+O(\zeta),
          \qquad &P \rightarrow P_{0,+},\cr
          r(x)\zeta+O(\zeta^2),&P \rightarrow P_{0,-},
         \end{cases}
         \quad\zeta=\xi.\label{3.32}
         \end{align}

         \end{lem3.4}
         \proof
          The existence of the
          asymptotic expansions of $\phi$ in terms of the appropriate
          local coordinates $\zeta=\xi^{-1}$
          near $P_{\infty_\pm}$ and $\zeta=\xi$ near $P_{0,\pm}$ is
          clear from its explicit
          expression in (\ref{3.10}). Next, we compute these explicit expansions coefficients
          in (\ref{3.27}) and (\ref{3.28}).
          Inserting each of the following asymptotic expansions
         \begin{align}
           \phi & =\phi_0\zeta^{-1}+\phi_1+O(\zeta), \ \ \textrm{as $P\rightarrow P_{\infty+}$,}\label{3.33a}\\
           \phi & =\phi_0\zeta+\phi_1\zeta^2+O(\zeta^2), \ \ \textrm{as $P\rightarrow P_{\infty-}$,}\label{3.33b}\\
            \phi & =\phi_0\zeta^{-1}+\phi_1+O(\zeta), \ \ \textrm{as $P\rightarrow P_{0,+}$,}\label{3.33c}\\
           \phi & =\phi_0\zeta+\phi_1\zeta^2+O(\zeta^2), \ \ \textrm{as $P\rightarrow P_{0,-}$}\label{3.33d}
         \end{align}
         into the Riccati-type equation (\ref{3.14}) and, upon
         comparing coefficients of powers of $\xi$, which determines the
         expansion coefficients of $\phi_k$ in (\ref{3.33a})-(\ref{3.33d}),
         one concludes (\ref{3.27}) and (\ref{3.28}). (\ref{3.29}), (\ref{3.30})
         are clear from (\ref{3.13a}), (\ref{3.27}) and (\ref{3.28}).
         (\ref{3.31}) and (\ref{3.32}) follow by (\ref{3.27})-(\ref{3.30}) and
         (\ref{3.19}).\qed\vspace{0.2cm}

         Next, we introduce the holomorphic differentials $\eta_\ell(P)$ on
         $\mathcal{K}_{n}$
        \begin{equation}\label{3.34}
         \eta_\ell(P)= \frac{\xi^{\ell-1}}{y(P)} d\xi,
         \qquad \ell=1,\ldots,n,
        \end{equation}
        and choose a homology basis $\{a_j,b_j\}_{j=1}^{n}$ on
        $\mathcal{K}_{n}$ in such a way that the intersection matrix of the
        cycles satisfies
        $$a_j \circ b_k =\delta_{j,k},\quad a_j \circ a_k=0, \quad
        b_j \circ   b_k=0, \quad j,k=1,\ldots, n.$$
       Associated with $\mathcal{K}_n$ one
        introduces an invertible
        matrix $E \in \textrm{GL}(n, \mathbb{C})$
         \begin{equation}\label{3.35}
          \begin{split}
          & E=(E_{j,k})_{n \times n}, \quad E_{j,k}=
           \int_{a_k} \eta_j, \\
          &  \underline{c}(k)=(c_1(k),\ldots, c_{n}(k)), \quad
           c_j(k)=(E^{-1})_{j,k},
           \end{split}
         \end{equation}
        and the normalized holomorphic differentials
        \begin{equation}\label{3.36}
          \omega_j= \sum_{\ell=1}^{n} c_j(\ell)\eta_\ell, \quad
          \int_{a_k} \omega_j = \delta_{j,k}, \quad
          \int_{b_k} \omega_j= \tau_{j,k}, \quad
          j,k=1, \ldots ,n.
        \end{equation}
        Apparently, the Riemann matrix $\tau=(\tau_{i,j})$ is symmetric and has a
        positive-definite imaginary part. Associated with $\tau$
        one defines the period lattice $L_n$ in $\mathbb{C}^n$
         by
       \begin{equation*}
        L_{n}=\{\underline{z}\in \mathbb{C}^{n}|
           ~\underline{z}=\underline{N}+\tau\underline{M},
           ~\underline{N},\underline{M}\in \mathbb{Z}^{n}\}.
       \end{equation*}
      The Riemann theta function associated with Riemann surface $\mathcal{K}_n$ and the homology basis $\{a_j,b_j\}_{j=1,\ldots,n}$
       is given by
       \begin{equation}\label{theta}
       \theta(\underline{z})=\sum_{\underline{n}\in\mathbb{Z}^n}\exp\Big(2\pi i(\underline{n},
       \underline{z})+\pi i(\underline{n},\tau\underline{n})\Big),~~\underline{z}\in\mathbb{C}^n,
       \end{equation}
        where $(\underline{A},\underline{B})=\sum_{j=1}^{n}\overline{A}_jB_j$ denotes the inner product in $\mathbb{C}^n.$
       Then the Jacobi variety $J(\mathcal{K}_n)$
       of $\mathcal{K}_n$ is defined by
       \begin{equation*}
         J(\mathcal{K}_n)=\mathbb{C}^n/L_n,
       \end{equation*}
       and the Abel maps are defined by
       \begin{equation}\label{3.37}
       \begin{split}
        \underline{A}_{Q_0}:&\mathcal{K}_{n} \rightarrow
       J(\mathcal{K}_{n}), \\
       &
       P \mapsto \underline{A}_{Q_0} (P)=(\underline{A}_{Q_0,1}(P),\ldots,
       \underline{A}_{Q_0,n} (P))\\
       &~~~~~~~~~~~~~~~~~
       =\left(\int_{Q_0}^P\omega_1,\ldots,\int_{Q_0}^P\omega_{n}\right)
        (\mathrm{mod}~L_{n})
       \end{split}
      \end{equation}
      and
       \begin{equation}\label{3.38}
       \begin{split}
      & \underline{\alpha}_{Q_0}:
       \mathrm{Div}(\mathcal{K}_{n}) \rightarrow
       J(\mathcal{K}_{n}),\\
      & ~~~~~~~~~\mathcal{D} \mapsto \underline{\alpha}_{Q_0}
       (\mathcal{D})= \sum_{P\in \mathcal{K}_{n}}
       \mathcal{D}(P)\underline{A}_{Q_0} (P)\\
       &~~~~~~~~~~~~~~~~~~~~~~~~~\triangleq(\alpha_{Q_0,1}(\mathcal{D}),\ldots,\alpha_{Q_0,n}(\mathcal{D})),
      \end{split}
       \end{equation}
       where $Q_0$ is a fixed base point
       and the same path is chosen from $Q_0$ to $P$
       in (\ref{3.37}) and (\ref{3.38}).

       Next, let $\Omega_{P_{0,-},P_{\infty+}}^{(3)},$
       be the normal differential of the third kind  holomorphic
       on $\mathcal{K}_n\backslash\{P_{0,-},P_{\infty+}\}$
       with simple
       poles at $P_{0,-}$ and $P_{\infty+}$, and residues
       $1$ and $-1,$ respectively.
       Explicitly,
       one writes $\Omega_{P_{0,-},P_{\infty+}}^{(3)}$ as
       \begin{equation}\label{3.39}
         \Omega_{P_{0,-},P_{\infty+}}^{(3)}=\left(\frac{y-1/2}{\xi}-\prod_{j=1}^n
         (\xi-\lambda^\prime_j)\right)
         \frac{d\xi}{2y},
       \end{equation}
       where the constants $\{\lambda_j^\prime\}_{j=1,\ldots,n}\subset\mathbb{C}$
       are uniquely determined by employing the normalization
       $$\int_{a_j}\Omega_{P_{0,-},P_{\infty+}}^{(3)}=0,\quad j=1,\ldots,n.$$
       The explicit formula (\ref{3.39}) then implies the following
       asymptotic expansion
       \begin{align}
        &\int_{Q_0}^P\Omega_{P_{0,-},P_{\infty+}}^{(3)}\underset{\zeta \rightarrow 0}{=}
          \left[
                 \begin{array}{c}
                   0 \\
                   \ln(\zeta) \\
                 \end{array}
               \right]+\omega_0^{0,\pm}+O(\zeta),
               \quad \omega_0^{0,\pm}\in\mathbb{C},\quad P\rightarrow P_{0,\pm}\nonumber\\
          & \label{3.40}\\
          &\int_{Q_0}^P\Omega_{P_{0,-},P_{\infty+}}^{(3)}\underset{\zeta \rightarrow 0}{=}
          \left[
                 \begin{array}{c}
                    -\ln(\zeta) \\
                     0\\
                 \end{array}
               \right]+\omega_0^{\infty,\pm}+O(\zeta),\quad\omega_0^{\infty,\pm}\in\mathbb{C},\quad P\rightarrow P_{\infty\pm}.\label{3.41}
               \end{align}
              Moreover, the Abelian diffrential of the second kind $\Omega_{P_{\infty\pm},1}^{(2)}$
               are chosen such that
               \begin{equation}\label{3.42}
                 \Omega_{P_{\infty\pm},1}^{(2)}\underset{\zeta\rightarrow 0}{=}[\zeta^{-3}+O(1)]d\zeta,\quad P\rightarrow P_{\infty\pm},
               \end{equation}
               \begin{equation}\label{3.43}
                 \int_{a_j} \Omega_{P_{\infty\pm},1}^{(2)}=0,\quad 1\leq j\leq n,
               \end{equation}
               \begin{equation}\label{3.44}
                 \underline{U}_0^{(2)}=(U_{0,1}^{(2)},\ldots,U_{0,n}^{(2)}),
                 \quad U_{0,j}^{(2)}=\frac{1}{2\pi i}\int_{b_j}\Omega_0^{(2)},\quad
                 \Omega_0^{(2)}=2(\Omega_{P_{\infty+},1}^{(2)}-\Omega_{P_{\infty-},1}^{(2)}),
               \end{equation}
               \begin{align}
                 &\int_{Q_0}^P\Omega_0^{(2)}\underset{\zeta\rightarrow 0}{=}
                 \mp[\zeta^{-2}+e_{0,0}+e_{0,1}\zeta+O(\zeta^2)],\quad P\rightarrow P_{\infty\pm},\label{3.45}\\
               & \int_{Q_0}^P\Omega_0^{(2)}\underset{\zeta\rightarrow 0}{=}
                  e_{0,\pm}+e_{1,\pm}\zeta+O(\zeta^2),\quad P\rightarrow P_{0,\pm}.\label{3.45a}
               \end{align}

          In the following it will be convenient to introduce
             the abbreviations
            \begin{eqnarray}\label{3.46}
               &&
           \underline{\xi}(P,\underline{Q})= \underline{\Xi}_{Q_0}
           -\underline{A}_{Q_0}(P)+\underline{\alpha}_{Q_0}
             (\mathcal{D}_{\underline{Q}}), \nonumber \\
          &&
           P\in \mathcal{K}_{n},\,
          \underline{Q}=(Q_1,\ldots,Q_{n})\in
          \mathrm{Sym}^{n}(\mathcal{K}_{n}),
         \end{eqnarray}
where $\underline{\Xi}_{Q_0}$ is
the vector of Riemann constants (cf.(A.45) \cite{15}).
It turns out that $\underline{z}(\cdot,\underline{Q}) $ is independent of the
choice of base point $Q_0$\,(cf.(A.52),\,(A.53) \cite{15}).\vspace{0.3cm}

      Given these preparations, the theta function representations of $\phi,\psi_1,\phi_2,q$
      and $r$ then read as follows.\vspace{0.25cm}

      \newtheorem{thm3.5}[lem3.1]{Theorem}
      \begin{thm3.5}
         Suppose $(\ref{2.1})$ and the $\underline{n}$th stationary Fokas-Lenells
         equation
        $(\ref{2.24})$ holds subject to the constraint $(\ref{3.2})$
        on an open interval $ \Omega \subseteq\mathbb{R}$.
        Moreover, let $P\in\mathcal{K}_n\backslash\{P_{0,-},P_{\infty+}\}$
        and $x\in\Omega.$  In
        addition, suppose that $\mathcal{D}_{\underline{\hat{\mu}}(x)}$, or
        equivalently, $\mathcal{D}_{\underline{\hat{\nu}}(x)}$ is nonspecial
        for $x\in \Omega$. Then, $\phi,\psi_1,\psi_2,q,r$ admit the following
        representations
        \begin{align}
       \phi(P,x)=&~C(x)\frac{\theta(\underline{\xi}(P,\hat{\underline{\nu}}(x)))}
       {\theta(\underline{\xi}(P,\hat{\underline{\mu}}(x)))}\exp\left(\int_{Q_0}^P
       \Omega_{P_{0,-},P_{\infty+}}^{(3)}\right),\label{3.46}\\
        \psi_1(P,x)=&~C(x,x_0)\frac{\theta(\underline{\xi}(P,\hat{\underline{\mu}}(x)))}
       {\theta(\underline{\xi}(P,\hat{\underline{\mu}}(x_0)))}
       \exp\left(-i(x-x_0)\int_{Q_0}^P\Omega_0^{(2)}\right),\label{3.47}\\
         \psi_2(P,x)=&~C(x)C(x,x_0)\frac{\theta(\underline{\xi}(P,\hat{\underline{\nu}}(x)))}
       {\theta(\underline{\xi}(P,\hat{\underline{\mu}}(x_0)))}\nonumber\\
       &~\times\exp\left(\int_{Q_0}^P
       \Omega_{P_{0,-},P_{\infty+}}^{(3)}-i(x-x_0)\int_{Q_0}^P\Omega_0^{(2)}\right),\label{3.47a}
        \end{align}
       where
       \begin{align}
        & C(x)=\frac{1}{q(x_0)}\frac{\theta(\underline{\xi}(P_{0,-},\hat{\underline{\mu}}(x)))}
       {\theta(\underline{\xi}(P_{0,-},\hat{\underline{\mu}}(x_0)))}
       \frac{\theta(\underline{\xi}(P_{0,+},\hat{\underline{\mu}}(x_0)))}
       {\theta(\underline{\xi}(P_{0,+},\hat{\underline{\nu}}(x)))}\nonumber\\
       &~~~~~~~~~\times e^{-i(x-x_0)(e_{0,-}-e_{0,+})-\omega_{0}^{0,+}},\label{cx}\\
       &C(x,x_0)=\frac{\theta(\underline{\xi}(P_{0,-},\hat{\underline{\mu}}(x_0)))}
       {\theta(\underline{\xi}(P_{0,-},\hat{\underline{\mu}}(x)))}e^{i(x-x_0)e_{0,-}}.\label{3.52}
       \end{align}
       The Abel map linearizes the divisors $\mathcal{D}_{\underline{\hat{\mu}}(x)}$
       and $\mathcal{D}_{\underline{\hat{\nu}}(x)}$
       in the sense that
       \begin{align}
         \underline{\alpha}_{Q_0}(\mathcal{D}_{\underline{\hat{\mu}}(x)})
         =&\underline{\alpha}_{Q_0}(\mathcal{D}_{\underline{\hat{\mu}}(x_0)})
         -i\underline{U}_0^{(2)}(x-x_0),\label{3.52A}\\
         \underline{\alpha}_{Q_0}(\mathcal{D}_{\underline{\hat{\mu}}(x)})
         =&\underline{\alpha}_{Q_0}(\mathcal{D}_{\underline{\hat{\mu}}(x_0)})
         -i\underline{U}_0^{(2)}(x-x_0).\label{3.52B}
       \end{align}
       Moreover, one derives
      \begin{align}
        &q(x)=q(x_0)\frac{\theta(\underline{\xi}(P_{0,-},\hat{\underline{\mu}}(x_0)))}
       {\theta(\underline{\xi}(P_{0,-},\hat{\underline{\mu}}(x)))}
       \frac{\theta(\underline{\xi}(P_{0,+},\hat{\underline{\mu}}(x)))}
       {\theta(\underline{\xi}(P_{0,+},\hat{\underline{\mu}}(x_0)))}e^{i(x-x_0)(e_{0,-}-e_{0,+})},
       \label{solutionq}\\
        &r(x)=r(x_0)\frac{\theta(\underline{\xi}(P_{0,-},\hat{\underline{\nu}}(x)))}
       {\theta(\underline{\xi}(P_{0,-},\hat{\underline{\nu}}(x_0)))}
       \frac{\theta(\underline{\xi}(P_{0,+},\hat{\underline{\nu}}(x_0)))}
       {\theta(\underline{\xi}(P_{0,+},\hat{\underline{\nu}}(x)))}e^{-i(x-x_0)(e_{0,-}-e_{0,+})},
       \label{solutionr}\\
       &q(x_0)r(x_0)=\frac{\theta(\underline{\xi}(P_{0,-},\hat{\underline{\nu}}(x_0)))}
       {\theta(\underline{\xi}(P_{0,-},\hat{\underline{\mu}}(x_0)))}
       \frac{\theta(\underline{\xi}(P_{0,+},\hat{\underline{\mu}}(x_0)))}
       {\theta(\underline{\xi}(P_{0,+},\hat{\underline{\nu}}(x_0)))}e^{\omega_0^{0,-}-\omega_{0}^{0,+}}.\label{qr}
      \end{align}

      \end{thm3.5}
      \proof
      First, we temporarily assume that
    \begin{equation}\label{3.48}
      \mu_j(x)\neq \mu_{j^\prime}(x), \quad \nu_k(x)\neq \nu_{k^\prime}(x)
      \quad \textrm{for $j\neq j^\prime, k\neq k^\prime$ and
      $x\in\widetilde{\Omega}$},
      \end{equation}
    for appropriate $\widetilde{\Omega}\subseteq\Omega$. Since by (\ref{3.11}), $\mathcal
  {D}_{P_{0,-}\underline{\hat{\nu}}}\sim
  \mathcal {D}_{P_{\infty+} \underline{\hat{\mu}}}$, and
 $(P_{0,-})^\ast \notin\{\hat{\nu}_1,\cdots,\hat{\nu}_n \}$
 by hypothesis, one can use Theorem A.31 in \cite{15} to
 conclude that $\mathcal {D}_{\underline{\hat{\mu}}}
 \in \textrm{Sym}^n(\mathcal {K}_n)$ is nonspecial. This
 argument is of course symmetric with respect to
 $\underline{\hat{\mu}}$ and $\underline{\hat{\nu}}$. Thus, $\mathcal
 {D}_{\underline{\hat{\mu}}}$ is nonspecial if and only
 if $\mathcal{D}_{\underline{\hat{\nu}}}$ is.

 Next we define the right-hand side of (\ref{3.47})
 to be $\widetilde{\psi}_1.$
 We intend to prove $\psi_1=\widetilde{\psi}_1,$ with $\psi_1$
 given by (\ref{3.13a}).
 For that purpose we first investigate the divisor of $\psi_1.$
 Since the zeros and
 poles can only come from zeros of $F_{\underline{n}}(\xi,x)$
 in (\ref{2.16}), one computes using (\ref{3.8}), the definition
 (\ref{3.10}) of $\phi,$ and the Dubrovin
 equations (\ref{3.23a}),
 \begin{align}\label{3.49}
   q_{x^\prime}(x^\prime)\xi\phi(P,x^\prime)\underset{P\rightarrow \hat{\mu}_j(x^\prime)}{=}&
   -q_x(x^\prime)\mu_j(x^\prime)\frac{2iy(P)}{-q_{x^\prime}(x^\prime)\mu_j(x^\prime)\prod_{k=1,k\neq j}^{n}(\mu_j(x^\prime)-\mu_k(x^\prime))}\nonumber\\
   &\times\frac{1}{\xi-\mu_j(x^\prime)}+O(1)\nonumber\\
   \underset{P\rightarrow \hat{\mu}_j(x^\prime)}{=}&\partial_{x^\prime}\ln(\xi-\mu_j(x^\prime))+O(1).
 \end{align}
 Together with (\ref{3.13a})
 this yields
 \begin{equation}\label{3.50}
 \begin{split}
  & \psi_1(P,x,x_0)=
   \begin{cases}
   (\xi-\mu_j(x))O(1),& \textrm{as\quad $P\rightarrow\hat{\mu}_j(x)\neq\hat{\mu}_j(x_0),$}\cr
   O(1),& \textrm{as\quad $P\rightarrow\hat{\mu}_j(x)=\hat{\mu}_j(x_0),$}\cr
   (\xi-\mu_j(x_0))^{-1}O(1)& \textrm{as\quad $P\rightarrow\hat{\mu}_j(x_0)\neq\hat{\mu}_j(x),$}
   \end{cases}\\
  & ~~~~~~~~~~~~~~~~~~~~~~~~~~~~~~~~~~~~~~~~~~~~~~ ~
  P=(\xi,y)\in\mathcal{K}_n,\,\,x,x_0\in\widetilde{\Omega},
 \end{split}
 \end{equation}
 with $O(1)\neq 1.$
 Consequently, $\psi_1$
 and $\widetilde{\psi}_1$ have identical zeros and poles
 on $\mathcal{K}_n\backslash\{P_{\infty\pm}\},$
  which are all simple by hypothesis (\ref{3.48}).
  Next, comparing
 the behavior of $\psi_1$
 and $\widetilde{\psi}_1$ near $P_{\infty\pm},$
 taking into account (\ref{3.13a}) and (\ref{3.45}),
 the expression (\ref{3.47})
 for $\widetilde{\psi}_1,$
 and (\ref{3.29}),
 then shows that  $\psi_1$
 and $\widetilde{\psi}_1$
 have
identical exponential behavior up to order
$O(1)$
near $P_{\infty\pm}.$ Thus, $\psi_1$
 and $\widetilde{\psi}_1$ share the
same singularities and zeros, and the Riemann-Roch-type uniqueness result
(cf. Lemma 3.4 \cite{12}) then proves that
 $\psi_1$
 and $\widetilde{\psi}_1$ coincide up to
normalization.
By (\ref{3.45a}) one infers from the right-hand side of (\ref{3.47})
that
\begin{align}
  &\widetilde{\psi}_1(P,x,x_0)\underset{\zeta\rightarrow 0}{=}C(x,x_0)\frac{\theta(\underline{\xi}(P_{0,-},\hat{\underline{\mu}}(x)))}
  {\theta(\underline{\xi}(P_{0,-},\hat{\underline{\mu}}(x_0)))}
  e^{-i(x-x_0)e_{0,-}}+O(\zeta)\,\,\nonumber\\
  &~~~~~~~~~~~~~~~~~~~~~~~~~~~~~~~~~~~~~
  ~~~~~~~~~~~~~~~~~~~~~~~~~~~~~~~~
  \textrm{as\,\,$P\rightarrow P_{0,-}$},\label{3.51}\\
  &\widetilde{\psi}_1(P,x,x_0)\underset{\zeta\rightarrow 0}{=}C(x,x_0)\frac{\theta(\underline{\xi}(P_{0,+},\hat{\underline{\mu}}(x)))}
  {\theta(\underline{\xi}(P_{0,+},\hat{\underline{\mu}}(x_0)))}
  e^{-i(x-x_0)e_{0,+}}+O(\zeta)\,\,\nonumber\\
  &~~~~~~~~~~~~~~~~~~~~~~~~~~~~~~~~~~~~~
  ~~~~~~~~~~~~~~~~~~~~~~~~~~~~~~~~
  \textrm{as\,\,$P\rightarrow P_{0,+}$}.\label{3.53}
\end{align}
A comparison of (\ref{3.30}) and (\ref{3.51}), (\ref{3.53}) then yields (\ref{3.52}),
(\ref{solutionq})
subject to (\ref{3.48}).
By (\ref{3.11}), one infers that
 $\phi(P,x)\exp(-\int_{Q_0}^P
 \Omega_{P_{0,-},P_{\infty+}}^{(3)})$
 must be of the type
 \begin{equation}\label{3.49}
   \phi(P,x)\exp\left(-\int_{Q_0}^P
 \Omega_{P_{0,-},P_{\infty+}}^{(3)}\right)=C(x)\frac{\theta(\underline{\xi}(P,\hat{\underline{\nu}}(x)))}
       {\theta(\underline{\xi}(P,\hat{\underline{\mu}}(x)))}
 \end{equation}
 for some function $C(x), x\in\mathbb{C}.$
 %Next we compute compute $C(x)$ we next consider the function $\psi_{1}(P,x,x_0)$
A comparison of (\ref{3.49}) and asymptotic relations (\ref{3.28})
then yields, with the help of (\ref{3.40}), the following expressions
\begin{align}
  \frac{1}{q(x)}=&C(x)\frac{\theta(\underline{\xi}(P_{0,+},\hat{\underline{\nu}}(x)))}
       {\theta(\underline{\xi}(P_{0,+},\hat{\underline{\mu}}(x)))}e^{\omega_{0}^{0,+}},\label{3.54}\\
     r(x)=&C(x)\frac{\theta(\underline{\xi}(P_{0,-},\hat{\underline{\nu}}(x)))}
       {\theta(\underline{\xi}(P_{0,-},\hat{\underline{\mu}}(x)))}e^{\omega_{0}^{0,-}},\label{3.55} \\
       q(x)r(x)=&\frac{\theta(\underline{\xi}(P_{0,-},\hat{\underline{\nu}}(x)))}
       {\theta(\underline{\xi}(P_{0,-},\hat{\underline{\mu}}(x)))}
       \frac{\theta(\underline{\xi}(P_{0,+},\hat{\underline{\mu}}(x)))}
       {\theta(\underline{\xi}(P_{0,+},\hat{\underline{\nu}}(x)))}e^{\omega_0^{0,-}-\omega_{0}^{0,+}}. \label{3.56}
\end{align}
Taking into account (\ref{3.52}),(\ref{solutionq}),(\ref{3.54})-(\ref{3.56}),
one easily derives (\ref{cx}), (\ref{solutionq})-(\ref{qr}). (\ref{3.47a}) follows by (\ref{3.19}), (\ref{3.46}) and
(\ref{3.47}).
Next we only prove the linearity of the Abel map with respect to $x$ in
(\ref{3.52A}) since the proof for (\ref{3.52B})
can be derived in an identical fashion.
 Using the Dubrovin equations (\ref{3.23a}), expression (\ref{3.36}),
and Lagrange's interpolation formula
$$\sum_{j=1}^n\frac{\mu_j^{\ell-1}}{\prod_{k=1,k\neq j}^{n}(\mu_j-\mu_\ell)}=\begin{cases}0&\ell\neq n\cr 1&\ell=n\end{cases},\,\mu_j\in\mathbb{C},\,\,\ell,j=1,\ldots,n,$$
one infers
\begin{align}\label{3.57}
   \partial_x\underline{\alpha}_{Q_{0}}(\mathcal{D}_{\underline{\hat{\mu}}(x)})
   =&\partial_x\Big(\sum_{j=1}^n\int_{Q_0}^{\hat{\mu}_j(x)}\underline{\omega} \Big)\nonumber\\
   =&\sum_{j=1}^n\mu_{j,x}(x)\sum_{\ell=1}^n\underline{c}(\ell)\frac{\mu_j^{\ell-1}(x)}{y(\hat{\mu}_j(x))}
   \nonumber\\
   =&\sum_{j=1}^n\sum_{\ell=1}^n\frac{-2i\mu_j^{\ell-1}(x)\underline{c}(\ell)}{\prod_{k=1,k\neq j}^{n}(\mu_j(x)-\mu_\ell(x))}\nonumber\\
   =&-2i\underline{c}(n)=-i\underline{U}_0^{(2)},
\end{align}
 which proves (\ref{3.52A}).
 The extension of all these results from $\widetilde{\Omega}$
 to $\Omega$ then simply
 follows from the continuity of $\underline{\alpha}_{Q_0}$
 and the hypothesis of $\mathcal{D}_{\underline{\hat{\mu}}(x)}$
being nonspecial on $\Omega.$
\qed

\section{Quasi-periodic Solutions}
In this section, we
extend the the algebro-geometric analysis of Section 2,3 to the
time-dependent FL hierarchy.

Throughout this section we assume (\ref{2.2}) holds.

The time-dependent algebro-geometric initial value problem of the
 FL hierarchy is to solve the time-dependent $\underline{r}$th FL flow with
 a stationary solution of the $\underline{n}$th equation as initial data in the
 hierarchy. More precisely, given $\underline{n}\in\mathbb{N}_0^2\backslash\{(0,0)\}$,
 based on the
 solution $q^{(0)},r^{(0)}$ of the $\underline{n}$th stationary HS equation
 $\textrm{s-FL}_{\underline{n}}(q^{(0)},r^{(0)})=0$ associated with $\mathcal{K}_n$ and a
 set of integration constants $\{c_{\ell,\pm}\}_{\ell=1,\ldots,n} \subset
 \mathbb{C}$, we want to build up a solution $q,r$ of the $\underline{r}$th FL flow
 $\mathrm{FL}_{\underline{r}}(q,r)=0$ such that $q(t_{0,\underline{r}})=q^{(0)},r(t_{0,\underline{r}})=r^{(0)}$ for some
 $t_{0,\underline{r}}\in\mathbb{R},\underline{r}\in\mathbb{N}_0^2\backslash\{(0,0)\}$.
 To emphasize that the integration
 constants in the definitions of the stationary and the time-dependent FL
equations are independent of each other, we indicate this by adding a tilde
on all the time-dependent quantities. Hence we shall employ the notation
 $\widetilde{V}_{\underline{r}},$ $\widetilde{F}_{\underline{r}},$ $\widetilde{G}_{\underline{r}},$
 $\widetilde{H}_{\underline{r}},$ $\tilde{f}_{s}$, $\tilde{g}_{s,\pm},$ $\tilde{h}_{s,\pm}$, $\tilde{c}_{s,\pm}$
 in order to distinguish them from $V_{\underline{n}},$ $F_{\underline{n}},$ $G_{\underline{n}},$ $H_{\underline{n}},$ $f_{s,\pm},$ $g_{s,\pm},$ $h_{s,\pm}$,
 $c_{s,\pm}$ with
 respect to $\xi$ in the following.
In addition, we mark
the individual $\underline{r}$th FL flow by a separate time variable $t_{\underline{r}}\in
\mathbb{R}$.

Summing up, we are interested in solutions $q,r$ of the time-dependent algebro-geometric initial
value problem
\begin{align}
     \textrm{$\widetilde{\text{FL}}$}_{\underline{n}}(q,r)=&
     \left(
       \begin{array}{c}
         q_{xt_{\underline{r}}}+ f_{r_{+}-1,+,x}-2iq_xg_{r_-,-}
             +2if_{r_--1,-} \\
         r_{xt_{\underline{r}}}-h_{r_+-1,+,x}
             +2ih_{r_--1,-}
            +2ir_xg_{r_-,-}
       \end{array}
     \right)=0,\nonumber\\
    % &~~~~~~~~~~~~~~~~~~~~~~~~~~~~~~
     %&~~\quad t_{\underline{n}}\in\mathbb{R},~\underline{n}=(n_-,n_+)\in\mathbb{N}_0^2,\nonumber\\
     (q,r)|_{t_{\underline{r}}=t_{0,\underline{r}}}&=(q^{(0)},r^{(0)}),\label{4.1}\\
 \textrm{s-FL}_{\underline{n}}(q^{(0)},r^{(0)})&=
   \left(
     \begin{array}{c}
       f_{n_+-1,+,x}-2iq_x^{(0)}g_{n_-,-}+2if_{n_--1,-}  \\
       -h_{n_+-1,+,x}+2ih_{n_--1,-}+2ir_x^{(0)}g_{n_-,-}  \\
     \end{array}
   \right)=0,\nonumber\\
    \label{4.2}
   \end{align}
for some $t_{0,\underline{r}}\in\mathbb{R},$
where $q=q(x,t_{\underline{r}}),r=r(x,t_{\underline{r}})$
satisfy (\ref{2.2})
and a fixed curve $\mathcal{K}_n$ is associated with the stationary
solution $q^{(0)},r^{(0)}$ in (\ref{4.2}). Here
\begin{equation*}
  \underline{n}=(n_{+},n_-)\in\mathbb{N}^2,\,\,
  \underline{r}=(r_{+},r_-)\in\mathbb{N}^2,\,\,n=2n_++2n_--1\in\mathbb{N}.
\end{equation*}
Noticing that
 the FL flows are isospectral, we further
 assume that (\ref{4.2}) holds not only for
 $t_{\underline{r}}=t_{0,\underline{r}}$, but also for all $t_{\underline{r}}\in
 \mathbb{R}$. In terms of Lax pairs this amounts to solving
the zero-curvature equations
    \begin{equation}\label{4.3}
        U_{t_{\underline{r}}}(\xi,x,t_{\underline{r}})-\widetilde{V}_{\underline{r},x}(\xi,x,t_{\underline{r}})
        +[U(\xi,x,t_{\underline{r}}),\widetilde{V}_{\underline{r}}(\xi,x,t_{\underline{r}})]=0,
    \end{equation}
    \begin{equation}\label{4.4}
        -V_{\underline{n},x}(\xi,x,t_{\underline{r}})+[U(\xi,x),V_{\underline{n}}(\xi,x,t_{\underline{r}})]=0,
    \end{equation}
    where
     \begin{equation}\label{5.5}
    \begin{split}
     & U(\xi,x,t_{\underline{r}})=
     \left(
  \begin{array}{cc}
    -iz & q_x(x,t_{\underline{r}})\xi \\
    r_x(x,t_{\underline{r}})\xi & iz \\
  \end{array}
\right),\, z=\xi^2,
     \\
     & V_{\underline{n}}(\xi,x,t_{\underline{r}})=
    \left(
      \begin{array}{cc}
      iG_{\underline{n}}(\xi,x,t_{\underline{r}}) & -F_{\underline{n}}(\xi,x,t_{\underline{r}}) \\
      H_{\underline{n}}(\xi,x,t_{\underline{r}}) & -iG_{\underline{n}}(\xi,x,t_{\underline{r}}) \\
      \end{array}
    \right),
      \\
     & \widetilde{V}_{\underline{r}}(\xi,x,{\underline{r}})=
       \left(
         \begin{array}{cc}
           i\widetilde{G}_{\underline{r}}(\xi,x,t_{\underline{r}}) & -\widetilde{F}_{\underline{r}}(\xi,x,t_{\underline{r}}) \\
           \widetilde{H}_{\underline{r}}(\xi,x,t_{\underline{r}}) & -i\widetilde{G}_{\underline{r}}(\xi,x,t_{\underline{r}}) \\
         \end{array}
       \right),
    \end{split}
  \end{equation}
  and
  \begin{align}
  G_{\underline{n}}(\xi,x,t_{\underline{r}})&=\sum_{\ell=0}^{n_-}\xi^{-2\ell}g_{n_--\ell,-}
  (x,t_{\underline{r}})+\sum_{\ell=1}^{n_+}
  \xi^{2\ell}g_{n_+-\ell,+}(x,t_{\underline{r}}),\label{4.6a}\\
  F_{\underline{n}}(\xi,x,t_{\underline{r}})&=\sum_{\ell=0}^{n_-}\xi^{-2\ell+1}f_{n_--\ell,-}(x,t_{\underline{r}})
  +\sum_{\ell=0}^{n_+}
  \xi^{2\ell}f_{n_+-\ell,+}(x,t_{\underline{r}}),\label{4.6b}\\
  H_{\underline{n}}(\xi,x,t_{\underline{r}})&=\sum_{\ell=0}^{n_-}\xi^{-2\ell+1}h_{n_--\ell,-}(x,t_{\underline{r}})
  +\sum_{\ell=0}^{n_+}
  \xi^{2\ell}h_{n_+-\ell,+}(x,t_{\underline{r}}),\label{4.6c}\\
  \widetilde{G}_{\underline{r}}(\xi,x,t_{\underline{r}})&=\sum_{s=0}^{r_-}\xi^{-2s}\tilde{g}_{r_--s,-}(x,t_{\underline{r}})
  +
  \sum_{s=1}^{r_+}
  \xi^{2s}\tilde{g}_{r_+-s,+}(x,t_{\underline{r}}),\label{4.6d}\\
  \widetilde{F}_{\underline{n}}(\xi,x,t_{\underline{r}})&=\sum_{s=0}^{r_-}\xi^{-2s+1}\tilde{f}_{r_--s,-}(x,t_{\underline{r}})
  +
  \sum_{s=0}^{r_+}
  \xi^{2s}\tilde{f}_{r_+-s,+}(x,t_{\underline{r}}),\label{4.6f}\\
  \widetilde{H}_{\underline{n}}(\xi,x,t_{\underline{r}})&=\sum_{s=0}^{r_-}\xi^{-2s+1}\tilde{h}_{r_--s,-}(x,t_{\underline{r}})
  +\sum_{s=0}^{r_+}\xi^{2s}\tilde{h}_{r_+-s,+}(x,t_{\underline{r}}),\label{4.6g}
  \end{align}
  for fixed $\underline{n},\underline{r}\in\mathbb{N}_0^2\backslash\{(0,0)\}$. Here $f_{\ell,\pm},$
 $g_{\ell,\pm},$ $h_{\ell,\pm},$
  $\tilde{f}_{s},$
 $\tilde{g}_{s}$, and $\tilde{h}_{s}$
 are defined as in
 (\ref{2.3})-(\ref{2.10}), with $q(x)$ replaced by $q(x,t_{\underline{r}}),$ etc, and with appropriate
 integration constants $c_{\ell,\pm},\ell\in\mathbb{N}$, and $\tilde{c}_{s,\pm},s\in\mathbb{N}$.
 Explicitly, (\ref{4.3}) and (\ref{4.4}) are equivalent to
  \begin{equation}\label{4.7}
  \begin{split}
  &0=-iG_{\underline{r},x}(\xi,x,t_{\underline{r}})+q_x(x,t_{\underline{r}})\xi H_{\underline{r}}(\xi,x,t_{\underline{r}})
           +r_x(x,t_{\underline{r}})\xi F_{\underline{r}}(\xi,x,t_{\underline{r}}),   \\
  &q_{xt_{\underline{r}}}(x,t_{\underline{r}})\xi= -F_{\underline{r},x}(\xi,x,t_{\underline{r}})-2izF_{\underline{r}}(\xi,x,t_{\underline{r}})+2iq_x(x,t_{\underline{r}})\xi G_{\underline{r}}(\xi,x,t_{\underline{r}}),  \\
      &r_{xt_{\underline{n}}}(x,t_{\underline{r}})\xi =H_{\underline{n},x}(\xi,x,t_{\underline{r}})-2ir_x(x,t_{\underline{r}})\xi G_{\underline{n}}(\xi,x,t_{\underline{r}})-2izH_{\underline{n}}(\xi,x,t_{\underline{r}}),
      \\
   &0=-iG_{\underline{n},x}(\xi,x,t_{\underline{r}})+\xi q_x(x,t_{\underline{r}}) H_{\underline{n}}(\xi,x,t_{\underline{r}})+\xi r_x(x,t_{\underline{r}}) F_{\underline{n}}(\xi,x,t_{\underline{r}}) , \\
    &0=F_{\underline{n},x}(\xi,x,t_{\underline{r}})+2izF_{\underline{n}}(\xi,x,t_{\underline{r}})-2i\xi q_x(x,t_{\underline{r}})G_{\underline{n}}(\xi,x,t_{\underline{r}}),  \\
     &0=-H_{\underline{n},x}(\xi,x,t_{\underline{r}})+2izH_{\underline{n}}(\xi,x,t_{\underline{r}})
     +2ir_x(x,t_{\underline{r}})\xi G_{\underline{n}}(\xi,x,t_{\underline{r}}).
 \end{split}
 \end{equation}
 Equation (\ref{4.7}) then yields
 \begin{equation}\label{x02}
   \frac{d}{dx}\textrm{det}(V_{\underline{n}}(\xi,x,t_{\underline{r}}))=\frac{d}{dx}
   \left(G_{\underline{n}}^2(\xi,x,t_{\underline{r}})
   +F_{\underline{n}}(\xi,x,t_{\underline{r}})H_{\underline{n}}(\xi,x,t_{\underline{r}})\right)=0,
 \end{equation}
  and meanwhile (cf. Lemma \ref{lemma4.2})
  \begin{equation}\label{x01}
    \frac{d}{dt_{\underline{r}}} \mathrm{det}(V_{\underline{n}}(\xi,x,t_{\underline{r}}))=\frac{d}{dt_r}
    \Big( G_{\underline{n}}(z,x,t_{\underline{r}})^2+F_{\underline{n}}(\xi,x,t_{\underline{r}})H_{\underline{n}}(\xi,x,t_{\underline{r}}) \Big)=0.
  \end{equation}
  Hence, $ G_{\underline{n}}(\xi)^2+zF_{\underline{n}}(\xi)H_{\underline{n}}(\xi)$ is independent of variables
  both $x$ and $t_{\underline{r}}$, which implies the basic identity (\ref{2.29})
   \begin{equation}\label{x02}
   G_{\underline{n}}(\xi,x,t_{\underline{r}})^2+ F_{\underline{n}}(\xi,x,t_{\underline{r}})H_{\underline{n}}(\xi,x,t_{\underline{r}})=R_{\underline{n}}(\xi)
   \end{equation}
   holds and the hyperelliptic curve
   $\mathcal{K}_n$ is still given by (\ref{2.32}).

   As in the stationary context (\ref{3.8}), (\ref{3.9}) we introduce
   \begin{equation}\label{4.8}
          \hat{\mu}_j(x,t_{\underline{r}})=(\mu_j(x,t_{\underline{r}}),
          -\mu_j(x,t_{\underline{r}})^{2n_-}G_{\underline{n}}(\mu_j(x,t_{\underline{r}}),x,t_{\underline{r}})),\quad j=1,\ldots,n,
        \end{equation}
   and
   \begin{equation}\label{4.9}
          \hat{\nu}_j(x,t_{\underline{r}})=(\nu_j(x,t_{\underline{r}}),\nu_j(x,t_{\underline{r}})^{2n_-}
          G_{\underline{n}}(\nu_j(x,t_{\underline{r}}),x,t_{\underline{r}})),\quad j=1,\ldots,n.
   \end{equation}

   In analogy to (\ref{3.10}), one defines the
   following meromorphic function $\phi(\cdot,x,t_{\underline{r}})$
   on $\cur,$
  \begin{align}\label{4.10}
          \phi(P,x,t_{\underline{r}})&=-\frac{i\xi^{-2n_-}y-iG_{\underline{n}}
          (\xi,x,t_{\underline{r}})}{F_{\underline{n}}(\xi,x,t_{\underline{r}})}\nonumber\\
                   &=\frac{H_{\underline{n}}(\xi,x,t_{\underline{r}})}{i\xi^{-2n_-}y+
                   iG_{\underline{n}}(\xi,x,t_{\underline{r}})},
        \end{align}
  with divisor of $\phi(\cdot,x,t_{\underline{r}})$ given by
        \begin{equation}\label{4.11}
          (\phi(\cdot,x,t_{\underline{r}}))=\mathcal{D}_{P_{0,-}\underline{\hat{\nu}}
          (x,t_{\underline{r}})}
          -\mathcal{D}_{P_{\infty+}\underline{\hat{\mu}}(x,t_{\underline{r}})}.
        \end{equation}
The time-dependent Baker-Ahiezer function $\psi$ is then defined in terms of $\phi$
by
  \begin{align}
          \psi(P,x,x_0,t_{\underline{r}},t_{0,\underline{r}}) = &\left(
                    \begin{array}{c}
                      \psi_1(P,x,x_0,t_{\underline{r}},t_{0,\underline{r}}) \\
                      \psi_2(P,x,x_0,t_{\underline{r}},t_{0,\underline{r}}) \\
                    \end{array}
                  \right),
          \nonumber \\
          \psi_1(P,x,x_0,t_{\underline{r}},t_{0,\underline{r}})
           =&\exp\Big(
          \int_{t_{0,\underline{r}}}^{t_{\underline{r}}}ds(i\widetilde{G}_{\underline{r}}(\xi,x_0,s)
          -\widetilde{F}_{\underline{r}}(\xi,x_0,s)\phi(P,x_0,s))
          \nonumber\\
          &+\int_{x_0}^xdx^\prime\left(-iz+q_x(x^{\prime},t_{\underline{r}})
          \xi\phi(P,x^\prime,t_{\underline{r}})\right)\Big),\label{4.12}\\
          \psi_2(P,x,x_0,t_{\underline{r}},t_{0,\underline{r}})=&\phi(P,x,t_{\underline{r}})
          \psi_1(P,x,x_0,t_{\underline{r}},t_{0,\underline{r}}),\nonumber\\
          &~~~~~P=(\xi,y)\in\cur\backslash\{P_{0,-},P_{\infty+}\}, (x,t_{\underline{r}})\in\mathbb{R}^2,\label{4.13}
        \end{align}
     with fixed $(x_0,t_{0,\underline{r}})\in\mathbb{R}^2.$

     The following lemma records basic properties of $\phi$ and $\psi$
     in analogy to the stationary case discussed in Lemma \ref{lemma3.1}.

     \newtheorem{lem4.1}{Lemma}[section]
     \begin{lem4.1}
     Assume $(\ref{2.2})$ and suppose that $(\ref{4.3}), (\ref{4.4})$ hold.

     $(i)$ Let $P=(\xi,y)\in\cur\backslash\{P_{0,-},P_{\infty+}\}$ and
     $(x,x_0,t_{\underline{r}},t_{0,\underline{r}})\in\mathbb{R}^4.$ Then $\phi$ satisfies
     \begin{equation}\label{4.14}
       \phi_x(P)=r_x\xi+2iz\phi(P)-q_x\xi\phi^2(P),
     \end{equation}
     and
     \begin{align}
      %\phi_x(P)=r_x\xi&+2iz\phi(P)-q_x\xi\phi^2(P),\label{4.14}\\
          (q_x\xi\phi(P))_{t_{\underline{r}}}=&\left(-\widetilde{F}_{\underline{r}}\phi(P)+i\widetilde{G}_{\underline{r}}\right)_x,
          \label{4.14a}\\
           %\left(iG_{\underline{r},x}-r_x\xi F_{\underline{r}}-2iG_{\underline{r}}\xi q_x\phi(P)\right)\\
           \phi_{t_{\underline{r}}}(P)=&(q_x\xi)^{-1}\Big(2iz\widetilde{F}_{\underline{r}}\phi(P)
           -\widetilde{F}_{\underline{r}}\phi_x(P)
           +i\widetilde{G}_{\underline{r},x}\Big)
          -2i\widetilde{G}_{\underline{r}}\phi(P)\nonumber\\
           =&\widetilde{H}_{\underline{r}}-2i\widetilde{G}_{\underline{r}}\phi(P)+
           \widetilde{F}_{\underline{r}}\phi^2(P),
           \label{4.14b}\\
       \phi(P)\phi(P^*) =&\frac{H_{\underline{n}}(\xi)}{F_{\underline{n}}(\xi)}, \label{4.15a}\\
          \phi(P)+\phi(P^*) =&\frac{2iG_{\underline{n}}(\xi)}{F_{\underline{n}}(\xi)},\label{4.15b}\\
           \phi(P)-\phi(P^*)=&-\frac{2i\xi^{-2n_-}y(P)}{F_{\underline{n}}(\xi)}.\label{4.15c}
        \end{align}

        $(ii)$
        Assuming $P=(z,y)\in\cur\backslash\{P_{0,\pm}\}$, then
        $\psi$ satisfies
           \begin{align}
           \psi_x(P)&=U(\xi)\psi(P), \label{4.16a} \\
           V_{\underline{n}}(\xi,x)\psi(P)&=i\xi^{-2n_-}y(P)\psi(P),\label{4.16b}\\
           \psi_{t_{\underline{r}}}(P)&=\widetilde{V}_{t_{\underline{r}}}(\xi)\psi(P).\label{4.16c}
          \end{align}
           and one derives
           \begin{align}
          \psi_1(P,x,x_0,t_{\underline{r}},t_{0,\underline{r}})
           =&\sqrt{\frac{F_{\underline{n}}(\xi,x,t_{\underline{r}})}{F_{\underline{n}}(\xi,x_0,t_{0,\underline{r}})}}
           \exp\Big(
           \int_{t_{0,\underline{r}}}^{t_{\underline{r}}}ds\Big(\frac{i\xi^{2n_-}y\widetilde{F}_{\underline{r}}(\xi,x_0,s)}
           {F_{\underline{n}}(\xi,x_0,s)}\Big)\nonumber\\
           &-\int_{x_0}^{x}dx^\prime\Big(\frac{q_x(x^\prime)\xi^{-2n_-+1}y(P)}
          {F_{\underline{n}}(\xi,x^\prime)}\Big)\Big),\label{y1}
          \end{align}
          and
          \begin{align}
          \psi_1(P,x,x_0,t_{\underline{r}},t_{0,\underline{r}})\psi_1(P^*,x,x_0,t_{\underline{r}},
          t_{0,\underline{r}}) & =\frac{F_{\underline{n}}(\xi,x,t_{\underline{r}})}{F_{\underline{n}}(\xi,x_0,t_{0,\underline{r}})}, \label{4.17a}\\
          \psi_2(P,x,x_0,t_{\underline{r}},t_{0,\underline{r}})
          \psi_2(P^*,x,x_0,t_{\underline{r}},t_{0,\underline{r}}) & =\frac{H_{\underline{n}}(\xi,x,t_{\underline{r}})}{F_{\underline{n}}(\xi,x_0,t_{0,\underline{r}})},
          \label{4.17b}\\
          \psi_1(P,x,x_0,t_{\underline{r}},t_{0,\underline{r}})
          \psi_2(P^*,x,x_0,t_{\underline{r}},t_{0,\underline{r}})&+\psi_1(P^*,x,x_0,t_{\underline{r}},
          t_{0,\underline{r}})\psi_2(P,x,x_0,t_{\underline{r}},t_{0,\underline{r}})\nonumber\\
          &=\frac{2iG_{\underline{n}}
          (\xi,x,t_{\underline{r}})}{F_{\underline{n}}(\xi,x_0,t_{0,\underline{r}})}.\label{4.17c}
        \end{align}
        In addition, as long as the zeros of $F_{\underline{r}}(\cdot,x,t_{\underline{r}})$
        are all simple for $(x,t_{\underline{r}})\in\Omega,\Omega\subseteq\mathbb{R}^2$ open and connected,
        $\psi(P,x_0,t_{\underline{r}},t_{0,\underline{r}})$
        is meromorphic on $\cur\backslash\{P_{0,\pm}\}$ for $(x,t_{\underline{r}}), (x_0,t_{0,\underline{r}})\in\Omega.$

     \end{lem4.1}
     \proof
     Equation (\ref{4.14}), (\ref{4.15a})-(\ref{4.15c}), (\ref{4.17a})-(\ref{4.17c})
     are proved as in Lemma \ref{lemma3.1}. To prove (\ref{4.14a}) and (\ref{4.14b})
     one first observes that
     \begin{equation}\label{O1}
       (\partial_x-2iz+2q_x\xi\phi-\frac{q_{xx}}{q_{x}})
       \left((q_x\phi)_{t_r}+\xi^{-1} (\widetilde{F}_{\underline{r}}\phi-i\widetilde{G}_{\underline{r}})_x\right)=0
     \end{equation}
     using (\ref{4.14}) and relations (\ref{4.7}) repeatedly.
     Thus,
     \begin{align}\label{O2}
       (q_x\phi)_{t_r}+\xi^{-1} (\widetilde{F}_{\underline{r}}\phi-i\widetilde{G}_{\underline{r}})_x
       =C\exp\left(\int_{x_0}^xdx^\prime(2iz-2q_x\xi\phi+\frac{q_{xx}}{q_{x}})\right),
     \end{align}
     where the left-hand side is meromorphic in a neighborhood of $P_{\infty+},$
     while the right-hand side is not meromorphic near $P_{\infty+}$ only if $C=0.$
     This proves (\ref{4.14a}). Equation (\ref{4.14b}) is an immediate consequence of
     (\ref{4.7}) and (\ref{4.14a}). Relations (\ref{4.16a})-(\ref{4.16c})
     are clear from (\ref{4.12}), (\ref{4.13}), (\ref{4.14}) and (\ref{4.14b}).
     (\ref{y1}) follows by (\ref{4.7}), (\ref{4.10}), (\ref{4.12}) and (\ref{x1}).
     %By (\ref{4.12}),
     That $\psi_1(\cdot,x,x_0,t_r,t_{0,r})$ is meromorphic on $\cur\backslash\{P_{0,\pm}\}$
     if $F_{\underline{n}}(\cdot,x,t_r)$
     has only simple zeros is a consequence of
     \begin{equation}\label{O3}
       -iz+q_x\xi\phi(P,x^\prime,t_{r})\underset{P\rightarrow \hat{\mu}_j(x^\prime,t_{\underline{r}})}{=}\partial_{x^\prime}\ln(F_{\underline{n}}(\xi,x^\prime,t_{\underline{r}}))+O(1)
     \end{equation}
     as $\xi\rightarrow\mu_j(x^\prime,t_{\underline{r}}),$
     using (\ref{4.7}), (\ref{4.8}) and (\ref{4.10})
     and
     \begin{equation}\label{O4}
       -\widetilde{F}_{\underline{r}}(\xi,x_0,s)\phi(P,x_0,s)\underset{P\rightarrow \hat{\mu}_j(x_0,s)}{=}\partial_{s}\ln(F_{\underline{n}}(\xi,x_0,s))+O(1),
     \end{equation}
     using (\ref{4.7}), (\ref{4.8}) and (\ref{x1}).
     This follows from (\ref{4.12})
     by restricting $P$
     to a sufficiently small neighborhood $\mathcal{U}_j(x_0)$
     of $\{\hat{\mu}_j(x_0,s)\in\cur|(x_0,s)\in\Omega,s\in [t_{0,\underline{r}},t_{\underline{r}}]\}$
     such that $\hat{\mu}_k(n_0,s)\in\mathcal{U}_j(x_0)$ for all $s\in [t_{0,\underline{r}}, t_{\underline{r}} ]$ and for all $k\in\{1,\ldots,n\}\backslash\{j\}$
     and by simultaneously restricting $P$ to a sufficiently small neighborhood $\mathcal{U}_j(t_{\underline{r}})$ of $\{\hat{\mu}_j(x^\prime,t_{\underline{r}})\in\cur|(x^\prime,t_{\underline{r}})\in\Omega, x^\prime\in [x_0, x ]\}$ such that $\hat{\mu}_k(x^\prime,t_{\underline{r}})\notin\mathcal{U}_j(t_{\underline{r}})$
     for all $x^\prime\in[x_0,x]$ and all $k\in\{1,\ldots,n\}\backslash\{j\}.$
     \qed\vspace{0.3cm}

     Next we consider the $t_{\underline{r}}$-dependence of
     $F_{\underline{n}},G_{\underline{n}},H_{\underline{n}}.$
     \newtheorem{lem4.2}[lem4.1]{Lemma}
     \begin{lem4.2}\label{lemma4.2}
     Assume $(\ref{2.2})$ and suppose that $(\ref{4.3}), (\ref{4.4})$ hold. Then
     \begin{align}
       F_{\underline{n},t_{\underline{r}}} & =2i(\widetilde{G}_{\underline{r}}F_{\underline{n}}-G_{\underline{n}}\widetilde{F}_{\underline{r}}), \label{x1}\\
       G_{\underline{n},t_{\underline{r}}}&
       =i(\widetilde{F}_{\underline{r}}H_{\underline{n}}-F_{\underline{n}}\widetilde{H}_{\underline{r}}),
       \label{x2}\\
       H_{\underline{n},t_{\underline{r}}}&
       =2i(G_{\underline{n}}\widetilde{H}_{\underline{r}}-\widetilde{G}_{\underline{r}}H_{\underline{n}})
       \label{x3}.
     \end{align}
     In addition, $(\ref{x1})$-$(\ref{x3})$ are equivalent to
    \begin{equation}\label{x4}
     -V_{\underline{n},t_{\underline{r}}}+[\widetilde{V}_{\underline{r}},V_{\underline{n}}]=0,
    \end{equation}
    and hence
    $(\ref{x01})$ holds.
     \end{lem4.2}
    \proof
    We proves (\ref{x1})
    by using (\ref{4.15c}) which shows that
    \begin{equation}\label{4.17}
      (\phi(P)-\phi(P^*))_{t_{\underline{r}}}=
      \frac{2i\xi^{-2n_-}yF_{\underline{n},t_{\underline{r}}}}{F_{\underline{n}}^2}.
    \end{equation}
    However, the left-hand side of (\ref{4.17}) also equals
    \begin{equation}\label{4.18}
      (\phi(P)-\phi(P^*))_{t_{\underline{r}}}=\frac{2i\xi^{-2n_-}y}{F_{\underline{n}}}
      (2i \widetilde{G}_{\underline{r}}-\frac{2i\widetilde{F}G_{\underline{n}}}{F_{\underline{n}}}),
    \end{equation}
    using (\ref{4.14b}), (\ref{4.15b}) and (\ref{4.15c}). Combing (\ref{4.17}) and (\ref{4.18})
     proves (\ref{x1}). Similarly, starting from (\ref{4.15b})
     \begin{equation}\label{4.19}
       (\phi(P)+\phi(P^*))_{t_{\underline{r}}}=2iF_{\underline{n}}^{-2}
       (G_{\underline{n},t_{\underline{r}}}F_{\underline{n}}-F_{\underline{n},t_{\underline{r}}}G_{\underline{n}})
     \end{equation}
     yields (\ref{x2}).
     (\ref{x3}) is a consequence of (\ref{4.7}), (\ref{x1}) and (\ref{x2}).
     Finally, differentiating $G_{\underline{n}}(\xi,x,t_{\underline{r}})^2+ F_{\underline{n}}(\xi,x,t_{\underline{r}})H_{\underline{n}}(\xi,x,t_{\underline{r}})$
     with respect to $t_{\underline{r}}$, and using (\ref{x1})-(\ref{x3})
     then yields $R_{\underline{n},t_{\underline{r}}}=0$, or equivalently, (\ref{x01}).
     %which indicates the $t_{\underline{r}}$-independence of $R_{\underline{n}}$ defined by (\ref{x02}).
    \qed\vspace{0.3cm}

    Next we turn to the Dubrovin-type equations, which governs the dynamics of
    of $\mu_j$ and $\nu_j$ with respect to variations of $x$ and $t_{\underline{r}}.$

    \newtheorem{lem4.3}[lem4.1]{Lemma}
    \begin{lem4.3}
     Suppose $(\ref{2.2})$, $(\ref{4.3})$, $(\ref{4.4})$
        on an open and connected interval $\widetilde{\Omega}_\mu\subseteq\mathbb{R}^2$.
        Suppose that the zeros $\{\mu_j(\cdot)\}_{j=0,\ldots,n}$
        of $\xi^{2n_--1}F_{\underline{n}}(\cdot)$ remain distinct and nonzero on $\widetilde{\Omega}_\mu.$ Then
        $\{\hat{\mu}_j(x)\}_{j=0,\ldots,n}$ defined by $(\ref{4.8})$, satisfies the
         following first-order system of differential equations
          \begin{align}
          \mu_{j,x}(x,t_{\underline{r}})  &=\frac{-2iy(\hat{\mu}_j(x,t_{\underline{r}}))}{\prod_{k=1,k\neq j}^{n}(\mu_j(x,t_{\underline{r}})-\mu_k(x,t_{\underline{r}}))}, \label{4.20}\\
          \mu_{j,t_{\underline{r}}}(x,t_{\underline{r}})  &=\frac{-2iy(\hat{\mu}_j(x,t_{\underline{r}}))\widetilde{F}_{\underline{r}}(\mu_j(x,t_{\underline{r}}))}
          {q_x(x,t_{\underline{r}})\mu_j(x,t_{\underline{r}})\prod_{k=1,k\neq j}^{n}(\mu_j(x,t_{\underline{r}})-\mu_k(x,t_{\underline{r}}))},\nonumber\\
          &~~~~~~~~~~~~~~~~~~~~~~~
          ~~~~~~~~~~~~~\ \ j=1,\ldots,n,\,\, (x,t_{\underline{r}})\in\widetilde{\Omega}_\mu \label{4.20a}.
          \end{align}
        %\begin{equation}\label{3.38}
    %       \mu_{j,x}= 2 \frac{ y(\hat{\mu}_j)}{\mu_j}
     %       \prod_{\scriptstyle k=0 \atop \scriptstyle k \neq j }^{n}
      %      (\mu_j(x)-\mu_k(x))^{-1}, \ \ j=0,\ldots,n, \ \ x\in\Omega_\mu.
       % \end{equation}
   Next, assume $\mathcal{K}_n$ to be nonsingular and introduce initial
   condition
   \begin{equation}\label{4.21}
   \{\hat{\mu}_j(x_0,t_{0,\underline{r}})\}_{j=1,\ldots,n}\subset\mathcal{K}_n
   \end{equation}
   for some $(x_0,t_{0,\underline{r}})\in\mathbb{R}^2,$ where $\mu_j(x_0,t_{0,\underline{r}})\neq 0,j=1,\ldots,n,$
   are assumed to be distinct. Then there exists an open interval
   $ \Omega_{\mu}\subseteq\mathbb{R},$ with $x_0\in \Omega_\mu,$
   such that the initial value problem $(\ref{4.20}$-$(\ref{4.21})$
   has a unique solution $\{\hat{\mu}_j\}_{j=1,\ldots,n}\subset\mathcal{K}_n$ satisfying
   \begin{equation}\label{4.22}
      \hat{\mu}_j \in C^\infty(\Omega_\mu,\mathcal{K}_{n}),
      \quad j=0,\ldots,n,
       \end{equation}
     and $\mu_j, j=1,\ldots,n,$ remain distinct and nonzero on $\Omega_{\mu}.$

     \noindent For the zeros $\{\nu_j(\cdot)\}_{j=1,\ldots,n}$
     of $\xi^{2n_--1}H_{\underline{n}}(\cdot)$
     similar statements hold with $\mu_j$ and $\Omega_\mu$
     replaced by $\nu_j$ and $\Omega_\nu,$ etc. In particular, $\{\hat{\nu}_j\}_{j=1,\ldots,n},$
     defined by $(\ref{4.9})$, satisfies the system
     %\begin{equation}\label{3.41}
      %     \nu_{l,x}=-2\frac{u_{xx}~y(\hat{\nu}_l)}{h_0 ~ \nu_l}
       %     \prod_{\scriptstyle k=1 \atop \scriptstyle k \neq l }^{n}
        %    (\nu_l(x)-\nu_k(x))^{-1}, \quad l=1,\ldots,n,\quad x\in\widetilde{\Omega}_\nu
        %\end{equation}
       \begin{align}
         \nu_{j,x}(x,t_{\underline{r}})  &=\frac{-2iy(\hat{\mu}_j(x,t_{\underline{r}}))}{\prod_{k=1,k\neq j}^{n}(\nu_j(x,t_{\underline{r}})-\nu_k(x,t_{\underline{r}}))}, \label{4.23b}\\
        \nu_{j,x}(x,t_{\underline{r}})  &=\frac{2iy(\hat{\nu}_j(x,t_{\underline{r}}))
        \widetilde{H}_{\underline{n}}(\nu_j(x,t_{\underline{r}}))}{r_x(x,t_{\underline{r}})\nu_j(x,t_{\underline{r}})\prod_{k=1,k\neq j}^{n}(\nu_j(x,t_{\underline{r}})-\nu_k(x,t_{\underline{r}}))},\nonumber\\
        &~~~~~~~~~~~~~~~~~~~~~~~~~~~~~~~~~~~~~~~~~~\ j=1,\ldots,n,\,\, x\in \Omega_\nu.\label{4.23c}
        \end{align}

    \end{lem4.3}
\proof
It suffices to prove (\ref{4.20a}) since the argument for (\ref{4.23b})
is analogous and that for (\ref{4.20}) and (\ref{4.23c}) has been given in the
proof of Lemma \ref{lemma3.2}. Inserting $\xi=\mu_j(x,t_{\underline{r}})$ into
(\ref{x1}), observing (\ref{4.8}), yields
\begin{align}\label{4.24}
  F_{\underline{n},t_{\underline{r}}}(\mu_j)&=-q_x\mu_j^{-2n_-+1}\mu_{j,x}\prod_{k=1,k\neq j}^n(\mu_j-\mu_k)=-2iG_{\underline{n}}(\mu_j)\widetilde{F}_{\underline{r}}(\mu_j),\nonumber\\
  &=2i\xi^{-2n_-}y(\hat{\mu}_j)\widetilde{F}_{\underline{r}}(\mu_j).
\end{align}
which indicates (\ref{4.20a}).
\qed\vspace{0.2cm}

Since the stationary trace formulas for $f_{\ell,\pm}$ and $h_{\ell,\pm}$
in terms of symmetric functions of the zeros
  $\mu_j$ and $\nu_\ell$ of $(\cdot)^{2n_--1}F_{\underline{n}}(\cdot)$ and $(\cdot)^{2n_--1}H_{\underline{n}}(\cdot)$ in Lemma \ref{lemma3.2} extend
  line by line to the corresponding time-dependent setting, we next record
  their $t_{\underline{r}}$-dependent analogs without proof. For simplicity we again
  confine ourselves to the simplest cases only.
\newtheorem{lem4.4}[lem4.1]{Lemma}
\begin{lem4.4}
Assume hypothesis $(\ref{2.2})$ and suppose that $(\ref{4.3})$ and $(\ref{4.4})$ hold.
Then,
\begin{align}
          &\frac{q_{xx}}{2iq_x} -\frac{1}{2}q_xr_x-c_{1,+} =-\sum_{j=1}^{n}\mu_j ,\label{4.21a} \\
          &\frac{r_{xx}}{2ir_x} +\frac{1}{2}q_xr_x+c_{1,+} =-\sum_{j=1}^{n}\nu_j ,\label{4.21b}\\
          &\frac{iq}{2q_x}=(-1)^n\prod_{j=1}^n\mu_j ,\label{4.21c}\\
          &\frac{ir}{2r_x}=(-1)^{n-1}\prod_{j=1}^n\nu_j.\label{4.21d}
 \end{align}

\end{lem4.4}
Next we turn to the asymptotic expansions of $\phi$ and $\psi$
in a neighborhood of $P_{\infty\pm}$ and $P_{0,\pm}.$

 \newtheorem{lem4.5}[lem4.1]{Lemma}
 \begin{lem4.5}
 Assume hypothesis $(\ref{2.2})$ and suppose that $(\ref{4.3})$ and $(\ref{4.4})$ hold.
 Moreover, let $P=(\xi,y)\in\cur\backslash\{P_{\infty\pm},P_{0,\pm}\}, \,(x,t_{{\underline{r}}})\in\mathbb{R}^2,\, (x,x_0,t_{0,\underline{r}},t_{\underline{r}})\in\mathbb{R}^4.$ Then
  \nopagebreak[4]
         \begin{align}
         \phi(P,x,t_{\underline{r}}) &\underset{\zeta \rightarrow 0}{=}
         \begin{cases} 2i[q_x(x,t_{\underline{r}})]^{-1}\zeta^{-1}+O(\zeta),
         \qquad &P \rightarrow P_{\infty+},\cr
         [ir_x(x,t_{\underline{r}})/2]\,\zeta+O(\zeta^3),&P \rightarrow P_{\infty-},
         \end{cases}
         \label{4.22}\\[0.15cm]
         \phi(P,x,t_{\underline{r}}) &\underset{\zeta \rightarrow 0}{=}
         \begin{cases} q(x,t_{\underline{r}})^{-1}\zeta^{-1}+O(\zeta),
         \qquad &P \rightarrow P_{0,+},\cr
         r(x,t_{\underline{r}})\,\zeta+O(\zeta^3),&P \rightarrow P_{0,-},
         \end{cases}
         \label{4.23}\\
         \psi_1(P,x,x_0,t_{\underline{r}},t_{0,\underline{r}}) &\underset{\zeta \rightarrow 0}{=}
          \begin{cases} e^{i(x-x_0)\zeta^{-2}+i(t_{\underline{r}}-t_{0,\underline{r}})\sum_{s=1}^{r_+}\tilde{c}_{r_+-s}\zeta^{-2s}+O(1)},
          \qquad &P \rightarrow P_{\infty+},\cr
            e^{-i(x-x_0)\zeta^{-2}-i(t_{\underline{r}}-t_{0,\underline{r}})\sum_{s=1}^{r_+}\tilde{c}_{r_+-s}\zeta^{-2s}+O(1)},&P \rightarrow P_{\infty-},
          \end{cases}
           \label{4.24}\\[0.15cm]
          \psi_1(P,x,x_0,t_{\underline{r}},t_{0,\underline{r}}) &\underset{\zeta \rightarrow 0}{=}
          \begin{cases} \frac{q(x,t_{\underline{r}})}{q(x_0,t_{0,\underline{r}})}(1+O(\zeta))
          e^{i(t_{\underline{r}}-t_{0,\underline{r}})\sum_{s=1}^{r_-}\tilde{c}_{r_--s}\zeta^{-2s}},
          \qquad &P \rightarrow P_{0,+},\cr
          (1+O(\zeta))e^{-i(t_{\underline{r}}-t_{0,\underline{r}})\sum_{s=1}^{r_-}\tilde{c}_{r_--s}\zeta^{-2s}},&P \rightarrow P_{0,-},
          \end{cases}
         \label{4.25}\end{align}
         with the local coordinates $\zeta=\xi^{-1}$ near $P_{\infty\pm}$
         and $\zeta=\xi$ near $P_{0,\pm}.$
        % \begin{align}
        %  \psi_2(P,x) &\underset{\zeta \rightarrow 0}{=}
        %  \begin{cases}\left(\frac{2i}{q_x(x)}\zeta^{-1}+O(\zeta)\right) e^{i(x-x_0)\zeta^{-2}+O(1)},
         % \qquad &P \rightarrow P_{\infty+},\cr
         %   \left(\frac{ir_x(x)}{2}\,\zeta+O(\zeta^3)\right)e^{-i(x-x_0)\zeta^{-2}+O(1)},&P \rightarrow P_{\infty-},
        %  \end{cases}
        %  \quad\zeta=\xi^{-1},\label{4.25}\\[0.15cm]
        %  \psi_2(P,x) &\underset{\zeta \rightarrow 0}{=}
        %  \begin{cases} \frac{1}{q(x_0)}\zeta^{-1}+O(\zeta),
        %  \qquad &P \rightarrow P_{0,+},\cr
       %   r(x)\zeta+O(\zeta^2),&P \rightarrow P_{0,-},
       %  \end{cases}
       %  \quad\zeta=\xi.\label{4.26}
       %  \end{align}

 \end{lem4.5}
\proof Since by the difinition of $\phi$
 in (\ref{4.10}) the time parameter can be viewed
 as an additional but fixed parameter, the asymptotic
 behavior of $\phi$ remains the same as in Lemma \ref{lemma3.1}.
Similarly, also the asymptotic behavior of $\psi_1(P,x,x_0,t_{\underline{r}},t_{\underline{r}})$
is derived in an identical fashion to that in Lemma \ref{lemma3.1}. This proves
(\ref{4.24}) and (\ref{4.25}) for $t_{0,\underline{r}}=t_{\underline{r}},$
that is,
\begin{align*}
\psi_1(P,x,x_0,t_{\underline{r}},t_{\underline{r}}) &\underset{\zeta \rightarrow 0}{=}
          \begin{cases} e^{i(x-x_0)\zeta^{-2}+O(1)},
          \qquad &P \rightarrow P_{\infty+},\cr
            e^{-i(x-x_0)\zeta^{-2}+O(1)},&P \rightarrow P_{\infty-},
          \end{cases}
          \\[0.15cm]
          \psi_1(P,x,x_0,t_{\underline{r}},t_{\underline{r}}) &\underset{\zeta \rightarrow 0}{=}
          \begin{cases} \frac{q(x,t_{\underline{r}})}{q(x_0,t_{\underline{r}})}(1+O(\zeta)),
          \qquad &P \rightarrow P_{0,+},\cr
          1+O(\zeta),&P \rightarrow P_{0,-}.
          \end{cases}
\end{align*}
It remains to investigate
\begin{equation}\label{4.26}
  \psi_1(P,x_0,x_0,t_{\underline{r}},t_{0,\underline{r}})
  =\exp\Big(\int_{t_{0,\underline{r}}}^{t_{\underline{r}}}ds(i\widetilde{G}_{\underline{r}}(\xi,x_0,s)
          -\widetilde{F}_{\underline{r}}(\xi,x_0,s)\phi(P,x_0,s))\Big).
\end{equation}
Next we compute the asymptotic expansions of the integrand in (\ref{4.26}).
Focusing on the homogeneous coefficients first, and then using the
relations
\begin{equation*}
  \widetilde{F}_{\underline{r}}\underset{\zeta\rightarrow 0}{=}\sum_{s=1}^{r_+}\tilde{c}_{r_+-s}\widehat{F}_{s,+}+O(1),\quad \widetilde{G}_{\underline{r}}\underset{\zeta\rightarrow 0}{=}\sum_{s=1}^{r_+}\tilde{c}_{r_+-s}\widehat{G}_{s,+}+O(1),
\end{equation*}
one finds
as $P\rightarrow P_{\infty\pm},$
\begin{align} \label{4.27}
  i \widetilde{G}_{\underline{r}}-\widetilde{F}_{\underline{r}}\phi& = i\widetilde{G}_{\underline{r}}+\widetilde{F}_{\underline{r}}F_{\underline{n}}^{-1}(i\xi^{-2n_-}y-iG_{\underline{n}})\nonumber\\
    &\underset{\zeta\rightarrow 0}{=}\frac{F_{\underline{n},t_{\underline{r}}}}{2 F_{\underline{n}}}+\frac{i\xi^{-2n_-}y\widetilde{F}_{\underline{r}}}{F_{\underline{n}}}\nonumber\\
   % &\underset{\zeta\rightarrow 0}{=}\frac{ q_{x,t_{\underline{r}}}}{ q_{t_{\underline{r}}}}+
   % (\hat{f}_{0,+})
   & \underset{\zeta\rightarrow 0}{=}
  \pm i\sum_{s=1}^{r_+}\tilde{c}_{r_+-s}\zeta^{-2s}+O(1).\,\,\,\,\,\zeta=1/\xi.
\end{align}
%Since
%\begin{equation*}
%  \widetilde{F}_{\underline{r}}\underset{\zeta\rightarrow 0}{=}\sum_{s=1}^{r_+}\tilde{c}_{r_+-s}\widehat{F}_{s,+}+O(1),\quad \widetilde{G}_{\underline{r}}\underset{\zeta\rightarrow 0}{=}\sum_{s=1}^{r_+}\tilde{c}_{r_+-s}\widehat{G}_{s,+}+O(1),
%\end{equation*}
%one infers from (\ref{4.27})
%\begin{equation}\label{4.28}
%  i\widetilde{G}_{\underline{r}}-\widetilde{F}_{\underline{r}}\underset{\zeta\rightarrow 0}{=}
 % \pm i\sum_{s=2}^{r_+}\tilde{c}_{r_+-s}\zeta^{-2s}+O(1).
%\end{equation}
Insertion of (\ref{4.27}) into (\ref{4.26})
then proves (\ref{4.24}) as $P\rightarrow P_{\infty\pm}.$
Similarly, as $P\rightarrow P_{0,\pm},$
\begin{align}\label{4.29}
 i \widetilde{G}_{\underline{r}}-\widetilde{F}_{\underline{r}}\phi &= i\widetilde{G}_{\underline{r}}+\widetilde{F}_{\underline{r}}F_{\underline{n}}^{-1}(i\xi^{-2n_-}y-iG_{\underline{n}})\nonumber\\
    &\underset{\zeta\rightarrow 0}{=}  i \widetilde{G}_{\underline{r}}+\widetilde{F}_{\underline{r}}(i-iG_{\underline{n}}\xi^{-2n_-}y)(F_{\underline{n}}\xi^{-2n_-}y)^{-1}\nonumber\\
    & \underset{\zeta\rightarrow 0}{=}
   \pm i\sum_{s=1}^{r_-}\tilde{c}_{r_--s}\zeta^{-2s}+
   \begin{cases}
   \frac{q_{t_{\underline{r}}}}{q}+O(1),&P\rightarrow P_{0,+},\cr
   O(\zeta)&P\rightarrow P_{0,-},
   \end{cases}
   ~~\zeta=\xi.
\end{align}
Insertion of (\ref{4.29}) into (\ref{4.26})
then proves (\ref{4.25}) as $P\rightarrow P_{0,\pm}.$\qed\vspace{0.25cm}

Next, we turn to the principal result of this section, the representation of
$\phi,\psi_1,q,r$ in terms of Riemann theta function associated with $\cur,$
assuming $\underline{n}=(n_-,n_+)\in\mathbb{N}_{0}^2\backslash\{(0,0)\}$ for the
remainder of this section.
In addition to (\ref{3.39}) and (\ref{3.42}), let $\Omega_{P_{\infty\pm},k}^{(2)}$
and $\Omega_{P_{0,\pm},k}^{(2)}$ be the normalized differentials of
the second kind with a unique pole at $P_{\infty\pm}$ and $P_{0,\pm}$, respectively,
and principal parts
\begin{align}
 \Omega_{P_{\infty\pm},k}^{(2)}\underset{\zeta\rightarrow 0}{=}(\zeta^{-2-k}+O(1))d\zeta,\,\, P\rightarrow P_{\infty\pm},\,\,\zeta=\xi^{-1},\,\,k\in\mathbb{N}_0,\label{4.30}\\
 \Omega_{P_{0,\pm},k}^{(2)}\underset{\zeta\rightarrow 0}{=}(\zeta^{-2-k}+O(1))d\zeta,\,\, P\rightarrow P_{0,\pm},\,\,\zeta=\xi^{-1},\,\,k\in\mathbb{N}_0,\label{4.31}
\end{align}
with vanishing $a$-periods,
\begin{equation*}
  \int_{a_j}\Omega_{P_{\infty\pm},k}^{(2)}=\int_{a_j}\Omega_{P_{0,\pm},k}^{(2)}=0,\quad j=1,\ldots,n.
\end{equation*}
Moreover, we define
\begin{align}\label{4.32}
  \widetilde{\Omega}_{\underline{r}}^{(2)}=& \Big(\sum_{s=1}^{r_-}2s\tilde{c}_{r_--s,-}
  (\Omega_{{P_{0,+},2s-1}}^{(2)}-\Omega_{{P_{0,-}},2s-1}^{(2)})\nonumber\\
  &+\sum_{s=1}^{r_+}2s\tilde{c}_{r_+-s,+}
  (\Omega_{{P_{\infty+}},2s-1}^{(2)}-\Omega_{{P_{\infty-}},2s-1}^{(2)})\Big)
\end{align}
and abbreviate
\begin{equation}\label{4.32a}
  \widetilde{\Omega}_{\underline{r}}^{\infty\pm}=\lim_{P\rightarrow P_{\infty\pm}}\Big(\int_{Q_0}^P\widetilde{\Omega}_{\underline{r}}^{(2)}\pm
  \sum_{s=1}^{r_+}\tilde{c}_{r_+-s}\zeta^{-2s}\Big),
\end{equation}
\begin{equation}\label{4.32b}
  \widetilde{\Omega}_{\underline{r}}^{0,\pm}=\lim_{P\rightarrow P_{0,\pm}}\Big(\int_{Q_0}^P\widetilde{\Omega}_{\underline{r}}^{(2)}\pm
  \sum_{s=1}^{r_-}\tilde{c}_{r_--s}\zeta^{-2s}\Big).
\end{equation}
The vector of $b$-periods of $\widetilde{\Omega}_{\underline{r}}^{(2)}$ is
denoted by
\begin{equation}\label{4.32c}
  \widetilde{\underline{U}}_{\underline{r}}^{(2)}
  =( \widetilde{U}_{\underline{r},1}^{(2)},\ldots, \widetilde{U}_{\underline{r},n}^{(2)}),\quad  \widetilde{U}_{\underline{r},j}^{(2)}=\frac{1}{2\pi i}\int_{b_j}\widetilde{\Omega}_{\underline{r}}^{0,\pm},\quad j=1,\ldots,n.
\end{equation}

\newtheorem{thm4.6}[lem4.1]{Theorem}
\begin{thm4.6}
Assume $(\ref{2.2})$ and suppose that $(\ref{4.3})$ and $(\ref{4.4})$ hold
subject to the constraint $(\ref{3.2})$
on $\Omega$, where
$\Omega\subseteq\mathbb{R}^2$ is open and connected.
In addition, let $P\in\cur\backslash\{P_{\infty\pm},P_{0,\pm}\}$, $(x,t_{\underline{r}})\in\mathbb{R}^2$ and $(x,x_0,t_{\underline{r}},t_{0,\underline{r}})\in\mathbb{R}^4.$
Moreover, suppose that $\mathcal{D}_{\underline{\hat{\mu}}(x,t_{\underline{r}})}$, or equivalently,
$\mathcal{D}_{\underline{\hat{\nu}}(x,t_{\underline{r}})}$, is nonspecial for $(x,t_{\underline{r}})\in\Omega.$ Then $\phi,\psi_1,q,r$
admit the following representations
 \begin{align}
       \phi(P,x,t_{\underline{r}})=&~C(x,t_{\underline{r}})
       \frac{\theta(\underline{\xi}(P,\hat{\underline{\nu}}(x,t_{\underline{r}})))}
       {\theta(\underline{\xi}(P,\hat{\underline{\mu}}(x,t_{\underline{r}})))}\exp\left(\int_{Q_0}^P
       \Omega_{P_{0,-},P_{\infty+}}^{(3)}\right),\label{4.33}\\
        \psi_1(P,x,x_0,t_{\underline{r}},t_{0,\underline{r}})=&~C(x,x_0,t_{\underline{r}},t_{0,\underline{r}})
        \frac{\theta(\underline{\xi}(P,\hat{\underline{\mu}}(x,t_{\underline{r}})))}
       {\theta(\underline{\xi}(P,\hat{\underline{\mu}}(x_0,t_{0,\underline{r}})))}\nonumber\\
       &~~~~\times
       \exp\Big(-i(x-x_0)\int_{Q_0}^P\Omega_0^{(2)}-i(t_{\underline{r}}-t_{0,\underline{r}})
       \int_{Q_0}^{P}\widetilde{\Omega}_{\underline{r}}^{(2)}\Big),\label{4.34}
        \end{align}
where
  \begin{align}
        & C(x,t_{\underline{r}})=\frac{1}{q(x_0,t_{0,\underline{r}})}
        \frac{\theta(\underline{\xi}(P_{0,-},\hat{\underline{\mu}}(x,t_{\underline{r}})))}
       {\theta(\underline{\xi}(P_{0,-},\hat{\underline{\mu}}(x_0,t_{0,\underline{r}})))}
       \frac{\theta(\underline{\xi}(P_{0,+},\hat{\underline{\mu}}(x_0,t_{0,\underline{r}})))}
       {\theta(\underline{\xi}(P_{0,+},\hat{\underline{\nu}}(x,t_{\underline{r}})))}\nonumber\\
       &~~~~~~~~~\times e^{-i(x-x_0)(e_{0,-}-e_{0,+})-i(t_{\underline{r}}-t_{0,\underline{r}})
       \widetilde{\Omega}_{\underline{r}}^{0,-}-\omega_{0}^{0,+}},\label{ctx}\\
       &C(x,x_0,t_{\underline{r}},t_{0,\underline{r}})=
       \frac{\theta(\underline{\xi}(P_{0,-},\hat{\underline{\mu}}(x_0,t_{0,\underline{r}})))}
       {\theta(\underline{\xi}(P_{0,-},\hat{\underline{\mu}}(x,t_{\underline{r}})))}
       e^{i(x-x_0)e_{0,-}+i(t_{\underline{r}}-t_{0,\underline{r}})
       \widetilde{\Omega}_{\underline{r}}^{0,-}}.\label{ctxx}
       \end{align}
       The Abel map linearizes the auxiliary divisors
       $\mathcal{D}_{\hat{\underline{\mu}}(x,t_{\underline{r}})},
       \mathcal{D}_{\hat{\underline{\nu}}(x,t_{\underline{r}})}$
       in the sense that
         \begin{align}
         \underline{\alpha}_{Q_0}(\mathcal{D}_{\underline{\hat{\mu}}(x,t_{\underline{r}})})
         =&\underline{\alpha}_{Q_0}(\mathcal{D}_{\underline{\hat{\mu}}(x_0,t_{0,\underline{r}})})
         -i\underline{U}_0^{(2)}(x-x_0)-i\widetilde{\underline{U}}_{\underline{r}}^{(2)}(t-t_{0,\underline{r}}),\label{3.52AA}\\
         \underline{\alpha}_{Q_0}(\mathcal{D}_{\underline{\hat{\nu}}(x,t_{\underline{r}})})
         =&\underline{\alpha}_{Q_0}(\mathcal{D}_{\underline{\hat{\nu}}(x_0,t_{0,\underline{r}})})
         -i\underline{U}_0^{(2)}(x-x_0)-i\widetilde{\underline{U}}_{\underline{r}}^{(2)}(t-t_{0,\underline{r}}).\label{3.52BB}
       \end{align}
  Moreover, one derives
      \begin{align}
       & q(x,t_{\underline{r}})= q(x_0,t_{0,\underline{r}})\frac{\theta(\underline{\xi}(P_{0,-},
        \hat{\underline{\mu}}(x_0,t_{0,\underline{r}})))}
       {\theta(\underline{\xi}(P_{0,-},\hat{\underline{\mu}}(x,t_{\underline{r}})))}
       \frac{\theta(\underline{\xi}(P_{0,+},\hat{\underline{\mu}}(x,t_{\underline{r}})))}
       {\theta(\underline{\xi}(P_{0,+},\hat{\underline{\mu}}(x_0,t_{0,\underline{r}})))}\nonumber\\
       &~~~~~~~~~~~\times
       e^{i(x-x_0)(e_{0,-}-e_{0,+})+i(t_{\underline{r}}-t_{0,\underline{r}})
       (\widetilde{\Omega}_{\underline{r}}^{0,-}-\widetilde{\Omega}_{\underline{r}}^{0,+})},
       \label{solutionqt}\\
       &r(x,t_{\underline{r}})= r(x_0,t_{0,\underline{r}})
       \frac{\theta(\underline{\xi}(P_{0,-},\hat{\underline{\nu}}(x,t_{\underline{r}})))}
       {\theta(\underline{\xi}(P_{0,-},\hat{\underline{\nu}}(x_0,t_{0,\underline{r}})))}
       \frac{\theta(\underline{\xi}(P_{0,+},\hat{\underline{\nu}}(x_0,t_{0,\underline{r}})))}
       {\theta(\underline{\xi}(P_{0,+},\hat{\underline{\nu}}(x,t_{\underline{r}})))}\nonumber\\
       &~~~~~~~~~~~\times e^{-i(x-x_0)(e_{0,-}-e_{0,+})-i(t_{\underline{r}}-t_{0,\underline{r}})
       (\widetilde{\Omega}_{\underline{r}}^{0,-}+\widetilde{\Omega}_{\underline{r}}^{0,+})},
       \label{solutionrt}\\
       &q(x_0,t_{0,\underline{r}})r(x_0,t_{0,\underline{r}})=
       \frac{\theta(\underline{\xi}(P_{0,-},\hat{\underline{\nu}}(x_0,t_{0,\underline{r}})))}
       {\theta(\underline{\xi}(P_{0,-},\hat{\underline{\mu}}(x_0,t_{0,\underline{r}})))}
       \frac{\theta(\underline{\xi}(P_{0,+},\hat{\underline{\mu}}(x_0,t_{0,\underline{r}})))}
       {\theta(\underline{\xi}(P_{0,+},\hat{\underline{\nu}}(x_0,t_{0,\underline{r}})))}\nonumber\\
       &~~~~~~~~~~~\times e^{\omega_0^{0,-}-\omega_{0}^{0,+}}.\label{qrt}
      \end{align}

\end{thm4.6}
\proof
As in the corresponding stationary case we
temporarily assume
\begin{equation}\label{4.35}
  \mu_j(x,t_{\underline{r}})\neq\mu_{j^\prime}(x,t_{\underline{r}}),\,\,\text{for}\,\,j\neq j^{\prime},\,\,
  (x,t_{\underline{r}})\in\widetilde{\Omega}
\end{equation}
for appropriate $\widetilde{\Omega}\subseteq\Omega$ and define the right-hand side of (\ref{4.34})
to be $\widetilde{\psi}_1.$ We
intend to prove $\widetilde{\psi}_1=\psi_1,$
where $\psi_1$ is given in (\ref{4.12}). For that purpose we first
investigate the local zeros and poles of $\psi_1$ and note
\begin{align}\label{4.36}
   q_{x^\prime}(x^\prime,t_{\underline{r}})\xi\phi(P,x^\prime,t_{\underline{r}})\underset{P\rightarrow \hat{\mu}_j(x^\prime,t_{\underline{r}})}{=}&
    \frac{2iy(\hat{\mu}_j(x,t_{\underline{r}}))}
   { \prod_{k=1,k\neq j}^{n}(\mu_j(x^\prime,t_{\underline{r}})-\mu_k(x^\prime,t_{\underline{r}}))}\nonumber\\
   &\times\frac{1}{\xi-\mu_j(x^\prime,t_{\underline{r}})}+O(1)\nonumber\\
   \underset{P\rightarrow \hat{\mu}_j(x^\prime,t_{\underline{r}})}{=}&\partial_{x^\prime}\ln(\xi-\mu_j(x^\prime,t_{\underline{r}}))+O(1).
 \end{align}
\begin{align}\label{4.37}
  -\widetilde{F}_{\underline{r}}(\xi,x_0,s)\xi\phi(P,x_0,s)\underset{P\rightarrow \hat{\mu}_j(x_0,s)}{=}&
    \frac{2iy(\hat{\mu}_j(x_0,s))}
   {\prod_{k=1,k\neq j}^{n}(\mu_j(x_0,s)-\mu_k(x_0,s))}\nonumber\\
   &\times\frac{1}{\xi-\mu_j(x_0,s)}\times
   \frac{\widetilde{F}_{\underline{r}}(x_0,s)}{q_{x}(x_0,s)}+O(1)\nonumber\\
   \underset{P\rightarrow \hat{\mu}_j(x_0,s)}{=}&\partial_{s}\ln(\xi-\mu_j(x_0,s))+O(1),
 \end{align}
using (\ref{4.8}), (\ref{4.10}), (\ref{4.20}) and (\ref{4.20a}). Thus
\begin{flalign}
&\psi_1(P,x,x_0,t_r,t_{0,\underline{r}})\nonumber\\
=&\begin{cases}
(\xi-\mu_j(x,t_r))O\left(1\right),& \text{as} \quad P\rightarrow \hat{\mu}_j(x,t_{\underline{r}})\neq \hat{\mu}_j(x_0,t_{0,\underline{r}}),\cr
O\left(1\right),&\text{as}\quad P\rightarrow \hat{\mu}_j(x,t_{\underline{r}})=\hat{\mu}_j(x_0,t_{0,\underline{r}}),\cr
\left(\xi-\mu_j(x_0,t_{0,\underline{r}})\right)^{-1}O\left(1\right),&\text{as} \quad P\rightarrow
\hat{\mu}_j(x_0,t_{0,\underline{r}})\neq \hat{\mu}_j(x,t_{r}),
\end{cases}\nonumber\\
&~~~~~~~~~~~~~~~~~~~~~~~~~~~~~~~~~~~~~~~~~~~~~~~
~~~~~~~~~~~~~~~~P=(\xi,y)\in\cur,\label{4.38}
\end{flalign}
with $O(1)\neq 0$ and hence $\psi_1$
and $\widetilde{\psi}_1$ have identical zeros and poles on
$\cur\backslash\{P_{\infty\pm},P_{0,\pm}\}$ which are all simple.
It remain to study the behavior of $\psi_1$ near $P_{\infty\pm},P_{0,\pm}.$
One infers from $(\ref{4.24})$, $(\ref{4.25}), (\ref{4.34})$ that
$\widetilde{\psi}_1$ and $\psi_1$ have the same essential singularities at
$P_{\infty\pm},P_{0,\pm}$
and the Riemann-Roch-type
uniqueness result \cite{15} proves that $\psi_1$ and $\widetilde{\psi}$
coincide up to
normalization.
This proves (\ref{4.34}) for some $C(x,x_0,t_{\underline{r}},t_{0,\underline{r}})\in C^{\infty}(\mathbb{R}^4).$ The expression (\ref{4.11})
for the divisor $\phi$ then yields
\begin{equation}\label{4.39}
\phi(P,x,t_{\underline{r}})=~C(x,t_{\underline{r}})
       \frac{\theta(\underline{\xi}(P,\hat{\underline{\nu}}(x,t_{\underline{r}})))}
       {\theta(\underline{\xi}(P,\hat{\underline{\mu}}(x,t_{\underline{r}})))}\exp\left(\int_{Q_0}^P
       \Omega_{P_{0,-},P_{\infty+}}^{(3)}\right),
\end{equation}
where $C(x,t_{\underline{r}})$ is in dependent of  $P\in\cur.$
Hence (\ref{4.23}) implies
\begin{align}
  \frac{1}{q(x,t_{\underline{r}})}=&C(x,t_{\underline{r}})\frac{\theta(\underline{\xi}
  (P_{0,+},\hat{\underline{\nu}}(x,t_{\underline{r}})))}
       {\theta(\underline{\xi}(P_{0,+},\hat{\underline{\mu}}(x,t_{\underline{r}})))}e^{\omega_{0}^{0,+}},\label{qq1}\\
     r(x,t_{\underline{r}})=&C(x,t_{\underline{r}})\frac{\theta(\underline{\xi}(P_{0,-},\hat{\underline{\nu}}(x,t_{\underline{r}})))}
       {\theta(\underline{\xi}(P_{0,-},\hat{\underline{\mu}}(x,t_{\underline{r}})))}e^{\omega_{0}^{0,-}},\label{qq2} \\
       q(x,t_{\underline{r}})r(x,t_{\underline{r}})=&\frac{\theta(\underline{\xi}(P_{0,-},\hat{\underline{\nu}}(x,t_{\underline{r}})))}
       {\theta(\underline{\xi}(P_{0,-},\hat{\underline{\mu}}(x,t_{\underline{r}})))}
       \frac{\theta(\underline{\xi}(P_{0,+},\hat{\underline{\mu}}(x,t_{\underline{r}})))}
       {\theta(\underline{\xi}(P_{0,+},\hat{\underline{\nu}}(x,t_{\underline{r}})))}e^{\omega_0^{0,-}-\omega_{0}^{0,+}}. \label{4.40}
\end{align}
The asymptotic behavior (\ref{4.25}) of $\psi_1$ near $P_{0,\pm}$
then yields
\begin{align}
 \widetilde{\psi}_1(P,x,x_0,t_{\underline{r}},t_{0,\underline{r}})\underset{\zeta\rightarrow 0}{=}&C(x,x_0,t_{\underline{r}},t_{0,\underline{r}})\frac{\theta(\underline{\xi}(P_{0,-},
  \hat{\underline{\mu}}(x,t_{\underline{r}})))}
  {\theta(\underline{\xi}(P_{0,-},\hat{\underline{\mu}}(x_0,t_{0,\underline{r}})))}\label{4.41}\\
  &\times e^{-i(x-x_0)e_{0,-}-i(t_{\underline{r}}-t_{0,\underline{r}})\widetilde{\Omega}_{\underline{r}}^{ 0,-}}+O(\zeta)~~ \textrm{~as~~$P\rightarrow P_{0,-}$},\nonumber\\
  \widetilde{\psi}_1(P,x,x_0,t_{\underline{r}},t_{0,\underline{r}})\underset{\zeta\rightarrow 0}{=}&C(x,x_0,t_{\underline{r}},t_{0,\underline{r}})
  \frac{\theta(\underline{\xi}(P_{0,+},\hat{\underline{\mu}}(x,t_{\underline{r}})))}
  {\theta(\underline{\xi}(P_{0,+},\hat{\underline{\mu}}(x_0,t_{0,\underline{r}})))}\label{4.42}\\
  &\times e^{-i(x-x_0)e_{0,+}-i(t_{\underline{r}}-t_{0,\underline{r}})
  \widetilde{\Omega}_{\underline{r}}^{0,+}}+O(\zeta)~~
  \textrm{~as~~$P\rightarrow P_{0,+}$}.\nonumber
\end{align}
A comparison of
(\ref{4.25}), (\ref{4.41}) and (\ref{4.42}) then yields (\ref{ctxx}) and (\ref{solutionqt}).
(\ref{ctx}) follows from (\ref{solutionqt}) and (\ref{qq1}). (\ref{solutionrt}) is a consequence
of (\ref{ctx}) and (\ref{qq2}).
The linearization property of the Abel map in (\ref{3.52AA}) and (\ref{3.52BB})
a standard investigation of the differentials $\Omega_i(x,x_0,t_{\underline{r}},t_{0,\underline{r}})=d\ln (\psi_i(\cdot,x,x_0,t_{\underline{r}},t_{0,\underline{r}}))$, $i=1,2$ (c.f. \cite{11}).
\qed

\section{$n$-Dark Solitons}
In this section, we will link
the quasi-periodic solutions of FL hierarchy derived in section 4 with the $n$-dark solitons through a limiting procedure.

It is known that the solutions obtained after degeneration
of the hyperelliptic spectral curve depend on the ramification points of $\cur$ and different choices
may lead
to different solutions such as solitons, cuspons or peakons, breathers, etc. in some other integrable models. To derive the $n$-dark solitons of FL hierarchy, we degenerate the hyperelliptic curve $\cur$ of genus $n$
into a genus zero algebraic curve by pinching all $a_j$-cycles of the
associated Riemann surface (cf.\,\cite{16}).
We assume that the ramification points $E_m$ are ordered according to
\begin{equation*}
  \textrm{Re($E_j$)}\leqslant\textrm{Re($E_k$)},\quad j<k, \quad j,k=0,\ldots,2n+1,
\end{equation*}
and consider the limit
\begin{equation}\label{5.1ab}
E_{2m-1}, E_{2m}\rightarrow \alpha_m,\quad m=1,\ldots,n,
\end{equation}
where $\alpha_m\neq\alpha_k$ for $m\neq k.$
Putting $E_{0}=-\beta, E_{2n+1}=\beta$ with $\beta>0,$ one finds
\begin{equation}\label{5.1}
  \cur\rightarrow \cur^{0}: y^2=(\xi^2-\beta^2)\prod_{j=1}^{n}(\xi-\alpha_j)^2,
\end{equation}
where $\beta\neq \alpha_j,j=1,\ldots,n.$
Then the holomorphic differentials $\omega_j$
(cf.(\ref{3.36})),
\begin{align}
  \omega_j & =\sum_{\ell=1}^{n}c_j(\ell)\xi^{\ell-1}\Big(\prod_{m=0}^{2n+1}(\xi-E_m)\Big)^{-1/2} \nonumber\\
 &\rightarrow \sum_{\ell=1}^{n}c_j^0(\ell)\xi^{\ell-1}\Big((\xi^2-\beta^2)\prod_{m=1}^{n}(\xi-\alpha_m)^2\Big)^{-1/2}.\label{5.2}
\end{align}
Using the normalization condition $\mathlarger{\int}_{a_k}\omega_j=\delta_{jk}$
and
\begin{align}\label{5.3}
\int_{a_k}\omega_j&\rightarrow\int_{a_k}\sum_{\ell=1}^{n}c_j^0(\ell)\xi^{\ell-1}\Big(\sqrt{\xi^2-\beta^2}
\prod_{m=1}^{n}(\xi-\alpha_m)\Big)^{-1}\nonumber\\
&= 2\pi i\varphi_j(\alpha_k)\Big(\sqrt{\alpha_k^2-\beta^2}
\prod_{m=1,m\neq k}^{n}(\alpha_k-\alpha_m)\Big)^{-1},
\end{align}
one concludes
\begin{equation}\label{5.4}
  \varphi_j(\alpha_k)=\frac{1}{2\pi i}\delta_{j,k}\sqrt{\alpha_k^2-\beta^2}
\prod_{m=1,m\neq k}^{n}(\alpha_k-\alpha_m).
\end{equation}
Here we employ the notation
\begin{equation}\label{5.4a}
  \varphi_j(\xi)=\sum_{\ell=1}^{n}c_j^0(\ell)\xi^{\ell-1}.
\end{equation}
Especially, one obtains
\begin{align}\label{5.5}
c_j^0(n)=\frac{\sqrt{\alpha_j^2-\beta^2}}{2\pi i},\quad
 \varphi_j(\xi)=c_j^0(n)\prod_{m=1,m\neq j}^{n}(\xi-\alpha_m)
\end{align}
from (\ref{5.4}).
Comparing the coefficients (\ref{5.4a}) and (\ref{5.5})
yields
\begin{flalign*}
c_j^0(n-1)&=-c_j^0(n)\sum_{m\neq j}\alpha_m
=-\frac{\sqrt{\alpha_j^2-\beta^2}}{2\pi i}\sum_{m\neq j}\alpha_m,\\
c_j^0(n-2)&=c_j^0(n)\sum\limits_{\begin{smallmatrix}
m<n,\\ m,n\neq j\end{smallmatrix}}\alpha_m\alpha_n=\frac{\sqrt{\alpha_j^2-\beta^2}}{2\pi i}\sum\limits_{\begin{smallmatrix}
m<n,\\ m,n\neq j\end{smallmatrix}}\alpha_m\alpha_n,
\quad\textrm{etc.}
\end{flalign*}
Then using (\ref{5.5}), one finds
\begin{equation}\label{5.6}
  \omega_j\rightarrow \omega_j^0=\frac{\sqrt{\alpha_j^2-\beta^2}}{2\pi i}\Big(\sqrt{\xi^2-\beta^2}
 (\xi-\alpha_j)\Big)^{-1}.
\end{equation}
The elements of Riemann matrix $\tau=(\tau_{jk})$
\begin{align}\label{5.7}
 \tau_{jk}&=\int_{b_j}\omega_k\nonumber\\
&\rightarrow 2\int_{\alpha_j}^{\beta}\omega_k^{0}=\frac{i}{\pi }\ln\Big|\frac{\eta_j+\eta_k}{\eta_j-\eta_k}\Big|\equiv\tau_{jk}^{0}
\end{align}
with $\eta_k=(\alpha_k-\beta)^{1/2}(\alpha_k+\beta)^{-1/2}.$
 So for the diagonal elements of $\tau$,
 $\textrm{Re($i\tau_{kk}$)}\rightarrow -\infty$ in the limit (\ref{5.1ab}).
Then one can rewrite the Riemann theta function (\ref{theta}) as
\begin{align}\label{thetad}
 &\theta(\underline{z})=\sum_{\underline{k}\in\mathbb{Z}^n}\exp\left(2 \pi i\sum_{j=1}^nk_jz_j+2\pi i\sum_{j<m}\tau_{jm}k_jk_m+\pi i\sum_{j=1}^{n}\tau_{jj}k_j^2\right)\nonumber\\
 &=\sum_{\underline{k}\in\mathbb{Z}^n}\exp\left(2 \pi i\sum_{j=1}^nk_j\Big(z_j+\frac{1}{2}\tau_{jj}\Big)+2\pi i\sum_{j<m}\tau_{jm}k_jk_m
 +\pi i\sum_{j=1}^n\tau_{jj}k_j(k_j-1)\right)\nonumber\\
 &\thicksim\sum_{\underline{k}\in\{0,1\}^n}\exp\left(2 \pi i\sum_{j=1}^nk_j\Big(z_j+\frac{1}{2}\tau_{jj}\Big)+2\pi i\sum_{j<m}\tau_{jm}^{0}k_jk_m\right),\quad \underline{k}=(k_1,\ldots,k_n).\nonumber\\
 \end{align}

 \newtheorem{thm5.1a}{Theorem}[section]
 \begin{thm5.1a}
The vectors $\underline{U}_0^{(2)},\widetilde{\underline{U}}_{\underline{r}}^{(2)}$ in (\ref{3.52AA}), (\ref{3.52BB}) have an alternative
description:
\begin{flalign}
 \underline{U}_0^{(2)} =&(U_{0,1}^{(2)},\ldots,U_{0,n}^{(2)}),\quad
  U_{0,j}^{(2)}=2\sum_{\ell=1}^{n}c_j(\ell)\hat{c}_{\ell+1-n}(\underline{E}),\label{a6.1}\\
  \widetilde{\underline{U}}_{\underline{r}}^{(2)}
  =&( \widetilde{U}_{\underline{r},1}^{(2)},\ldots, \widetilde{U}_{\underline{r},n}^{(2)}),\nonumber\\
 \widetilde{U}_{\underline{r},j}^{(2)} =&2(-1)^n\prod_{m=0}^{2n+1}E_m^{-1/2}
   \sum_{s=1}^{r_-}\tilde{c}_{r-s,-}\sum_{\ell=1}^{2s}c_j(\ell)\hat{c}_{2s-\ell}(\underline{E}^{-1})\nonumber\\
 &+2\sum_{s=1}^{r_+}\tilde{c}_{r_+-s,+}\sum_{\ell=1}^{n}c_j(\ell)\hat{c}_{\ell+2s-n-1}(\underline{E}).\label{a6.3}
\end{flalign}
Accordingly, in the limit (\ref{5.1ab})
\begin{flalign}
\underline{U}_{0,j}^{(2)}\rightarrow  [\underline{U}_{0,j}^{(2)}]^0
 =&2\sum_{\ell=1}^{n}c_j^0(\ell)\hat{c}_{\ell+1-n}(\underline{E}^0),\label{a6.4}\\
\widetilde{U}_{\underline{r},j}^{(2)}\rightarrow    [\widetilde{U}_{\underline{r},j}^{(2)}]^0=&
2(-1)^{n+1}\beta^2\prod_{m=1}^{n}\alpha_m^{-1}
   \sum_{s=1}^{r_-}\tilde{c}_{r-s,-}\sum_{\ell=1}^{2s}c_j^0(\ell)\hat{c}_{2s-\ell}((\underline{E}^0)^{-1})\nonumber\\
 &+2\sum_{s=1}^{r_+}\tilde{c}_{r_+-s,+}\sum_{\ell=1}^{n}c_j^0(\ell)\hat{c}_{\ell+2s-n-1}(\underline{E}^0)\label{a6.5}
\end{flalign}
where we use the notation
\begin{flalign}
\underline{E}^0=(\beta,-\beta,\alpha_1,\ldots,\alpha_n).
\end{flalign}

 \end{thm5.1a}
 \proof
 We only prove (\ref{a6.3}) and the proof for (\ref{a6.1}) is similar (or cf. \cite{15}).
 One computes
 \begin{flalign}\label{p1}
  \omega_j&=\sum_{\ell=1}^{n}c_j(\ell)\eta_\ell=\sum_{\ell=1}^nc_j(\ell)\frac{\xi^{\ell-1}}{y(\xi)}d\xi\nonumber\\
  &=\sum_{\ell=1}^nc_j(\ell)\zeta^{\ell-1}\sum_{k=0}^{\infty}\bar{c}_k(\underline{E})\zeta^kd\zeta\nonumber\\
  &=(-1)^n\prod_{m=0}^{2n+1}E_m^{-1/2}\sum_{k=0}^\infty\left(\sum_{\ell=1}^{k+1}c_j(\ell)\bar{c}_{k+1-\ell}(\underline{E})\zeta^k\right)d\zeta
  \end{flalign}
  where
  \begin{equation*}
    \bar{c}_j(\underline{E})=\hat{c}_j(\underline{E}^{-1}).
  \end{equation*}
  Then the Bilinear Riemann Relation shows that
  %(cf.\cite{15}, the proof of (\ref{a2.45}) and (\ref{a2.46}))
\begin{flalign}
\frac{1}{2\pi i}\int_{b_j}\Omega^{(2)}_{P_{\infty+},2s-1}&=\frac{1}{2s}
\sum_{\ell=1}^{n}c_j(\ell)\hat{c}_{\ell+2s-n-1}(\underline{E}),\label{a2.45}\\
\frac{1}{2\pi i}\int_{b_j}\Omega^{(2)}_{P_{\infty-},2s-1}&=-\frac{1}{2s}
\sum_{\ell=1}^{n}c_j(\ell)\hat{c}_{\ell+2s-n-1}(\underline{E}),\label{a2.46}\\
\frac{1}{2\pi i}\int_{b_j}\Omega^{(2)}_{P_{0,+},2s-1}&=\frac{1}{2s}(-1)^n\prod_{m=0}^{2n+1}E_m^{-1/2}
\sum_{\ell=1}^{2s}c_j(\ell)\bar{c}_{2s-\ell}(\underline{E}),\\
\frac{1}{2\pi i}\int_{b_j}\Omega^{(2)}_{P_{0,-},2s-1}&=-\frac{1}{2s}(-1)^n\prod_{m=0}^{2n+1}E_m^{-1/2}
\sum_{\ell=1}^{2s}c_j(\ell)\bar{c}_{2s-\ell}(\underline{E}),
\end{flalign}
and hence
 \begin{flalign}
 \widetilde{\underline{U}}_{\underline{r},j}^{(2)}
 =&2(-1)^n\prod_{m=0}^{2n+1}E_m^{-1/2}\sum_{s=1}^{r_-}\tilde{c}_{r-s,-}\sum_{\ell=1}^{2s}c_j(\ell)
 \bar{c}_{2s-\ell}(\underline{E})\nonumber\\
 &+2\sum_{s=1}^{r_+}\tilde{c}_{r_+-s,+}\sum_{\ell=1}^{n}c_j(\ell)\hat{c}_{\ell+2s-n-1}(\underline{E})
 \in\mathbb{R},
~~~j=1,\ldots,n.
 \end{flalign}
 Finally,
the expressions (\ref{a6.4}) (\ref{a6.5}) follow by (\ref{a6.1}), (\ref{a6.3}).
 \qed\vspace{0.2cm}

In the following we calculate the limit values of the constants $\widetilde{\Omega}_{\underline{r}}^{0,+},e_{0,+}.$
We introduce the notations
\begin{flalign}
  &\mathcal{L}_n=\frac{(2n-1)!!}{2^n n!},\quad \mathcal{M}_0^m=1, \nonumber\\
 &\mathcal{M}_1^{m}=-\sum_{j=1}^{2s}m_j,\quad \mathcal{M}_n^m=(-1)^n\sum_{1\leq i_1<\ldots<i_n\leq 2s}m_{i_1}\ldots m_{i_n},\nonumber\\
 & m_j\in\mathbb{C},\quad  1\leq j\leq 2s, \quad s\in\mathbb{N},
\end{flalign}
and
the Abel differentials $\Omega_{P_{0,\pm},2s-1}^{(2)},\Omega_{P_{\infty\pm},2s-1}^{(2)}$ of second kind are
explicitly defined by
\begin{flalign}
\Omega_{P_{\infty\pm},2s-1}^{(2)}=&\pm\frac{\sum_{j=0}^{2s}
c_j(\underline{E})\xi^{n+2s-j}}{y(\xi)}d\xi+\sum_{i=1}^{n}c_{j,\pm}^{(2s-1)}\omega_j,\nonumber\\
 =&\pm\frac{\prod_{j=1}^{n+2s}(\xi-m_j^s)}{y(\xi)}d\xi,\\
\Omega_{P_{0,\pm},2s-1}^{(2)}=&\pm\frac{(-1)^{n+1}\prod_{m=0}^{2n+1}E_m^{1/2}\sum_{j=0}^{2s}c_j(\underline{E})
\xi^{-(2s+1)+j}}{y(\xi)}d\xi\nonumber\\
&+\sum_{i=1}^{n}\tilde{c}_{j,\pm}^{(2s-1)}\omega_j,\nonumber\\
 =&\pm\frac{\xi^{-(2s+1)}\prod_{j=1}^{n+2s}(\xi-\widetilde{m}_j^s)}{y(\xi)}d\xi,
\end{flalign}
where the constants $c_{j,\pm}^{(2s-1)},\tilde{c}_{j,\pm}^{(2s-1)},m_j,\widetilde{m}_j$ are determined by the normalization
conditions
\begin{flalign}
 \int_{a_j}\Omega_{P_{\infty\pm},2s-1}^{(2)}=0,\quad j=1,\ldots,n,\label{20136901}\\
 \int_{a_j}\Omega_{P_{0,\pm},2s-1}^{(2)}=0,\quad j=1,\ldots,n.\label{20136902}
\end{flalign}
Then we get
\begin{flalign}
0&=\int_{a_j}\Omega_{P_{\infty\pm},2s-1}^{(2)}=\int_{a_j}
\frac{\prod_{j=1}^{n+2s}(\xi-m_j^s)}{y(\xi)}d\xi\nonumber\\
&\overset{(\ref{5.1ab})}{\rightarrow}\int_{a_j}
\frac{\prod_{j=1}^{n+2s}(\xi-[m_j^s]^0)}{\sqrt{\xi^2-\beta^2}\prod_{j=1}^n(\xi-\alpha_j)}d\xi.
\end{flalign}
Hence $\alpha_j, j=1,\ldots,n$ are the roots of polynomials $\prod_{j=1}^{n+2s}(\xi-[m_j^s]^0)$
using a standard residue formula. Keeping $s\in\mathbb{N}$ fixed and assuming $[m_{2s+j}^s]^0=\alpha_j, j=1,\ldots,n,$ and
$\mathcal{M}_p=0$ for $p\in\mathbb{Z}\backslash [1,2s],$
one obtains
that
\begin{flalign}
 \int_{a_j}\Omega_{P_{\infty\pm},2s-1}^{(2)}&\overset{(\ref{5.1ab})}{\rightarrow}\pm \int_{a_j}\frac{\prod_{j=1}^{2s}(\xi-[m_j^s]^0)}{\sqrt{\xi^2-\beta^2}}d\xi\nonumber\\
 &\overset{\zeta\rightarrow 0}{=}\mp \int_{a_j}\zeta^{-(2s+1)}\Big(\sum_{i=1}^{2s} \mathcal{M}_i^{[m^s]^0}\zeta^i\Big)\Big(\sum_{j=0}^{\infty}\mathcal{L}_j\beta^{2j}\zeta^{2j}\Big)d\zeta\nonumber\\
 &=\mp \int_{a_j}\zeta^{-(2s+1)}\sum_{t=0}^{\infty}
 \Big(\sum_{j=0}^{\infty}\mathcal{M}_{t-2j}^{[m^s]^0}\mathcal{L}_j\beta^{2j}\Big)\zeta^t d\zeta,\,\nonumber\\
 &=0, \quad\xi=\zeta^{-1},\,P\rightarrow P_{\infty\pm}.\label{a5.28}
\end{flalign}
This leads to
\begin{flalign}\label{a5.29}
 \sum_{j=0}^{\infty}\mathcal{M}_{t-2j}^{[m^s]^0}\mathcal{L}_j\beta^{2j}=0,\quad t=1,\ldots,2s.
\end{flalign}
Therefore, $\mathcal{M}_{j}^{[m^s]^0}, j=1,\ldots, [\frac{t}{2}],$
can be derived from the recursion relation above. Explicitly,
\begin{flalign}
&\mathcal{M}_1^{[m^s]^0}=0,\nonumber\\
&\mathcal{M}_2^{[m^s]^0}=-\mathcal{L}_1\beta^2\nonumber\\
&\mathcal{M}_3^{[m^s]^0}=0,\nonumber\\
&\mathcal{M}_4^{[m^s]^0}=(\mathcal{L}_1^2-\mathcal{L}_2)\beta^4\nonumber,\,\,\textrm{etc.}
\end{flalign}
Solving the algebraic equation
$
\prod_{j=0}^{2s}\mathcal{M}_j^{[m^s]^0} z^{2s-j}=0,
$
one gets the explicit expressions for $[m_j^s]^0, j=1,\ldots,2s$, which depend on the parameter $\beta.$
Similarly,
\begin{flalign}
 \int_{a_j}\Omega_{P_{0,\pm},2s-1}^{(2)}&\overset{(\ref{5.1ab})}{\rightarrow} \pm \int_{a_j}\frac{\xi^{-(n+2s)}\prod_{j=1}^{2s}(\xi-[\widetilde{m}_j^s]^0)}{\sqrt{\xi^2-\beta^2}}d\xi\nonumber\\
 &\overset{\zeta\rightarrow 0}{=}\pm \int_{a_j}\zeta^{-(2s+1)}\Big(\sum_{i=1}^{2s} \mathcal{M}_i^{[\widetilde{m}^s]^0}\zeta^i\Big)\Big(\sum_{j=0}^{\infty}\mathcal{L}_j\beta^{2j}\zeta^{2j}\Big)d\zeta\nonumber\\
 &=\pm \frac{\prod_{m=1}^{2s}[\widetilde{m}_j^s]^0}{\sqrt{-\beta^2}}\int_{a_j}\zeta^{-(2s+1)}\sum_{t=0}^{\infty}
 \Big(\sum_{j=0}^{\infty}\mathcal{M}_{t-2j}^{[\widetilde{m}^s]^0}\mathcal{L}_j\beta^{-2j}\Big)\zeta^t d\zeta,\,\nonumber\\
 &=0, \quad\xi=\zeta, \quad P\rightarrow P_{0,\pm},\label{a5.30}
\end{flalign}
gives rise to
\begin{flalign}\label{a5.31}
 \sum_{j=0}^{\infty}\mathcal{M}_{t-2j}^{[\widetilde{m}^s]^0}\mathcal{L}_j\beta^{-2j}=0,\quad t=1,\ldots,2s.
\end{flalign}
and the explicit constants for $[\widetilde{m}_j^s]^0, j=1,\ldots,2s.$

In summary, we have the following conclusions for the constants $e_{0,\pm},\widetilde{\Omega}_{\underline{r}}^{0,\pm}$ and related limiting process.

\newtheorem{thm5.2a}[thm5.1a]{Theorem}
 \begin{thm5.2a}
 In the limit (\ref{5.1ab}),
 \begin{flalign}
 &\Omega_{P_{\infty\pm},2s-1}^{(2)}\rightarrow [\Omega_{P_{\infty\pm},2s-1}^{(2)}]^0=\pm
 \frac{\prod_{j=1}^{2s}(\xi-[m_j^s]^0)}{\sqrt{\xi^2-\beta^2}}d\xi,\label{a5.32}\\
 &\Omega_{P_{0,\pm},2s-1}^{(2)}\rightarrow [\Omega_{P_{0,\pm},2s-1}^{(2)}]^0=\pm
 \frac{\xi^{-(n+2s)}\prod_{j=1}^{2s}(\xi-[\widetilde{m}_j^s]^0)}{\sqrt{\xi^2-\beta^2}}d\xi,\label{a5.33}
 \end{flalign}
 where
 $[m_j^s]^0,[\widetilde{m}_j^s]^0$
 are $2s$ roots of polynomials
\begin{flalign}\label{a5.34}
&\prod_{j=0}^{2s}\mathcal{M}_j^{[m^s]^0} z^{2s-j}=0,\quad
\prod_{j=0}^{2s}\mathcal{M}_j^{[\widetilde{m}^s]^0} z^{2s-j}=0,
\end{flalign}
and $\mathcal{M}_j^{[m^s]^0},\mathcal{M}_j^{[\widetilde{m}^s]^0}$
are defined by (\ref{a5.29}) and (\ref{a5.31}), respectively.
Let $Q_0=(\beta,0)\in\cur$ and
$Q_1=(\beta_1,y(\beta_1))$ be a point
near $P_{0,+}$. Then
\begin{flalign}
\widetilde{\Omega}_{\underline{r}}^{0,+}\overset{(\ref{5.1ab})}{\rightarrow}
[\widetilde{\Omega}_{\underline{r}}^{0,+}]^0=&-4\sum_{s=1}^{r_-}s\widetilde{c}_{r_--s,-}
\frac{\prod_{m=1}^{2s}[\widetilde{m}_j^s]^0}
{\sqrt{-\beta^2}}\sum_{t=0}^{\infty}\sum_{j=0}^{\infty}\frac{1}{t+1}
\mathcal{M}_{t-2j}^{([m^s]^0)^{-1}}\mathcal{L}_j\beta^{-2j}
\beta_1^{t+1}\nonumber\\
&+4\sum_{s=1}^{r_+}s\widetilde{c}_{r_+-s,+}\frac{\prod_{j=1}^{2s}[m_j^s]^0}{\sqrt{-\beta^2}}
\sum_{t=0}^{\infty}\sum_{j=0}^{\infty}\frac{1}{t+1}\mathcal{M}_{t-2j}^{[m^s]^0}\mathcal{L}_j\beta^{-2j}
(\frac{1}{\beta_1})^{t+1},\label{a5.35}\\
 \widetilde{\Omega}_{\underline{r}}^{0,-}\overset{(\ref{5.1ab})}{\rightarrow}
[\widetilde{\Omega}_{\underline{r}}^{0,-}]^0=&-[\widetilde{\Omega}_{\underline{r}}^{0,+}]^0,\label{a5.36}\\
e_{0,+}\overset{(\ref{5.1ab})}{\rightarrow}
[e_{0,+}]^0=& 2\frac{\prod_{j=1}^{2}[m_j^1]^0}{\sqrt{-\beta^2}}
\sum_{t=0}^{\infty}\sum_{j=0}^{\infty}\frac{1}{t+1}\mathcal{M}_{t-2j}^{[m^s]^0}\mathcal{L}_j\beta^{-2j}
(\frac{1}{\beta_1})^{t+1},\label{a5.37a}\\
e_{0,-}\overset{(\ref{5.1ab})}{\rightarrow}
[e_{0,-}]^0=&-[e_{0,+}]^0.\label{a5.38a}
\end{flalign}

 \end{thm5.2a}
\proof (\ref{a5.32})-(\ref{a5.34}) follow by
(\ref{a5.28})-(\ref{a5.31}). We only need to consider
the asymptotic behavior near $P_{0,+}, P_{\infty+}$
since
\begin{equation}\label{a5.37}
\int_{Q_0}^P\Omega_{P_{\infty\pm},2s-1}^{(2)}=-\int_{Q_0}^{P^*}\Omega_{P_{\infty\pm},2s-1}^{(2)},\quad
\int_{Q_0}^P\Omega_{P_{0,\pm},2s-1}^{(2)}=-\int_{Q_0}^{P^*}\Omega_{P_{0,\pm},2s-1}^{(2)},
\end{equation}
($Q_0$ is a branch point).
Using (\ref{a5.28})-(\ref{a5.34}), one obtains
\begin{flalign}
\int_{Q_0}^P[\Omega_{P_{\infty\pm},2s-1}^{(2)}]^0 =&\pm\int_{Q_0}^P
\frac{\prod_{j=1}^{2s}(\xi-[m_j^s]^0)}{\sqrt{\xi^2-\beta^2}}d\xi \nonumber \\
\overset{\zeta\rightarrow 0}{=}&\mp\frac{1}{2s}\zeta^{-2s}\pm\sum_{k=1}^\infty
\sum_{j=0}^{\infty}\frac{1}{k}\mathcal{M}_{2s+k-2j}^{[m^s]^0}\mathcal{L}_j\beta^{2j-k}+O(\zeta),\nonumber\\
&~~~~~~~~~~~~~~~~~~~~~~~~~~~\quad\xi=\zeta^{-1},\,P\rightarrow P_{\infty+},\\
\int_{Q_0}^P[\Omega_{P_{\infty\pm},2s-1}^{(2)}]^0 =&\pm\int_{Q_0}^P
\frac{\prod_{j=1}^{2s}(\xi-[m_j^s]^0)}{\sqrt{\xi^2-\beta^2}}d\xi \nonumber \\
\overset{\zeta\rightarrow 0}{=}&  \pm \frac{\prod_{j=1}^{2s}[m_j^s]^0}{\sqrt{-\beta^2}}
\sum_{t=0}^{\infty}\sum_{j=0}^{\infty}\frac{1}{t+1}\mathcal{M}_{t-2j}^{[m^s]^0}\mathcal{L}_j\beta^{-2j}
(\frac{1}{\beta_1})^{t+1}\nonumber\\
 &-\int_{\beta}^{\beta_1}\frac{\prod_{j=1}^{2s}(\xi-[m_j^s]^0)}
 {\sqrt{\xi^2-\beta^2}}d\xi+O(\zeta),\nonumber\\
&~~~~~~~~~~~~~~~~~~~~~~~~~~~\quad\xi=\zeta,\,P\rightarrow P_{0,+},\label{a5.37}\\
\int_{Q_0}^P[\Omega_{P_{0,\pm},2s-1}^{(2)}]^0=&\pm \int_{Q_0}^P\frac{\xi^{-(2s+1)}\prod_{j=1}^{2s}(\xi-[\widetilde{m}_j^s]^0)}{\sqrt{\xi^2-\beta^2}}d\xi\nonumber\\
\overset{\zeta\rightarrow 0}{=}& \mp \sum_{t=0}^{\infty}\frac{1}{t+2}(\frac{1}{\beta})^{t+2}+O(\zeta),\nonumber\\
&~~~~~~~~~~~~~~~~~~~~~~~~~~~\quad\xi=\zeta^{-1},\,P\rightarrow P_{\infty+},\\
\int_{Q_0}^P[\Omega_{P_{0,\pm},2s-1}^{(2)}]^0=&\pm \int_{Q_0}^P\frac{\xi^{-(2s+1)}\prod_{j=1}^{2s}(\xi-[\widetilde{m}_j^s]^0)}{\sqrt{\xi^2-\beta^2}}d\xi\nonumber\\
\overset{\zeta\rightarrow 0}{=}& \mp \frac{1}{2s}\zeta^{-2s}\mp \frac{\prod_{m=1}^{2s}[\widetilde{m}_j^s]^0}
{\sqrt{-\beta^2}}\sum_{t=0}^{\infty}\sum_{j=0}^{\infty}\frac{1}{t+1}
\mathcal{M}_{t-2j}^{([m^s]^0)^{-1}}\mathcal{L}_j\beta^{-2j}
\beta_1^{t+1}\nonumber\\
&+\int_{\beta}^{\beta_1}\frac{\xi^{-(2s+1)}\prod_{j=1}^{2s}
(\xi-[\widetilde{m}_j^s]^0)}{\sqrt{\xi^2-\beta^2}}d\xi+O(\zeta),\nonumber\\
&~~~~~~~~~~~~~~~~~~~~~~~~~~~\quad\xi=\zeta,\,P\rightarrow P_{0,+}.\label{a5.39}
\end{flalign}
(\ref{a5.35}) is a consequence of (\ref{4.32}), (\ref{4.32b}),(\ref{a5.37}) and (\ref{a5.39}).
(\ref{a5.37a}) follows from (\ref{3.44}), (\ref{3.45a}) and (\ref{a5.37}).
Also (\ref{a5.36}) and (\ref{a5.38a}) hold by (\ref{a5.37}).

\qed

 %Next we choose $Q_0=(\beta,0)$ as a base point of Abel map $\underline{\alpha}_{Q_0}(\cdot)$ and
 Let
 $\tau_{Q_0 P_{0,+}}$ be
 the integration path from $Q_0$ to $P_{0,+}$ which completely lies on the sheet 1 (containing $P_{0,+}$) of
 the Riemann surface $\cur$. Then
 \begin{flalign}
  \underline{\alpha}_{Q_0,j}(P_{0,+})=&\int_{\small \tau_{Q_0 P_{0,+}}}\omega_j\nonumber\\
  \rightarrow & \frac{\sqrt{\alpha_j^2-\beta^2}}{2\pi i}\int_{\beta}^0\frac{1}{\sqrt{\xi^2-\beta^2}
 (\xi-\alpha_j)}d\xi \nonumber\\
 =&\frac{i}{2\pi}\ln\Big|
 \frac{i+\eta_j}{i-\eta_j}\Big|
 \equiv  [\underline{\alpha}_{Q_0,j}(P_{0,+})]^0,
 \end{flalign}
 or more generally,
 for any $P=(\xi_0,y(\xi_0)),\xi_0\in\mathbb{C}$ on sheet 1,
 \begin{flalign}
  \underline{\alpha}_{Q_0,j}(P)=&\int_{Q_0}^P\omega_j\nonumber\\
  \rightarrow & \frac{\sqrt{\alpha_j^2-\beta^2}}{2\pi i}\int_{\beta}^{\xi_0}\frac{1}{\sqrt{\xi^2-\beta^2}
 (\xi-\alpha_j)}d\xi \nonumber\\
 =&\frac{i}{2\pi}\ln\Big|
 \frac{\sqrt{\frac{\xi_0-\beta}{\xi_0+\beta}}+\eta_j}{\sqrt{\frac{\xi_0-\beta}{\xi_0+\beta}}-\eta_j}\Big|
 \equiv  [\underline{\alpha}_{Q_0,j}(P)]^0.
 \end{flalign}
To obtain reasonable solutions
 we assume
 \begin{flalign}
   \underline{\Xi}_{Q_0,j}=\frac{1}{2}B_{jj}+\varepsilon_j
 \end{flalign}
 to hold where $\varepsilon_j, j=1,\ldots,n$
 are supposed to be chosen arbitrarily
 but to be invariant with respect to variations of $E_j$, for example,
 \begin{flalign}
 \varepsilon_j=
 \frac{1}{2}-2\pi i\sum_{\ell=1,\ell\neq j}^n\textrm{Res}_{\xi=\alpha_\ell }\Big(\int_{\beta}^\xi\frac{1}{\sqrt{(\xi^2-\beta^2)((\xi^\prime)^2-\beta^2)}
 (\xi-\alpha_\ell)(\xi^\prime-\alpha_j)}d\xi^\prime\Big).
 \end{flalign}
Similar to the Cauchy problem discussed in section 4, we are interested in soliton solutions
$q,r$ of
\begin{flalign}\label{a6.6}
  \textrm{$\widetilde{\text{FL}}$}_{\underline{r}}(q,r)=0,\quad &(q,r)|_{t_{\underline{r}}=t_{0,\underline{r}}}=(q^{(0)},r^{(0)}),\quad
  \end{flalign}
with $q^{(0)},r^{(0)}$ satisfying
\begin{flalign}
\textrm{s-FL}_{\underline{n}}(q^{(0)},r^{(0)})&=0,
\end{flalign}
or equivalently,
\begin{flalign}
 q^{(0)}(x)=&q(x_0)\frac{\textrm{det}(\delta_{ik}+\frac{2\eta_j}
 {\eta_j+\eta_k}e^{\pi i(\Lambda_j^0(P_{0,+})+\Lambda_k^0(P_{0,+}))})}
  {\textrm{det}(\delta_{ik}+\frac{2\eta_j}{\eta_j+\eta_k}e^{\pi i(\Lambda_j^0(P_{0,-})+\Lambda_k^0(P_{0,-}))})}
  e^{i(x-x_0)([e_{0,-}]^0-[e_{0,+}]^0)},\label{5.71}\\
r^{(0)}(x)=&r(x_0)\frac{\textrm{det}(\delta_{ik}+\frac{2\eta_j}
  {\eta_j+\eta_k}e^{\pi i(\Lambda_j^0(P_{0,+})+\Lambda_k^0(P_{0,+}))})}
   {\textrm{det}(\delta_{ik}+\frac{2\eta_j}{\eta_j+\eta_k}e^{\pi i
   (\Lambda_j^0(P_{0,-})+\Lambda_k^0(P_{0,-}))})}
   e^{-i(x-x_0)([e_{0,-}]^0-[e_{0,+}]^0)}.
\end{flalign}
%\begin{align}
  %\textrm{$\widetilde{\text{FL}}$}_{\underline{n}}(q,r)=&
  %   \left(
 %      \begin{array}{c}
  %       q_{xt_{\underline{r}}}+ f_{r_{+}-1,+,x}-2iq_xg_{r_-,-}
  %           +2if_{r_--1,-} \\
  %       r_{xt_{\underline{r}}}-h_{r_+-1,+,x}
  %           +2ih_{r_--1,-}
  %          +2ir_xg_{r_-,-}
  %     \end{array}
  %   \right)=0,\label{5.70}\\
    % &~~~~~~~~~~~~~~~~~~~~~~~~~~~~~~
     %&~~\quad t_{\underline{n}}\in\mathbb{R},~\underline{n}=(n_-,n_+)\in\mathbb{N}_0^2,\nonumber\\
  %   &(q,r)|_{t_{\underline{r}}=t_{0,\underline{r}}}=(q^{(0)},r^{(0)}),\\
% \textrm{s-FL}_{\underline{n}}(q^{(0)},r^{(0)})&=
%  \left(
  %   \begin{array}{c}
   %    f_{n_+-1,+,x}-2iq_x^{(0)}g_{n_-,-}+2if_{n_--1,-}  \\
   %    -h_{n_+-1,+,x}+2ih_{n_--1,-}+2ir_x^{(0)}g_{n_-,-}  \\
   % \end{array}
 % \right)=0, \label{5.71Q}
%\end{align}
%with
%\begin{align}
% q^{(0)}(x)=&q(x_0)\frac{\textrm{det}(\delta_{ik}+\frac{2\eta_j}
 % {\eta_j+\eta_k}e^{\pi i(\Lambda_j^0(P_{0,+})+\Lambda_k^0(P_{0,+}))})}
 %  {\textrm{det}(\delta_{ik}+\frac{2\eta_j}{\eta_j+\eta_k}e^{\pi i(\Lambda_j(P_{0,-})+\Lambda_k(P_{0,-}))})}
 %  e^{i(x-x_0)(e_{0,-}-e_{0,+})},\label{5.71}\\
% r^{(0)}(x)=&r(x_0)\frac{\textrm{det}(\delta_{ik}+\frac{2\eta_j}
 %  {\eta_j+\eta_k}e^{\pi i(\Lambda_j^0(P_{0,+})+\Lambda_k^0(P_{0,+}))})}
 %  {\textrm{det}(\delta_{ik}+\frac{2\eta_j}{\eta_j+\eta_k}e^{\pi i
  % (\Lambda_j^0(P_{0,-})+\Lambda_k^0(P_{0,-}))})}
  % e^{-i(x-x_0)(e_{0,-}-e_{0,+})}.\label{5.72}
 %\end{align}
Here and thereafter, we denote
\begin{align*}
 \Lambda_j^0(P)=&-\varepsilon_j
  +[\alpha_{Q_0,j}(P)]^0
  -[\alpha_{Q_0,j}(\mathcal{D}_{\underline{\hat{\mu}}(x_0)})]^0-i [\underline{U}_{0,j}^{(2)}]^0(x-x_0),\\
 % \Lambda_j(P)=&-1/2+\sum_{\ell=1,\ell\neq j}^{n}\int_{a_\ell}\omega_\ell(P)\int_{Q_0}^P\omega_j
 % +A_{Q_0,j}(P)\\
 % &-\alpha_{Q_0,j}(\mathcal{D}_{\underline{\hat{\mu}}(x_0,t_{0,\underline{r}})})-i U_{0,j}^{(2)}(x-x_0)
 % -i\widetilde{U}_{\underline{r},j}^{(2)}(t_{\underline{r}}-t_{0,\underline{r}}).
  \Lambda_j(P)=&-\varepsilon_j
  +[\alpha_{Q_0,j}(P)]^0
  -[\alpha_{Q_0,j}(\mathcal{D}_{\underline{\hat{\mu}}(x_0,t_{0,\underline{r}})})]^0-i [\underline{U}_{0,j}^{(2)}]^0(x-x_0)\\
  &-
  i[\widetilde{U}_{\underline{r},j}^{(2)}]^0(t_{\underline{r}}-t_{0,\underline{r}}),
\end{align*}
for $\forall P\in\cur\backslash\{Q_0\}$.
Then we have the following result.
\newtheorem{thm5.1}[thm5.1a]{Theorem}
\begin{thm5.1}
Assume $(\ref{2.1}),(\ref{2.2})$ and suppose that $(\ref{4.1})$ and $(\ref{4.2})$ hold
with respect to the constraint $(\ref{3.2})$
on $\Omega$, where
$\Omega\subseteq\mathbb{R}^2$ is open and connected.
Moreover, suppose that $\mathcal{D}_{\underline{\hat{\mu}}(x,t_{\underline{r}})}$, or equivalently,
$\mathcal{D}_{\underline{\hat{\nu}}(x,t_{\underline{r}})}$, is nonspecial for $(x,t_{\underline{r}})\in\Omega.$ Then for the Cauchy problem of FL hierarchy $(\ref{a6.6})$
we obtain the following $n$-dark soliton solutions
\begin{align}
 q(x,t_{\underline{r}})=&q(x_0,t_{0,\underline{r}})\frac{\textrm{\emph{det}}(\delta_{ik}+\frac{2\eta_j}
  {\eta_j+\eta_k}e^{\pi i(\Lambda_j(P_{0,+})+\Lambda_k(P_{0,+}))})}
   {\textrm{\emph{det}}(\delta_{ik}+\frac{2\eta_j}{\eta_j+\eta_k}e^{\pi i(\Lambda_j(P_{0,-})+\Lambda_k(P_{0,-}))})}\nonumber\\
   &\times
   e^{i(x-x_0)([e_{0,-}]^0-[e_{0,+}]^0)+i(t_{\underline{r}}-t_{0,\underline{r}})
       ([\widetilde{\Omega}_{\underline{r}}^{0,-}]^0+[\widetilde{\Omega}_{\underline{r}}^{0,+}]^0)},\label{5.75}\\
 r(x,t_{\underline{r}})=&r(x_0,t_{0,\underline{r}})\frac{\textrm{\emph{det}}(\delta_{ik}+\frac{2\eta_j}
   {\eta_j+\eta_k}e^{\pi i(\Lambda_j(P_{0,+})+\Lambda_k(P_{0,+}))})}
   {\textrm{\emph{det}}(\delta_{ik}+\frac{2\eta_j}{\eta_j+\eta_k}e^{\pi i
   (\Lambda_j(P_{0,-})+\Lambda_k(P_{0,-}))})}\nonumber\\
   &\times
   e^{-i(x-x_0)([e_{0,-}]^0-[e_{0,+}]^0)-i(t_{\underline{r}}-t_{0,\underline{r}})
       ([\widetilde{\Omega}_{\underline{r}}^{0,-}]^0+[\widetilde{\Omega}_{\underline{r}}^{0,+}]^0)}.\label{5.76}
 \end{align}

\end{thm5.1}
\proof
It suffices to consider the limit (\ref{5.1ab}) of (\ref{solutionqt}), (\ref{solutionrt}).
Using (\ref{thetad}), the symmetric property $\theta(\underline{z})=\theta(-\underline{z})$ and
the formula \cite{22}
\begin{align*}
&\sum_{\underline{k}\in\{0,1\}^n}\exp\left(2 \pi i\sum_{j=1}^nk_jz_j+2\pi i\sum_{j<m}\tau_{jm}^{0}k_jk_m\right)=\textrm{det}{B},\\
&~~~~~~~B=(b_{ik})_{n\times n},\quad b_{ik}=\delta_{ik}+\frac{2\eta_i}{\eta_i+\eta_k}e^{\pi i(z_i+z_k)},
\end{align*}
one concludes (\ref{5.75}), (\ref{5.76}).
\qed

\newtheorem{rmk5.2}[thm5.1a]{Remark}
\begin{rmk5.2}
$(i)$
Taking fixed $\underline{r}=(1,1)$
and varying $\underline{n}\in\mathbb{N}^2\backslash\{(0,0)\},$
one finally derives the $n$-dark soliton solutions of FL equation $(\ref{1.1})$,
which are consistent with those by Darboux transformation method $\cite{8}$.

\noindent $(ii)$
In Theorem 5.1,
taking some fixed $\underline{r}\in\mathbb{N}^2\backslash\{(0,0)\},$
and varying $\underline{n}\in\mathbb{N}^2\backslash\{(0,0)\},$
we obtains the $n$-dark soliton solutions of the $\underline{r}$th equation in the FL hierarchy.

\noindent $(iii)$ The $n$-dark solitons of FL hierarchy in fact depend on $2n+2$ parameters $\beta, \alpha_1,\ldots,\alpha_n, \beta_1,\varepsilon_1,$ $\ldots,\varepsilon_n.$
\end{rmk5.2}

\section*{Acknowledgments}
The work described in this paper
was supported by grants from the National Science
Foundation of China (Project No.11271079), Doctoral Programs Foundation of
the Ministry of Education of China, and the Shanghai Shuguang Tracking Project (project 08GG01).


\begin{thebibliography}{99}


\bibitem{01}M. J. Ablowitz, D. J. Kaup, A. C. Newell and H. Segur, The inverse
scattering transform-Fourier analysis for nonlinear problems,
Stud.Appl.Math. 53, 249-315, (1974).

\bibitem{20} M. S. Alber, F. Fedorov, N. Yu, Algebraic geometrical solutions
for certain evolution equations and Hamiltionian flows on nonlinear subvarieties
of generalized Jacobians, Inverse Problems, 17, 1017-1042, (2001).


\bibitem{16} E. D. Belokolos, A. I. Bobenko, V.Z. Enol’skii, A.R. Its, and V.B.
Matveev, Algebro-Geometric Approach to Nonlinear Integrable Equations,
Springer, Berlin, (1994).

\bibitem{19} B. A. Dubrovin, Completely integrable Hamiltonian systems associated
with matrix operators and Abelian varieties, Funct. Anal. Appl. 11,
265-277, (1977).







\bibitem{1} A. S. Fokas, On a class of physically important integrable equations, Physica. D. 87, 145-150, (1995).










\bibitem{12} F. Gesztesy and R. Ratneseelan, An alternative approach
to algebro-geometric solutions of the AKNS hierarchy, Rev. Math. Phys.
10 345--391 (1998).
\bibitem{15} F. Gesztesy and H. Holden, Soliton Equations and their
Algebro-Geometric Solutions, Cambridge University Press, Cambridge,
(2003).





\bibitem{11a} J. S. He, S. W. Xu, and K. Porsezian, Rogue Waves of the Fokas-Lenells Equation,
J. Phys. Soc. Jpn. 81, 124007, (2012).

\bibitem{ii} Y. Hou, P. Zhao, E. G. Fan, Z. J. Qiao, Algebro-geometric solutions
for the Degasperis-Procesi hierarchy, SIAM  J. Math. Anal. (to appear).
\bibitem{ij} Y. Hou, E. G. Fan, P. Zhao, The algebro-geometric solutions for the Hunter-Saxton hierarchy,
Z. Angew. Math. Phys. (to appear).

\bibitem{16a}A. R. Its,  A. V. Rybin, and M. A. Sail, Exact integration
of nolinear schr$\ddot{o}$dinger equation, Thore. i. Mat. Fiz, 74, N.1, 29-45, (1988).

\bibitem{5} A. Kundu, Two-fold integrable hierarchy of nonholonomic deformation of the derivative nonlinear Schr$\ddot{o}$dinger
and the Lenells-Fokas equation, J. Math. Phys. 51, 022901,  (2010).
\bibitem{6} A. Kundu, Integrable twofold hierarchy of perturbed equations and application to optical soliton dynamics,
   Theo. Math. Phys. 167, 800-810, (2011).

\bibitem{21} C. Kalla, C. Klein, On the numerical evaluation of
algebro-geometric solutions to integrable equations, Nonlinearity, 25, 569, (2012).





\bibitem{18} I. M. Krichever, Integration of nonlinear equations by the methods of
algebraic geometry, Funct.Anal.Appl. 11, 12-26, (1977).

\bibitem{2} J. Lenells, Exactly solvable model for nonlinear pulse propagation in optical fibers, Stud. Appl. Math. 123, 215-232,(2009).

\bibitem{3} J. Lenells, A. S. Fokas, On a novel integrable generalization of the nonlinear Schr$\ddot{o}$dinger equation,
Nonlinearity. 22, 11-27, (2009).
\bibitem{4} J. Lenells, Dressing for a novel integrable generalization of the nonlinear Schr$\ddot{o}$dinger Equation,
J. Nonlinear. Sci. 20, 709-722, (2010).

\bibitem{17} D. Mumford, Tata Lectures on Theta II, Birkh$\ddot{a}$user, Boston, (1984).


\bibitem{22} Y. Matsuno, Multiperiodic and multisoliton solutions of a nonlocal nonlinear
 Schr$\ddot{o}$dinger equation for envelope waves, Psys. Lett. A, 278, 53, (2000).
\bibitem{7}Y. Matsuno, A direct method of solution for the Fokas-Lenells derivative nonlinear
Schr$\ddot{o}$dinger equation: I. Bright soliton solutions, J. Phys. A: Math. Theor. 45, 235202, (2011).
\bibitem{9}Y. Matsuno, A direct method of solution for the Fokas-Lenells derivative nonlinear
Schr$\ddot{o}$dinger equation: II. Dark soliton solutions, J. Phys. A: Math. Theor. 45, 475202, (2012).







\bibitem{11} S. Novikov, S. V. Manakov, L. P. Pitaevskii, and V. E. Zakharov, Theory of soltions,
Consultants Bureau, New York, (1984).







\bibitem{22a} E. Previato, Hyperelliptic curves and solitons, Ph.D. thesis, Harvard, (1983).
\bibitem{8}V. E. Vekslerchik, Lattice representation and dark solitons of the Fokas-Lenells equation, Nonlinearity, 24, 1165-1175, (2011).
\bibitem{10} O. C. Wright. III, Some homoclinic connections of a novel integrable generalized nonlinear Schr$\ddot{o}$dinger equation,
    Nonlinearity, 22, 2633-2643, (2009).













\end{thebibliography}
\end{document}